\documentclass[a4paper,11pt]{article}
\usepackage{graphicx}
\pdfoutput=1 

\usepackage{jcappub} 
\usepackage{mathrsfs}
\usepackage[T1]{fontenc} 
\usepackage{bbold}
\usepackage{url}
\usepackage{hyperref}
\usepackage[english]{babel}
\usepackage[utf8]{inputenc}
\usepackage{amsmath}
\usepackage{color}
\usepackage{amsfonts}
\usepackage{amssymb}
\usepackage{mathtools}
\usepackage{placeins}
\usepackage{float}
\usepackage{tabularx}
\usepackage{adjustbox}
\usepackage{caption}
\usepackage{subcaption}
\begin{document}
\title{\boldmath A novel way of constraining the $\alpha$-attractor chaotic inflation through Planck data}
\author{Arunoday Sarkar, Chitrak Sarkar}
\author{and Buddhadeb Ghosh}
\affiliation{Centre of Advanced Studies, Department of Physics, The University of Burdwan,\\Burdwan 713 104, India}


\emailAdd{sarkararunoday7@gmail.com}
\emailAdd{chitraksarkar@gmail.com}
\emailAdd{bghosh@phys.buruniv.ac.in}

\abstract{Defining a scale of $k$-modes of the quantum fluctuations during inflation through the dynamical horizon crossing condition $k = aH$ we go from the physical $t$ variable to $k$ variable and solve the equations of cosmological first-order perturbations self consistently, with the chaotic $\alpha$-attractor type potentials. This enables us to study the behaviour of $n_{s}$, $r$, $n_{t}$ and $N$ in the $k$-space. Comparison of our results in the low-$k$ regime with the Planck data puts constraints on the values of the $\alpha$ parameter through microscopic calculations. Recent studies had already put model-dependent constraints on the values of $\alpha$ through the hyperbolic geometry of a Poincar\'{e} disk: consistent with  both  the maximal supergravity model $\mathcal{N}=8$ and the minimal supergravity model $\mathcal{N}=1$, the constraints on the values of $\alpha$ are $\frac{1}{3}$, $\frac{2}{3}$, 1, $\frac{4}{3}$, $\frac{5}{3}$, 2, $\frac{7}{3}$. The minimal $\mathcal{N}=1$ supersymmetric cosmological models with $B$-mode targets, derived from these  supergravity models, predicted the values of $r$ between $10^{-2}$ and $10^{-3}$. Both in the $E$-model and the $T$-model potentials, we have obtained, in our calculations, the values of $r$ in this range for all the constrained values of $\alpha$ stated above, within      $68\%$ CL. Moreover, we have calculated $r$ for some other possible values of $\alpha$ both in low-$\alpha$ limit, using the formula $r=\frac{12\alpha}{N^{2}}$, and in the high-$\alpha$ limit, using the formula $r=\frac{4n}{N}$, for $n=2$ and $4$. With all such values of $\alpha$, our calculated results match with the Planck-2018 data with 68\% or near 95\% CL. }

\maketitle
\flushbottom

\section{Introduction}
\label{sec:intro}
The microscopic origin of the $\alpha$-attractor potentials can be traced back to the works of S. Ferrara, R. Kallosh, A. Linde and M. Porrati \cite{Ferrara:2013rsa} for the $\mathcal{N}=1$ minimal supergravity model of vector multiplet inflaton and its chiral version by R. Kallosh, A. Linde, D. Roest \cite{Kallosh:2013yoa}. Two choices  of the chaotic inflaton potential for superconformal generalization \cite{Kallosh:2013pby} of the model  corresponding to two values of the deformation parameter were proposed \cite{Kallosh:2013maa,Kallosh:2015lwa,Carrasco:2015pla} : (I) the $E$-model potential 
\begin{equation}
    V(\phi)=V_{0}(1-e^{-\sqrt{\frac{2}{3\alpha}}\phi})^n
    \label{fig:pot_E_model }
\end{equation}
and (II) the $T$-model potential 
\begin{equation}
    V(\phi)=V_{0}\tanh^n\frac{\phi}{\sqrt{6\alpha}}.
    \label{fig:pot_T_model}
\end{equation}
Here $\alpha$ is the inverse curvature of $SU(1,1)/U(1)$ K\"ahler manifold, 
\begin{equation}
    \alpha =-\frac{2}{3\mathcal{R}_{K}}.
    \label{eq:geometry}
\end{equation}
Later on, and recently, the $\alpha$-attractor potentials have been extensively applied in the study of cosmological inflation \cite{Kallosh:2013hoa,Carrasco:2015rva,Alho:2017opd,Maeda:2018sje,Chojnacki:2021fag,Rodrigues:2021olg} and also in preheating \cite{Krajewski:2018moi} and reheating \cite{Kanfon:2020yxx, Shojaee:2020xyr}. 
Also, the quintessential versions of the $\alpha$-attractors have been applied to explain the late time acceleration of the universe \cite{Dimopoulos:2017zvq,Garcia-Garcia:2018hlc,AresteSalo:2021lmp,AresteSalo:2021wgb} in the predictive study of the stage-IV galaxy surveys \cite{Akrami:2017cir,Garcia-Garcia:2019ees,Akrami:2020zxw}, as well as in the study of inflation with primodial black holes \cite{Dalianis:2018frf,Mahbub:2019uhl} and the superconformal $E$-models in brane inflation \cite{Sabir:2019wel}. The constraints of the single-field $\alpha$-attractor models are also being studied \cite{Canas-Herrera:2021sjs} in conjunction with the expected results of the upcoming CMBR experiments.\par
Constraining the value of $\alpha$ is one of the prime tasks in the frontier of future experiments. It is crucial indeed, as the parameter $\alpha$ is not only connected to a specific model, but also to the various physical aspects, such as, the geometry of the K\"ahler manifold (as in (\ref{eq:geometry})), behaviour of the boundary of moduli space \cite{Kallosh:2014rga}, the kinetic pole structure \cite{Galante:2014ifa}, asymptotic freedom of inflaton field \cite{Kallosh:2016gqp}, hyperbolic geometry \cite{Carrasco:2015rva,Kallosh:2015zsa,Carrasco:2015uma}, maximal supersymmetry \cite{Kallosh:2017ced}, modified gravity \cite {Odintsov:2016vzz}, and string theory (such as \cite{Scalisi:2018eaz,Kallosh:2017wku}). Therefore, the task of finding $\alpha$ is very important for its multi-funtional diversity (as discussed above). Besides the above points, the connection between $\alpha$ and the tensor-to-scalar ratio $r$ is most important (discussed in Section \ref{sec:mode analysis}). According to \cite{Kallosh:2019hzo,Kallosh:2019eeu,Canas-Herrera:2021sjs}, all upcoming experiments are going to improve the upper limit of $r$ through possible detection of the $B$-modes along with the other CMB parameters, which in turn will validate many single- and multi-field models including the $\alpha$-attractors. Although Planck-2018 \cite{Akrami:2018odb, Aghanim:2018eyx} has put an upper bound $\alpha\lesssim 10$ ($T$-modoel), $\alpha\lesssim 19$ ($E$-model), upcoming experiments will probably constrain $\alpha$ more precisely that  will provide more convincing evidences for all the $\alpha$-related phenomena and internal geometry of space-time. For instance, recent studies in $\mathcal{N}$=$1$ \cite{Ferrara:2016fwe} and $\mathcal{N}$=8 \cite{Kallosh:2017ced} supergravity models suggested some specific values of $\alpha$ \textit{viz.,} $\frac{1}{3}$, $\frac{2}{3}$, 1, $\frac{4}{3}$, $\frac{5}{3}$, 2, $\frac{7}{3}$.  

\par Motivated by these ideas, in this paper, we examine the values of $\alpha$ vis-\`{a}-vis the Planck-2018 data, by observing the transition from its lower to higher limit in the $n_s-r$ plane through the $k$-mode analysis by dynamical horizon exit method. By this method, we can make a transition from the $t$ variable to the $k$ variable through the horizon exit condition $k=aH=\dot{a}$, while  the Hubble sphere shrinks during inflation. In this way we shall consider only those  modes which will exist the horizon and these will be relevant for explaining the Planck data.  This point will be elaborated further in Section $\ref{sec:mode analysis}$. Considering the first order perturbation in the inflaton field and the metric and the quantum fluctuations in the inflaton field we set up non-linear coupled differential equations and solve them self-consistently in the sub-Planckian region to obtain the $k$-dependent power spectra, spectral indices and the number of e-folds. Our calculations have been done mainly through Wolfram Mathematica 12.\par The paper is organised in this way.
In Section \ref{sec:chaotic attractors}, we briefly review the chaotic $\alpha$-attractors relevant for our present study. We present in Section \ref{sec:mode analysis} the formalism and the calculational framework. Section \ref{sec:results} contains results and discussion. Finally, we conclude our analysis in Section \ref{sec:conclusions} by highlighting some future prospects.
\section{Chaotic \texorpdfstring{$\alpha$}{a}-attractors }
\label{sec:chaotic attractors}
Long ago, Linde \cite{Linde:1983gd,Linde:1986fd} suggested an inflationary scenario, based on chaotic initial conditions in the very early universe. He also argued that this scenario is quite realizable in the $\mathcal{N}=1$ supergravity theory, among other theories.\par Kung and Brandenberger \cite{Brandenberger:1990wu}  and Brandenberger, Feldman and Kung \cite{Brandenberger_1991} made Fourier analysis of the inflaton field and found that a single long wavelength mode can trigger chaotic inflation and is responsible for late-time evolution of the universe, whereas the energy density in short wavelength modes decay as radiation. In this sense, the chaotic inflation is an attractor phenomenon in the space of initial conditions, which is independent of perturbations in short wavelength matter field and gravitational field.\par
The idea of chaotic inflation further evolved \cite{Kallosh:2014xwa,Linde:2017pwt} within the framework of the $\alpha$-attractor models of supergravity. The $\alpha$-attractor is the most generic universal class of superconformal chaotic attractors in $\mathcal{N}=1$ minimal supergravity for chiral inflaton multiplet, having two attracting points corresponding to two extreme limits of $\alpha$ \cite{Linde:2014nna}.  
\begin{figure}[H]
\begin{subfigure}{0.50\textwidth}
  \centering
   \includegraphics[width=70mm,height=50mm]{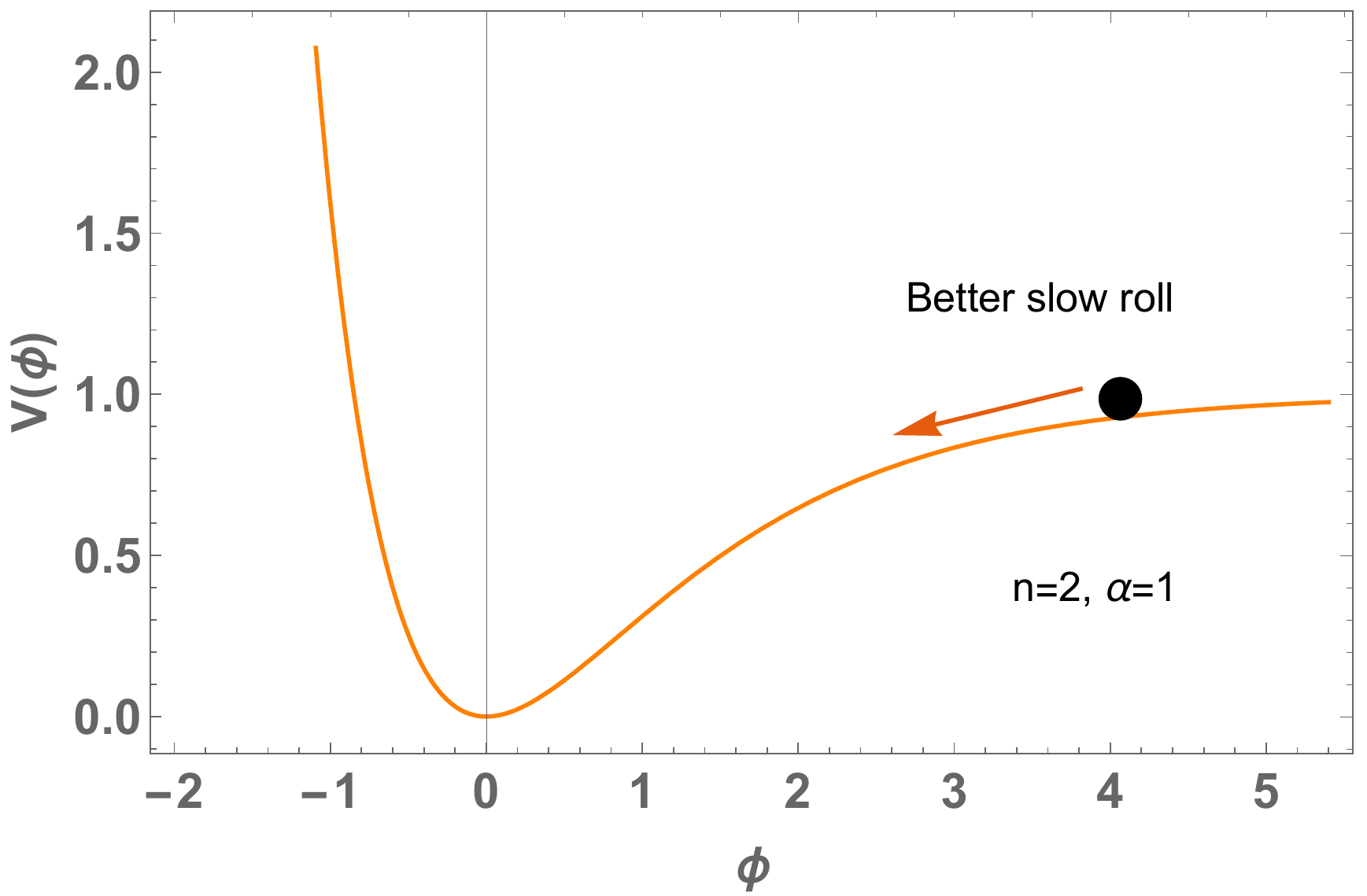}
   \subcaption{}
   \label{fig: sample_E_1}
\end{subfigure}%
\begin{subfigure}{0.50\textwidth}
  \centering
   \includegraphics[width=70mm,height=50mm]{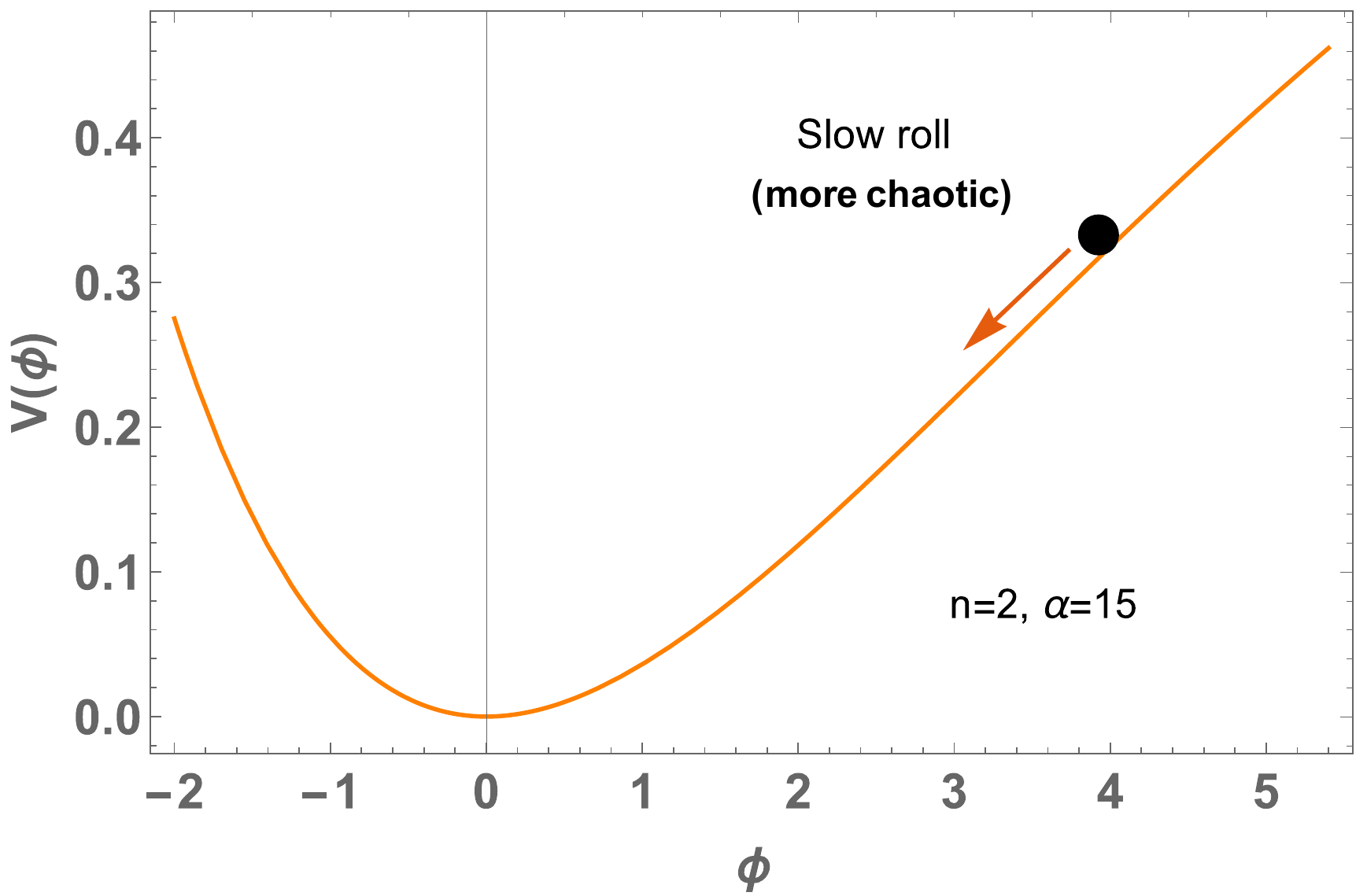}
   \subcaption{}
   \label{fig:sample_E_2}
\end{subfigure}
    \caption{The $\alpha$-attractor $E$-model potentials with  small and  large values of $\alpha$, for $V_{0}$=1. In both the cases inflation continues down the valley until the reheating starts. The inflaton field rolls down the valley slowly because of the Hubble friction (\cite{Linde:2014nna}). The potential energy at the same values of $\phi$, \textit{viz}., $\phi=4$, i) $V(\phi)=0.96$ for $\alpha=1$ (Figure \ref{fig: sample_E_1}) ii) $V(\phi)=0.33$ for $\alpha=15$ (Figure \ref{fig:sample_E_2}). At this value of $\phi$, potential energy dominates more in Figure \ref{fig: sample_E_1} than Figure \ref{fig:sample_E_2}. Thus, smaller value of $\alpha$ conforms to better slow roll. However, larger value of $\alpha$ means a more chaotic case.}
    \label{fig: example E model}
    \end{figure}
    \begin{figure}[H]
    \begin{subfigure}{0.50\textwidth}
  \centering
   \includegraphics[width=70mm,height=50mm]{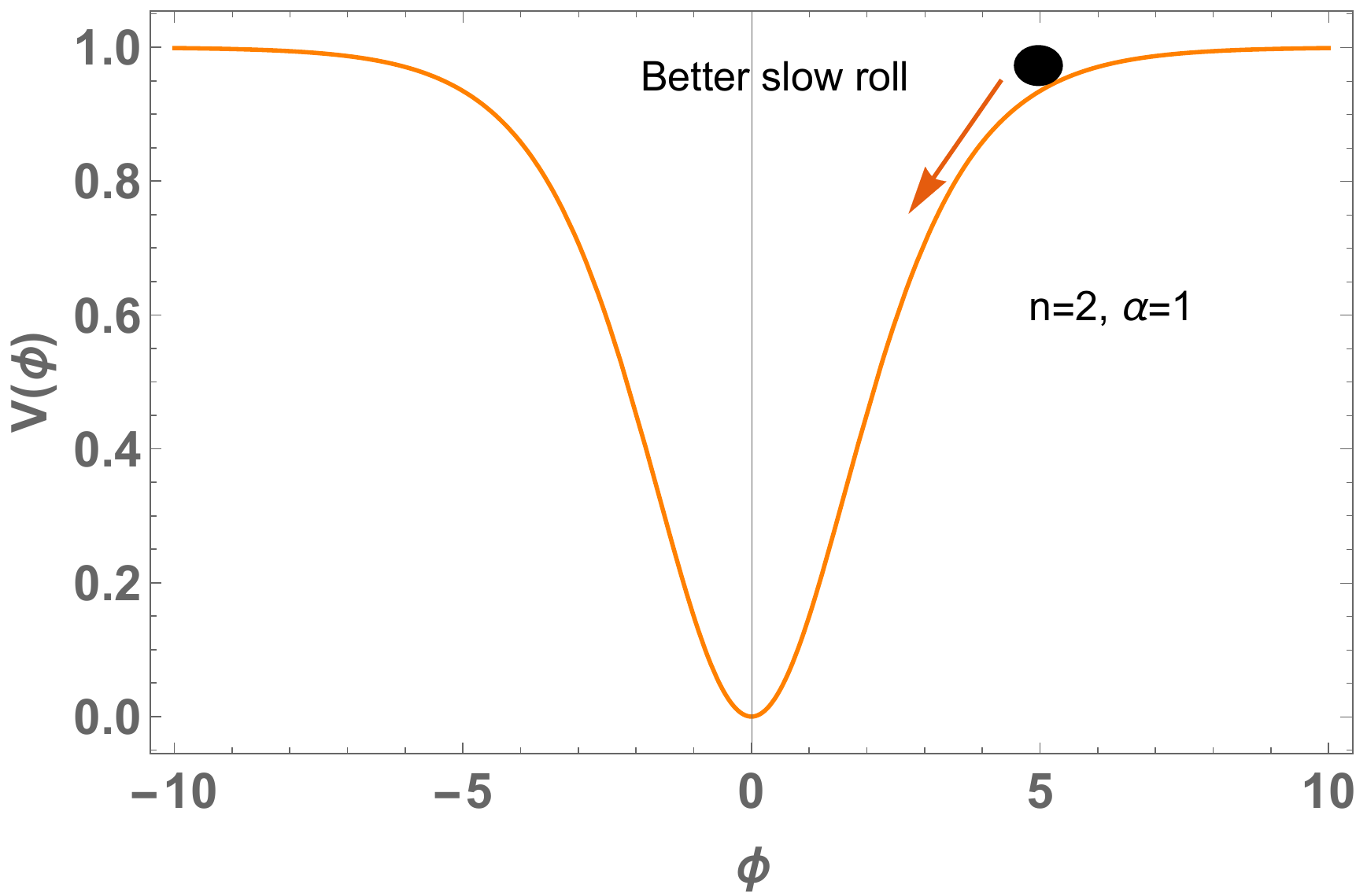}
   \subcaption{}
   \label{fig:2a}
\end{subfigure}
\begin{subfigure}{0.50\textwidth}
  \centering
   \includegraphics[width=70mm,height=50mm]{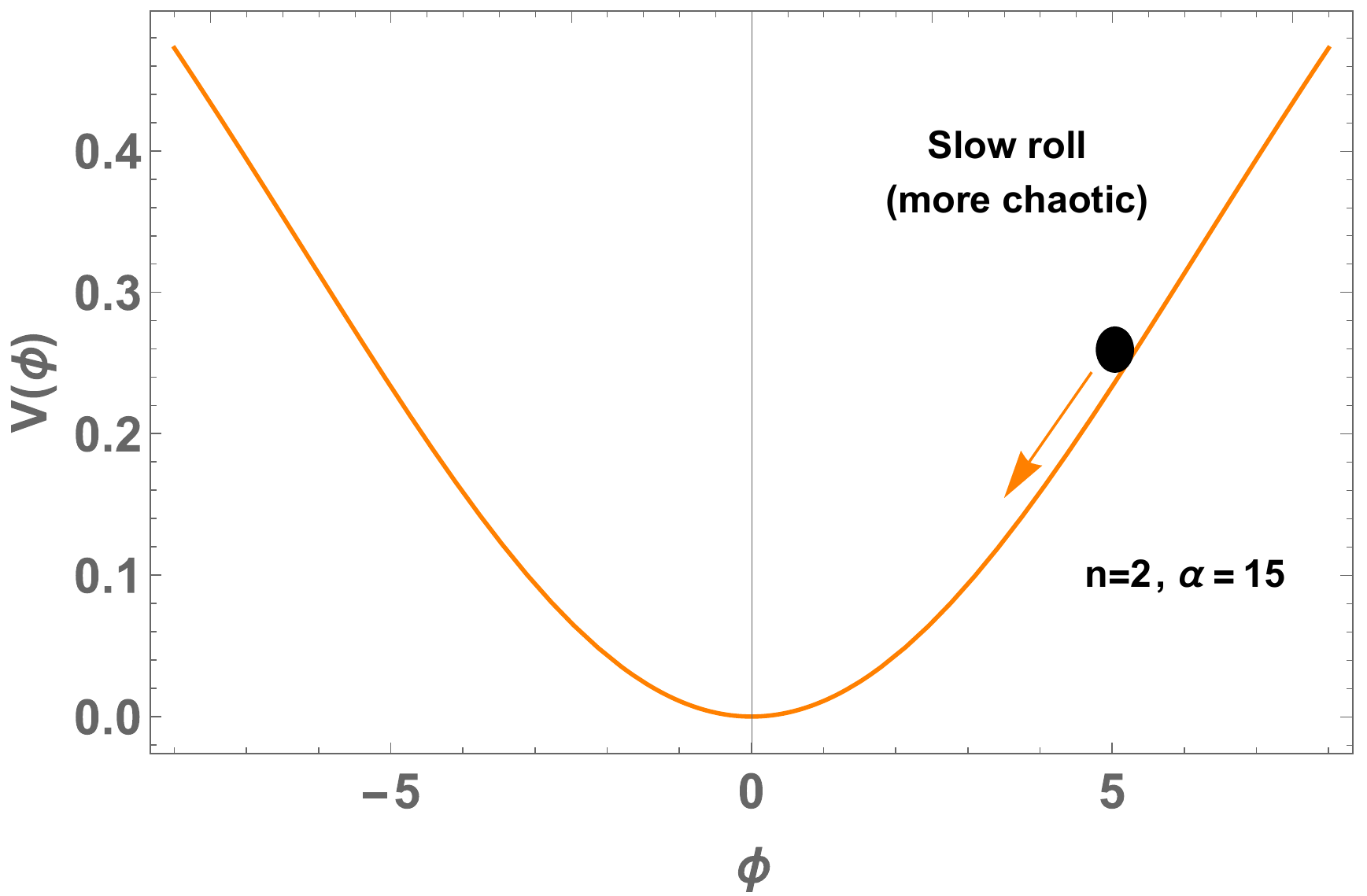}
   \subcaption{}
    \label{fig:2b}
\end{subfigure}
     \caption{ Comparison of the $\alpha$-attractor $T$-model potentials for small and large values of $\alpha$, for $V_{0}$=1. As in the case of the $E$-model, a small value of $\alpha$ (Figure \ref{fig:2a}) corresponds to a better slow roll situation where as a large value of $\alpha$ (Figure \ref{fig:2b}) corresponds to more chaotic case.}
    \label{fig:example T model}
\end{figure}
In Figures \ref{fig: example E model} and \ref{fig:example T model}, we show the chaotic $\alpha$-attractor potentials in the $E$-model and the $T$-model respectively in the sub-Planckian region. As will be shown by our calculations, the smaller values of $\alpha$ can explain the Planck data in the $n_{s}$-$r$ plane with $68\%$ CL.

\section{Sub-Planckian mode analysis of linear cosmological perturbations}
\label{sec:mode analysis}
In this section, with a brief introduction to the basic formalism of the cosmological perturbation theory \cite{Baumann:2009ds}, we shall set up the coupled nonlinear differential equations for the inflaton and the metric perturbations (see \cite{Sarkar:2020sbx}, for details). We shall work with the physical time $t$ in the natural system of units where $c=\hbar = 1$ and will go from the $t$ variable to the $k$ variable using the equation, $k=aH$. 
\paragraph{The perturbed metric in spatially flat gauge:}
Following Scalar-Vector-Tensor (SVT) decomposition \cite{Kurki-Suonio,Lifshitz:1945du,Bertschinger:2001is} the relevant metric in this gauge ($\Psi$=E=0) can be written as,
\begin{equation}
   ds^2 = g_{\mu\nu}dx^\mu dx^\nu = -(1+2{\Phi})dt^2 +2a(t)\partial_i B dx^i dt+a^2 (t)\delta_{ij} dx^i dx^j.
   \label{eq:metric}
\end{equation}
\paragraph{The perturbed energy-momentum tensor:}
All the components of the perturbed energy-momentum tensor up to first order can be expressed in matrix form as,
\begin{equation}
{T^\mu }_{\nu}=
\begin{pmatrix}
-\rho& \delta{\partial _1} q & \delta{\partial _2 }q &\delta{\partial _3} q  \\  
-a^{-2}\delta{\partial ^1} q+a^{-1}(p^{(0)}+\rho^{(0)}){\partial^1}B & p & 0 & 0\\ 
-a^{-2}\delta{\partial ^2} q+a^{-1}(p^{(0)}+\rho^{(0)}){\partial^2}B & 0 & p & 0\\ 
-a^{-2}\delta{\partial ^3} q+a^{-1}(p^{(0)}+\rho^{(0)}){\partial^3}B & 0 & 0 & p\\
\end{pmatrix}.
\label{eq:energy_momentum_tensor}
\end{equation}
\paragraph{Coupled density-momentum equations:}
By solving the Einstein field equations using (\ref{eq:metric}) and (\ref{eq:energy_momentum_tensor}), one can obtain two coupled time evolution equations of density and momentum incorporating the Bardeen potentials \cite{Bardeen:1980kt} $\Phi_B$ and $\Psi_B$,
\begin{equation}
\delta\dot\rho+3H(\delta{p}+\delta\rho)=\frac{{\vec{k}}^2}{a^2}\delta{q}+(p^{(0)}+{\rho}^{(0)})\frac{{\vec{k}}^2}{a^2} \left(\frac{\Phi_B}{H}\right)
\label{eq:density time evolution equation}
\end{equation}
and
\begin{equation}
\delta\dot{q}+3H\delta{q}=-\delta{p}-(p^{(0)}+{\rho}^{(0)})\left[\Phi_B+\frac{d}{dt}\left(\frac{\Phi_B}{H}\right)\right].
\label{eq:momentum time evolution equation}
\end{equation}
 These equations in the Fourier space originate from the linear metric perturbations.
\paragraph{Classical evolution equation of inflaton perturbation minimally coupled to gravity:}
 We consider that the evolution of matter perturbations given by (\ref{eq:density time evolution equation}) and (\ref{eq:momentum time evolution equation}) originates from single field inflaton perturbation over classical metric background with the idea that the observed large scale structures of the universe are the magnified versions of spatially distributed quantum nuggets of this field, formed by gravitational instabilities. Therefore we can now write the classical equations for linearly independent mode functions of the unperturbed ($\phi^{(0)}$) and perturbed part ($\delta\phi$) of the inflaton field in the expanding background as,
 \begin{equation}
\ddot\phi^{(0)}(k,t)+3H\dot\phi^{(0)}(k,t)+\frac{\partial V(\phi^{(0)}(k,t))}{\partial \phi^{(0)}(k,t)}=0 ,
\label{eq:unperturbed_inflaton}
\end{equation}
\begin{equation}
\begin{split}
\delta\ddot\phi(k,t)+3H\delta\dot\phi(k,t)+\frac{{\partial}^{2}V(\phi^{(0)}(k,t))}{\partial{\phi^{(0)}(k,t)}^{2}}\delta\phi(k,t)+\frac{k^2}{a^2}\delta\phi(k,t) &\\=\dot\phi^{(0)}(k,t)\left[(\dot\Phi_B)(k,t)+\frac{d^2}{dt^2}\left(\frac{(\Phi_B)(k,t)}{H}\right)+\frac{k^2}{a^2}
\left(\frac{(\Phi_B)(k,t)}{H}\right)\right]\\-2\left[(\Phi_B)(k,t)+\frac{d}{dt}\left(\frac{(\Phi_B)(k,t)}{H}\right)\right]\frac{\partial V(\phi^{(0)}(k,t))}{\partial \phi^{(0)}(k,t)}.
\end{split}
\label{eq:perturbed_inflaton}
\end{equation}
In (\ref{eq:unperturbed_inflaton}) and (\ref{eq:perturbed_inflaton}) we have dropped the three-vector notation on $k$ as these two equations depend only on the magnitude of $\vec{k}$ \cite{Baumann:2009ds, mukhanov2005physical}.
\paragraph{Quantization of the inflaton field:}In order to study the quantum modes of the perturbations we will now, proceed to quantise the inflaton field. Now this field will behave as an operator when it is quantized canonically,
\begin{equation}
\hat{\phi}(t,\vec{X})=\int \frac{d^3k}{(2\pi)^3}[\phi(k,t)\hat{a}(\vec{k})e^{i\vec{k}.\vec{x}}+{{\phi}^*}(k,t){\hat{a}^\dagger}(\vec{k})e^{-i\vec{k}.\vec{x}}],
\label{eq:mode_operator}
\end{equation}
with the commutation,
\begin{equation}
[\hat{a}(\vec{k}), {\hat{a}^\dagger}(\vec{k'})]=(2\pi)^3\delta(\vec{k}-\vec{k'}).
\end{equation}
The mode-dependent inflaton field operator can be expressed as,
\begin{equation}
\hat{\phi}(\vec{k},t)=\phi(k,t)\hat{a}_{\vec{k}}+\phi^{*}(-k,t){\hat{a}}^{\dagger}_{-{\vec{k}}}.
\label{eq:operator_inflaton}
\end{equation}
\paragraph{The power spectra:}
A two-point correlation function in a quasi de-Sitter universe is the vacuum expectation value of a product of first-order perturbations in the quantum mode functions (\ref{eq:operator_inflaton}), as given by,
\begin{equation}
\left\langle\delta\phi(\vec{k},t)\delta\phi(\vec{k'},t)\right\rangle =\frac{4\pi^3}{k^3}\delta(\vec{k}+\vec{k'})H^2.
\end{equation}
In the spatially flat gauge, the curvature perturbation $\mathcal{R}$ is related to the inflaton perturbation $\delta\phi$ by, 
\begin{equation}
\mathcal{R}=H\frac{\delta\phi}{\dot\phi}.
\end{equation} 
The quantum fluctuations in the inflaton field are microscopically induced in the curvature perturbation and thus the curvature correlation function can be written as,
\begin{equation}
\left\langle\mathcal{R}(\vec{k},t)\mathcal{R}(\vec{k'},t)\right\rangle=\frac{H^2}{\dot\phi^2}\left\langle\delta\phi(\vec{k},t)\delta\phi(\vec{k'},t)\right\rangle
\end{equation} 
which can also be written as,
\begin{equation}
\left\langle\mathcal{R}(\vec{k},t)\mathcal{R}(\vec{k'},t)\right\rangle=\frac{16\pi^5}{k^3}\delta(\vec{k}+\vec{k'})\Delta_\mathcal{R}(t)
\end{equation} 
where $\Delta_\mathcal{R}(t)$ is a dimensionless curvature power spectrum,
\begin{equation}
\Delta_\mathcal{R}(t)=\frac{H^4}{(2\pi)^2{\dot{\phi}^2}}.
\label{eq:curvature_power}
\end{equation} 
Similarly, for tensor perturbation we have a dimensionless power spectrum for single polarization state,
\begin{equation}
\Delta_h(t)=\frac{1}{{M_{Pl}}^2}\left(\frac{H}{\pi}\right)^2.
\label{eq:tensor_power}
\end{equation} 
(For two polarization states, the customary notation is $\Delta_t$, which is equal to $2\Delta_h$.)
\paragraph{Horizon crossing of the sub-Planckian modes and the mode-dependent power spectra:}
The horizon crossing is a dynamical process, as different modes cross the horizon at different times satisfying the condition $k=aH$. Here $a$ is a time-dependent dynamical quantity. Thus, like the time scale, we can define a scale of the momentum $k$ crossing the horizon and thereby we can change over from the $t$-scale to the $k$-scale.
This way, the time dependence of the power spectra can be converted into momentum dependence if we use the dynamical horizon crossing condition, whereby we include all the sub-Planckian $k$-modes crossing the shrinking horizon during the inflationary period at slightly different instants. The fact that different modes exit the horizon at different times, is buried in the quantum fluctuations and this will be our theme in the subsequent analysis of the mode-dependent cosmological parameters.\par Thus, we get two momentum dependent power spectra (scalar and tensor) (see (\ref{eq:curvature_power}) and (\ref{eq:tensor_power}) and the subsequent comment) as,
\begin{equation}
\Delta_s(k)= \frac{1}{8\pi^2{M_{Pl}}^2}{\left(\frac{H^2}{\epsilon}\right)}_{k=aH}
\label{eq:scalar_power}
\end{equation}
and
\begin{equation}
\Delta_t(k)=\frac{2}{\pi^2{M_{Pl}}^2}{\left(H^2\right)}_{k=aH}
\label{eq:tensor_power2}
\end{equation} 
where, in (\ref{eq:scalar_power}), we have identified (in spatially flat gauge) the curvature perturbation with the scalar perturbation and used the relation,
\begin{equation} 
{\dot\phi}^2=2{M^2_{Pl}} \epsilon H^2.
\end{equation} Here $\epsilon$ is the slow-roll parameter and $M_{Pl}( = \frac{1}{\sqrt{8\pi G}})$ is the reduced Planck mass. We shall work in the Planck units where $M_{Pl}=1$. In this  system of units, the unit of $k$ is Mpc$^{-1}$.
\paragraph{Mode equations in $k$-space:}
 We first change the $t$ derivatives in (\ref{eq:unperturbed_inflaton}) and (\ref{eq:perturbed_inflaton}) into $k$ derivatives exploiting the dynamical horizon exit constraint.\par  The condition,
\begin{equation}
k=aH=a\frac{\dot{a}}{a}=\dot{a}
\label{eq:horizon_exit}
\end{equation}
will imply that each point in  $k$-space will correspond to a  different expansion rate. Similarly, each value of
\begin{equation}
\dot{k}=\ddot{a}=-\frac{1}{2}\left(p+\frac{\rho}{3}\right)a
\label{eq:kdot_equation}
\end{equation}
will correspond to a different acceleration rate.\par Now the inflationary condition,
\begin{equation}
\dot{\phi}^2<<V(\phi)
\label{eq:slow_roll_condition1}
\end{equation} allows us to write the approximate density and pressure as,
\begin{equation}
\rho\approx V(\phi), 
\label{eq:slow_roll_condition2}
\end{equation}
\begin{equation}
p \approx- V(\phi)
\label{eq:slow_roll_condition3}
\end{equation} 
and the slow-roll parameter as,
\begin{equation}
\epsilon\approx \epsilon_{V},
\label{eq:final_epsilon}
\end{equation} where, $\epsilon_{V}$ is the potential slow-roll parameter.
Thus from (\ref{eq:kdot_equation}), we get,
\begin{equation}
\dot{k}=\frac{1}{3}aV(\phi).
\label{eq:slow_condition_4}
\end{equation}
Retaining upto the first-order term in the Taylor series expansion of $V(\phi)$, we have, 
\begin{equation}
V(\phi)=V(\phi^{(0)})\left( 1+\frac{\partial}{\partial\phi^{(0)}}\ln({V(\phi^{(0)}})\delta\phi\right) .
\label{eq:slow_roll_condition5}
\end{equation}
The scale factor is,
\begin{equation}
a=\frac{k}{H}=k\sqrt{\frac{3}{\rho}}=k\sqrt{3}{(V(\phi))}^{-\frac{1}{2}}.
\label{eq:slow_roll_condition6}
\end{equation}

Thus we get from (\ref{eq:slow_condition_4} - \ref{eq:slow_roll_condition6}),
\begin{equation}
\dot{k}=k\sqrt{\frac{V(\phi^{(0)})}{3}}\left( 1+\frac{\partial}{\partial\phi^{(0)}}\ln\sqrt{V(\phi^{(0)})}\delta\phi\right) 
\end{equation}
and
\begin{equation}
H=\sqrt{\frac{V(\phi^{(0)})}{3}}\left( 1+\frac{\partial}{\partial\phi^{(0)}}\ln\sqrt{V(\phi^{(0)})}\delta\phi\right).
\label{eq:final_Hubble}
\end{equation}
We now write the equations of  linear perturbations in $k$-space (from (\ref{eq:unperturbed_inflaton}) and (\ref{eq:perturbed_inflaton})) as,
\begin{equation}
k^2{\phi}''^{(0)}+k^2G_1{{\phi}'^{(0)}}^2+4k{\phi}'^{(0)} +k^2(G_1\delta\phi'+G_2{\phi}'^{(0)}\delta\phi){\phi}'^{(0)}+6G_1(1-2G_1\delta\phi)=0
\label{eq:mode_equation1}
\end{equation}
\begin{equation}
\begin{split}
k^2\delta\phi''+(k^2 G_1{\phi}'^{(0)}+4k)\delta\phi'+[1+6(G_2+2G_1^2)]\delta\phi= \Phi_B''[k^3{\phi}'^{(0)}]
+\Phi_B'[2k^2{\phi}'^{(0)}-k^3G_1{{\phi}'^{(0)}}^2-12kG_1]\\
 +\Phi_B[k{\phi}'^{(0)}-k^2{{\phi}'^{(0)}}^2G_1-k^3{\phi}'^{(0)}(G_2{{\phi}'^{(0)}}^2+G_1{\phi}''^{(0)})-12G_1+12k{G_1}^2{\phi}'^{(0)}]
\end{split}
\label{eq:mode_equation2}
\end{equation}
where,
\begin{equation}
G_n=\frac{\partial^n}{\partial\phi^{(0)n)}}\ln\sqrt{V(\phi^{(0)})}, \quad n=1,2
\end{equation} and the primes denote the derivatives in $k$-space.
 Trading $\Phi$ and $B$ for the Bardeen potential $\Phi_B$ \cite{Baumann:2009ds} and using the Horizon crossing condition, an evolution equation for the latter can be derived from the perturbed Einstein's equations, under the spatially flat gauge and with no anisotropic stress, as,
\begin{equation}
H\partial_t\left(\Phi_B +\frac{\delta \rho}{2H^2}\right)-3H^2\left(\Phi_B + \frac{\delta \rho}{2H^2}\right)=3\left(\delta \dot{q} +\frac{1}{2}\delta p\right).
\end{equation}
Then we get the equation for the $\Phi_B$ in $k$-space as,
\begin{equation}
\begin{split}
k^4{{\phi}'^{(0)}}^2 {\Phi''}_B+(2k^4 {\phi}'^{(0)}{\phi}''^{(0)}-2k^3 {{\phi}'^{(0)}}^2-k^4G_1 {{\phi}'^{(0)}}^3-2k){\Phi'}_B
+(6+k^3(2-3kG_1{\phi'}^{(0)}){\phi''}^{(0)}{\phi'}^{(0)}\\-k^2(k^2G_2{{\phi'}^{(0)}}^2-3kG_1{\phi'}^{(0)}+4){{\phi'}^{(0)}}^2)\Phi_B=6k(k{\phi''}^{(0)}+kG_1{{\phi'}^{(0)}}^2+{\phi'}^{(0)}+G_2{\phi'}^{(0)})\delta\phi+k^3{\phi'}^{(0)}\delta\phi''\\+k(k^2{\phi''}^{(0)}+2k{\phi'}^{(0)}+6G_1)\delta \phi'.
\end{split}
\label{eq:mode_equation3}    
\end{equation}

Eqs.( \ref{eq:mode_equation1}), (\ref{eq:mode_equation2}) and (\ref{eq:mode_equation3}) are coupled nonlinear differential equations  for the $k$-space evolution of the unperturbed inflaton field, its perturbation, the perturbation in the potential therefrom and the metric perturbation. The metric perturbation, $\Phi_B$ and its derivatives appear on the r.h.s of (\ref{eq:mode_equation2}), thus making a connection between the quantum fluctuation and the geometrical perturbation. This aspect will be reflected in the $k$ dependence of the calculated power spectra and the spectral indices through the solutions of (\ref{eq:mode_equation1}), (\ref{eq:mode_equation2}) and (\ref{eq:mode_equation3}). In order to a have the solution, there should be a proper matching of the quantum fluctuations and the geometrical perturbations. This matching is sensitively  controlled by the factors  $G_{1}$ and $G_{2}$ which contain the derivatives of logarithm of the  potentials. 
\paragraph{Mode-dependent cosmological parameters:}
The number of e-folds corresponding to a potential $V(\phi)$ is defined as,
\begin{equation}
N (\phi) = \int_{\phi_\text{end}}^{\phi} \frac{V(\Bar{\phi})}{\frac{dV(\Bar{\phi})}{d\Bar{\phi}}} d\Bar{\phi}.
\label{eq:e-folds}
\end{equation}
The above integration can be evaluated under the  approximation $\phi_\text{end}\ll\phi$, for $\alpha$-attractor $E$-model, $V(\phi)=V_0 (1-e^{-\sqrt{\frac{2}{3\alpha}}\phi})^n$ and $T$-model, $V(\phi)=V_0 \tanh^n(\frac{\phi}{\sqrt{6\alpha}})$ and we get,
\begin{equation}
N_E(\phi) = \frac{3\alpha}{2n}e^{\sqrt{\frac{2}{3\alpha}}\phi},
\end{equation}
\begin{equation}
N_T(\phi)=\frac{3\alpha}{2n}\cosh{\frac{2\phi}{\sqrt{6\alpha}}}.
\end{equation}
From these we can now write the mode-dependent number of e-folds as,
\begin{equation}
N_E (k)=N_E(\phi(k)) = \frac{3\alpha}{2n}e^{\sqrt{\frac{2}{3\alpha}}\phi(k)},
\label{eq:mode e-folds1}
\end{equation}
\begin{equation}
N_T (k)=N_T(\phi(k))=\frac{3\alpha}{2n}\cosh{\frac{2\phi(k)}{\sqrt{6\alpha}}}
\label{eq:mode e-folds2}
\end{equation}
where $\phi(k) = \phi^{(0)}(k)+\delta\phi(k)$.
Using (\ref{eq:mode e-folds1}) and (\ref{eq:mode e-folds2}), one can express mode-dependent first ($\epsilon_V$) and second ($\eta_V$) slow-roll parameters for both $E$ and $T$ models as,
\begin{equation}
\epsilon_V (N(k)) \approx \frac{3\alpha}{4N(k)^2}
\label{eq:first_slow_roll_parameter}
\end{equation} and 
\begin{equation}
\eta_V (N(k)) \approx \left(\frac{3\alpha}{2N(k)^2}-\frac{1}{N(k)}\right).
\label{eq:second_slow_roll_parameter}
\end{equation}
 The mode-dependent scalar and tensor spectral indices respectively are,
\begin{equation}
n_s(k)=1+\frac{d\ln\Delta_s(k)}{d\ln k},
\end{equation}
\begin{equation}
n_t(k)=\frac{d\ln\Delta_t(k)}{d\ln k}
\end{equation}
and the tensor-to-scalar ratio is,
\begin{equation} 
r(k)=\frac{\Delta_t(k)}{\Delta_s(k)}.
\label{eq:tensor_scalar_ratio1}
\end{equation}
We can relate  the mode-dependent spectral indices and the tensor-to-scalar ratio to the corresponding slow-roll parameters as \cite{Baumann:2009ds},
\begin{equation}
n_s(k)=1+2\eta_V(k)-6\epsilon_V(k),
\label{eq:scalar_index}
\end{equation}
\begin{equation}
n_t(k)=-2\epsilon_V(k)
\label{eq:tensor_index}
\end{equation}
and
\begin{equation}
r(k)=16\epsilon_V(k).
\label{eq:tensor_scalar_ratio}
\end{equation}
Using (\ref{eq:first_slow_roll_parameter}) and (\ref{eq:second_slow_roll_parameter}) in (\ref{eq:scalar_index}),  (\ref{eq:tensor_index}) and (\ref{eq:tensor_scalar_ratio}) up to leading order of $1/N$ \\($\mathcal{ O}(1/N^2)\ll1$) we get,
\begin{equation}
n_s (N(k)) = 1-\frac{2}{N(k)},
\label{eq:final_scalar_index}
\end{equation}
\begin{equation}
n_t(N(k)) = -\frac{3\alpha}{2N(k)^2}
\label{eq:final_tensor_index}
\end{equation}
and
\begin{equation}
r(k) = \frac{12\alpha}{N(k)^2}.
\label{eq:final_tensor_scalar_ratio}
\end{equation}
Eqs. (\ref{eq:final_scalar_index}) and (\ref{eq:final_tensor_scalar_ratio}) are the mode-dependent versions of the corresponding universal equations for the $\alpha$-attractors \cite{Carrasco:2015rva,Kallosh:2013yoa} (both for $E$ and $T$ models). In the large $\alpha$ limit, (\ref{eq:final_tensor_scalar_ratio}) reduces to \cite{Kallosh:2013yoa,Kallosh:2014rga}
\begin{equation}
    r(k)=\frac{4n}{N(k)},
\label{eq:high_alpha}    
\end{equation}
for the potentials in Eqs. (\ref{fig:pot_E_model }) and (\ref{fig:pot_T_model}).
\par In the next section, we will display the mode dependence of power spectra, e-folds and spectral indices, through the numerical solutions of $\phi^{(0)}(k)$ and $\delta\phi(k)$ from (\ref{eq:mode_equation1}), (\ref{eq:mode_equation2}) and (\ref{eq:mode_equation3}).
\section{Results and discussions}
\label{sec:results}
\subsection{Validity of perturbative caculations}
\label{sec:Validity}

\begin{figure}[H]
\begin{subfigure}{0.5\textwidth}
  \centering
   \includegraphics[width=75mm,height=80mm]{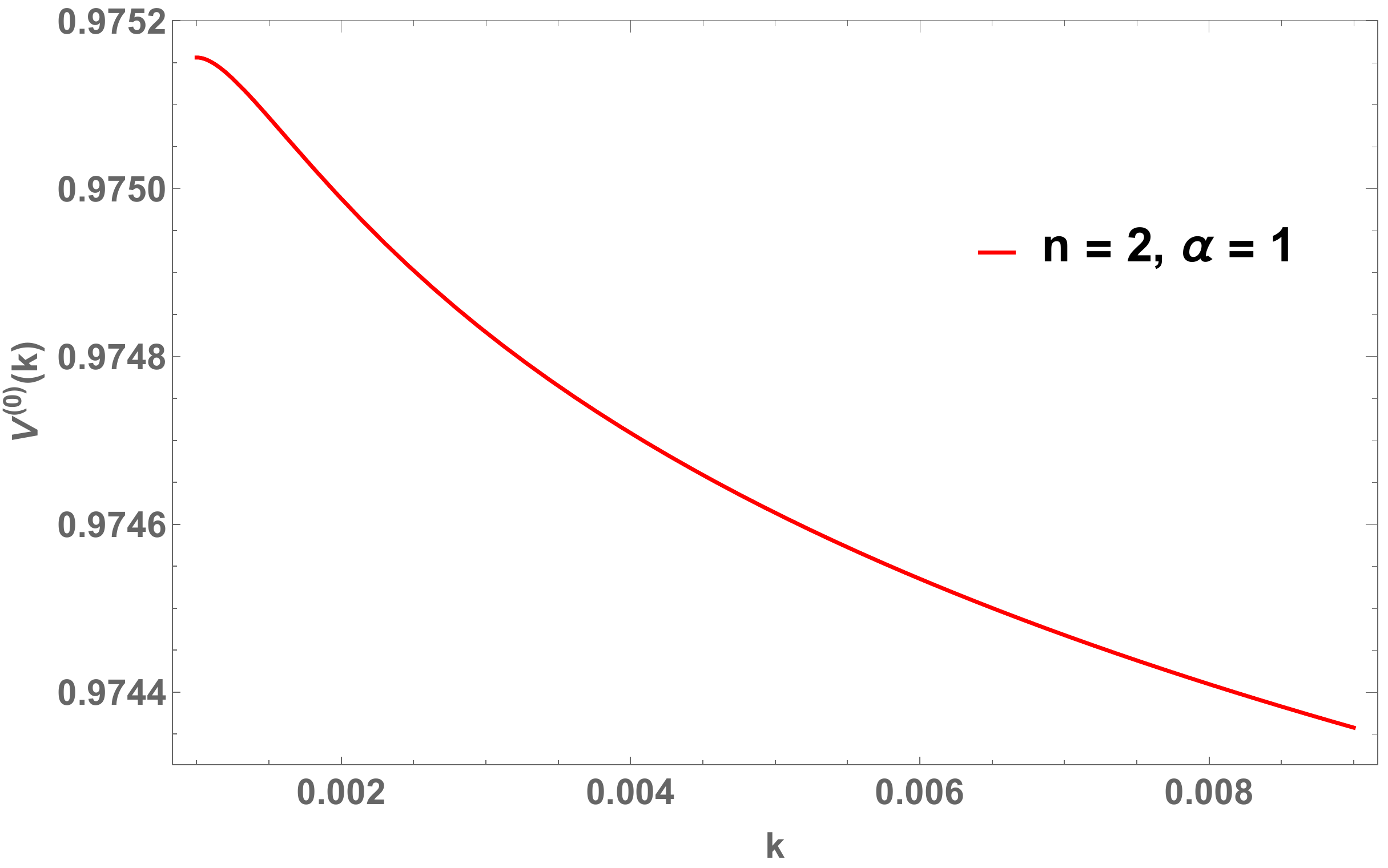} 
   \subcaption{}
   \label{fig:unperturbed_pot_E_1}
\end{subfigure}%
\begin{subfigure}{0.5\textwidth}
  \centering
   \includegraphics[width=75mm,height=80mm]{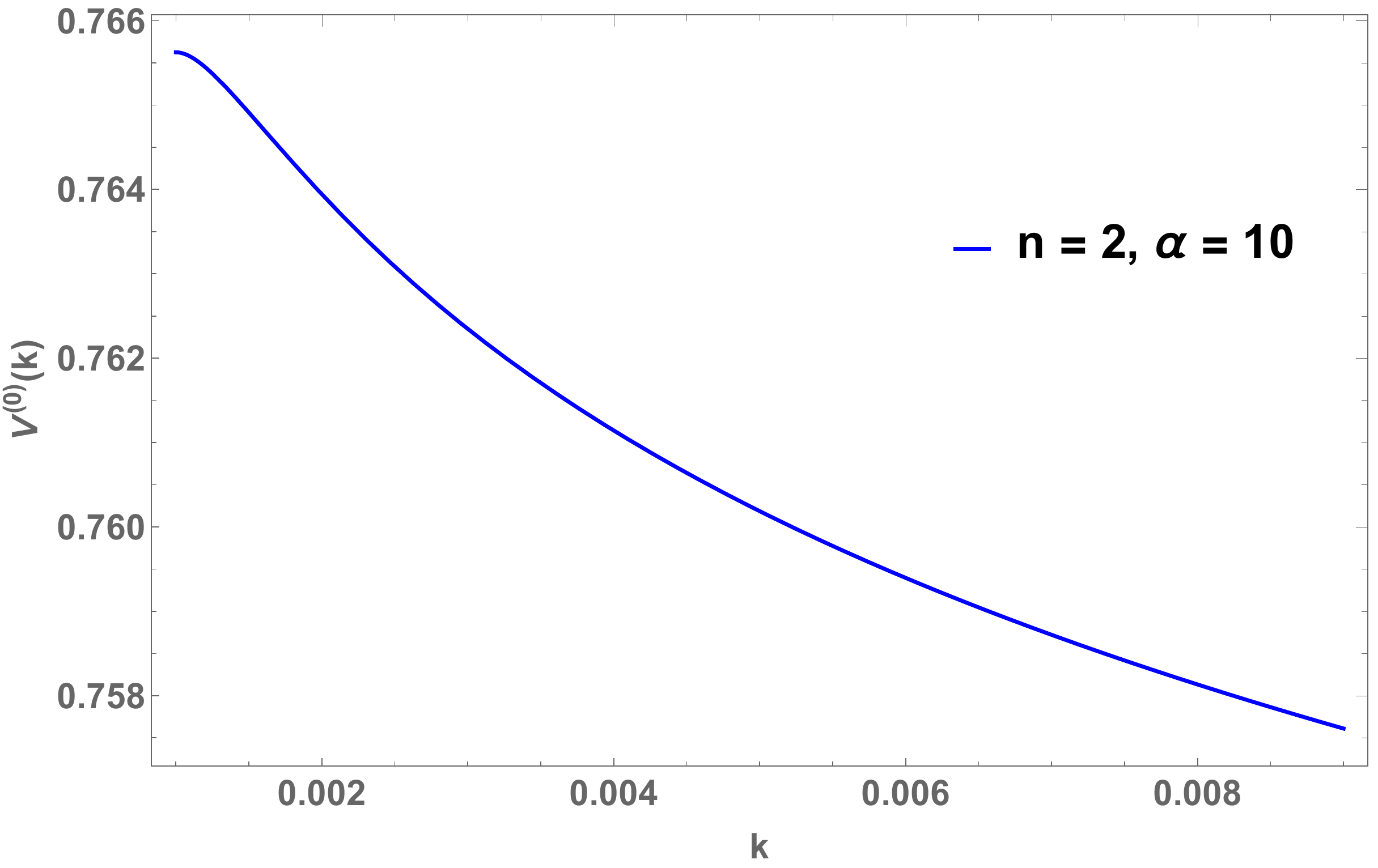}
   \subcaption{}
\end{subfigure}%

\begin{subfigure}{1.0\textwidth}
  \centering
   \includegraphics[width=80mm,height=80mm]{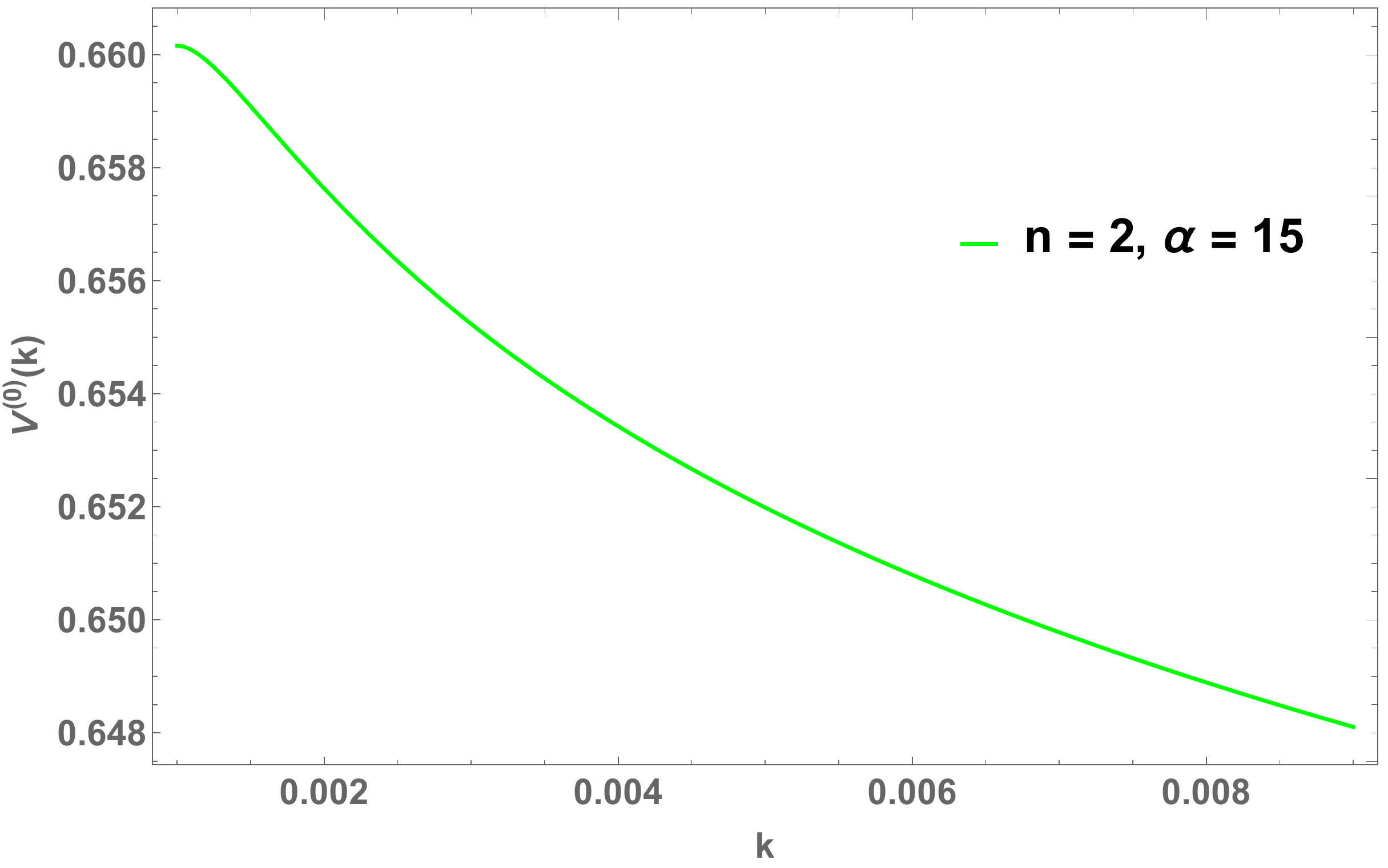}
   \subcaption{}
   \label{fig:unperturbed_pot_E_2}
\end{subfigure}%
\caption{Unperturbed $\alpha$-attractor potentials, $V^{(0)}(k)$, for the $E$-model, for one value of $n$ and three values of $\alpha$. The potential decreases with increase in $\alpha$, for a given value of $k$.}
    \label{fig: unperturbed part of the E_potential}
\end{figure}

\begin{figure}[H]
\begin{subfigure}{0.5\textwidth}
  \centering
   \includegraphics[width=75mm,height=80mm]{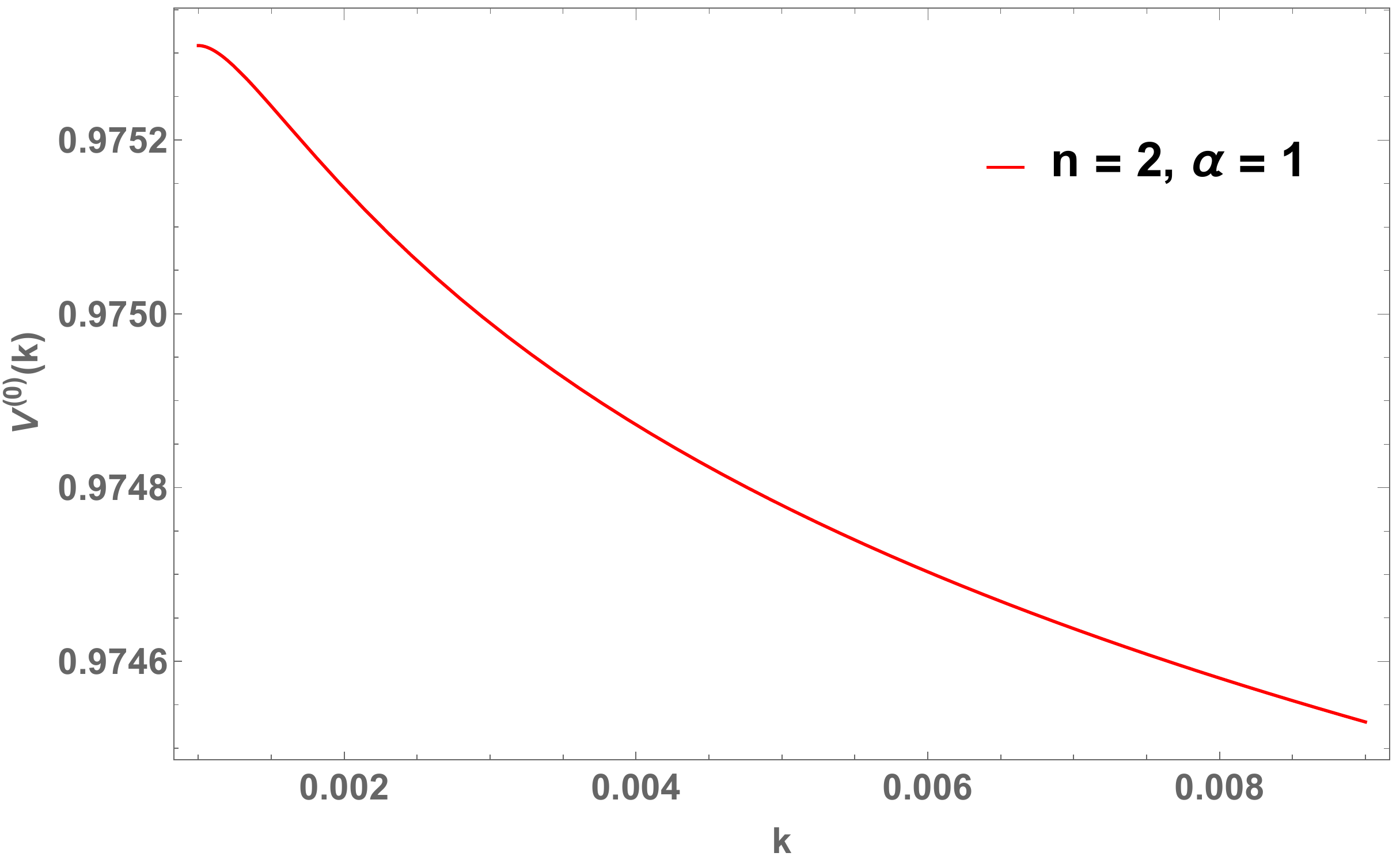}
   \subcaption{}
   \label{fig:unperturbed_pot_T_1}
\end{subfigure}%
\begin{subfigure}{0.5\textwidth}
  \centering
   \includegraphics[width=75mm,height=80mm]{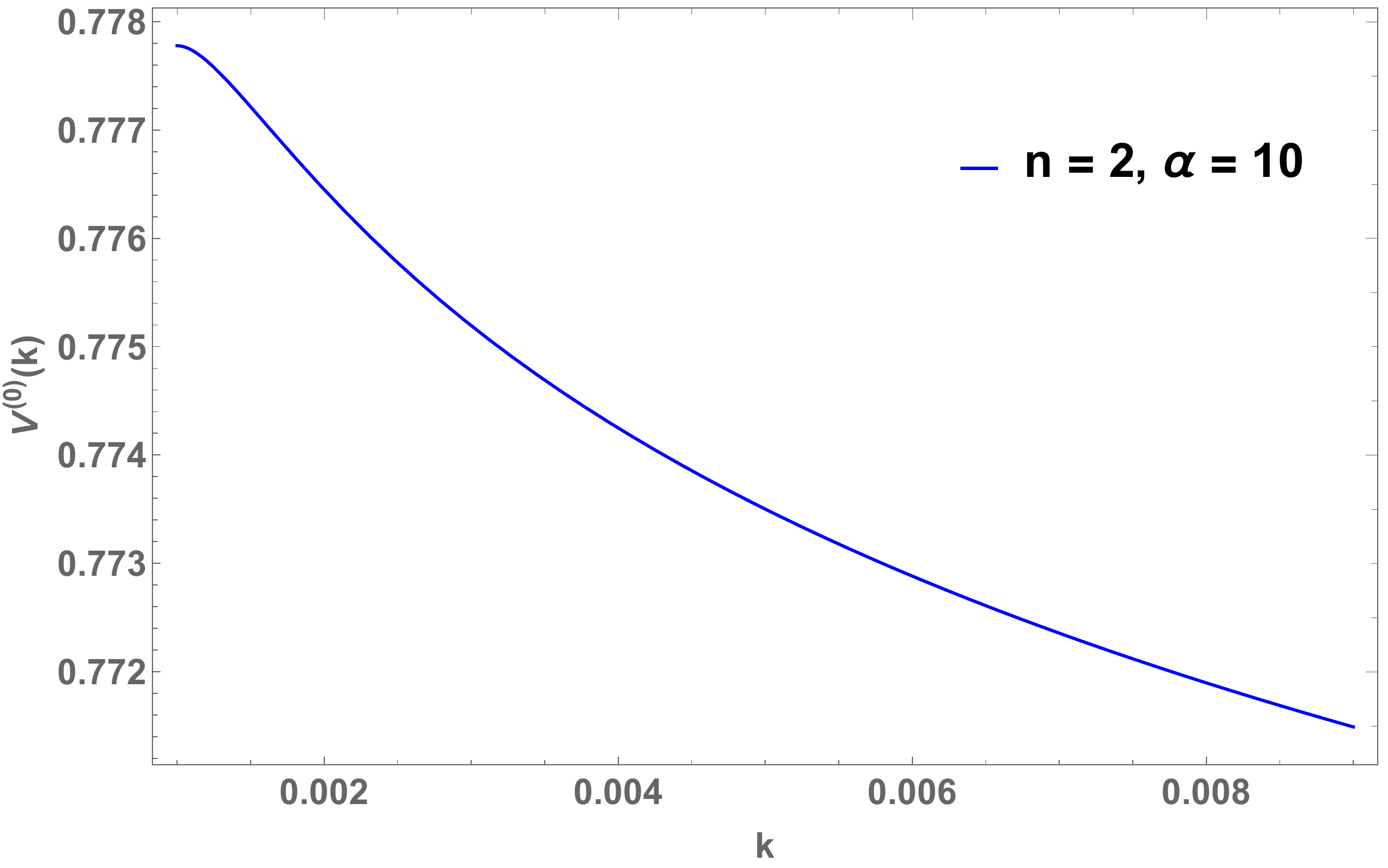}
   \subcaption{}
\end{subfigure}%

\begin{subfigure}{1\textwidth}
  \centering
   \includegraphics[width=80mm,height=80mm]{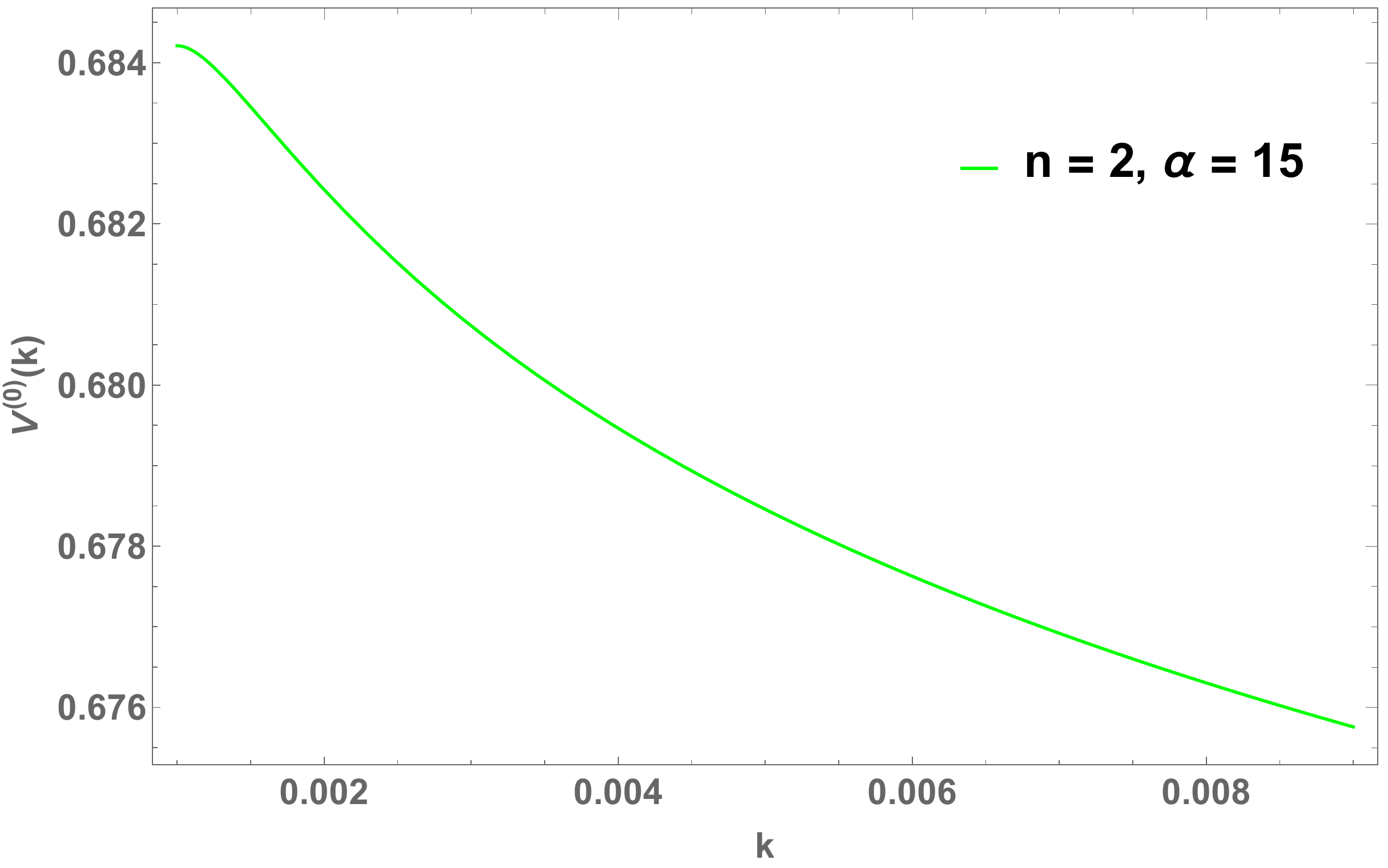}
   \subcaption{}
   \label{fig:unperturbed_pot_T_2}
\end{subfigure}%
    \caption{Unperturbed $\alpha$-attractor potentials, $V^{(0)}(k)$, for the $T$-model, for one value of $n$ and three values of $\alpha$. The potential decreases with increase in $\alpha$, for a given value of $k$.}
    \label{fig: unperturbed part of the T_potential}
\end{figure}

\begin{figure}[H]
\begin{subfigure}{0.5\textwidth}
  \centering
   \includegraphics[width=75mm,height=80mm]{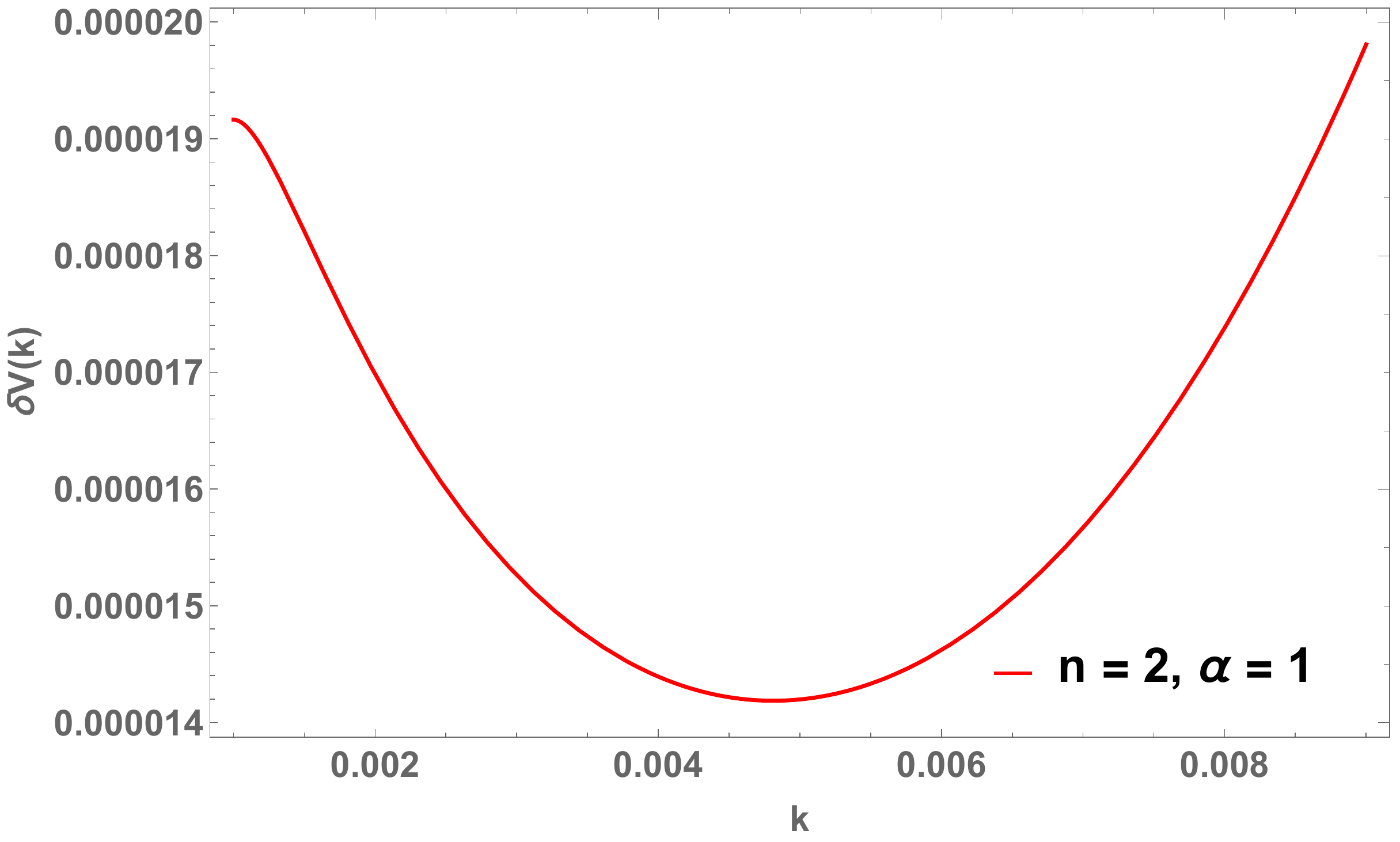} 
   \subcaption{}
   \label{fig:perturbed_pot_E_1}
\end{subfigure}%
\begin{subfigure}{0.5\textwidth}
  \centering
   \includegraphics[width=75mm,height=80mm]{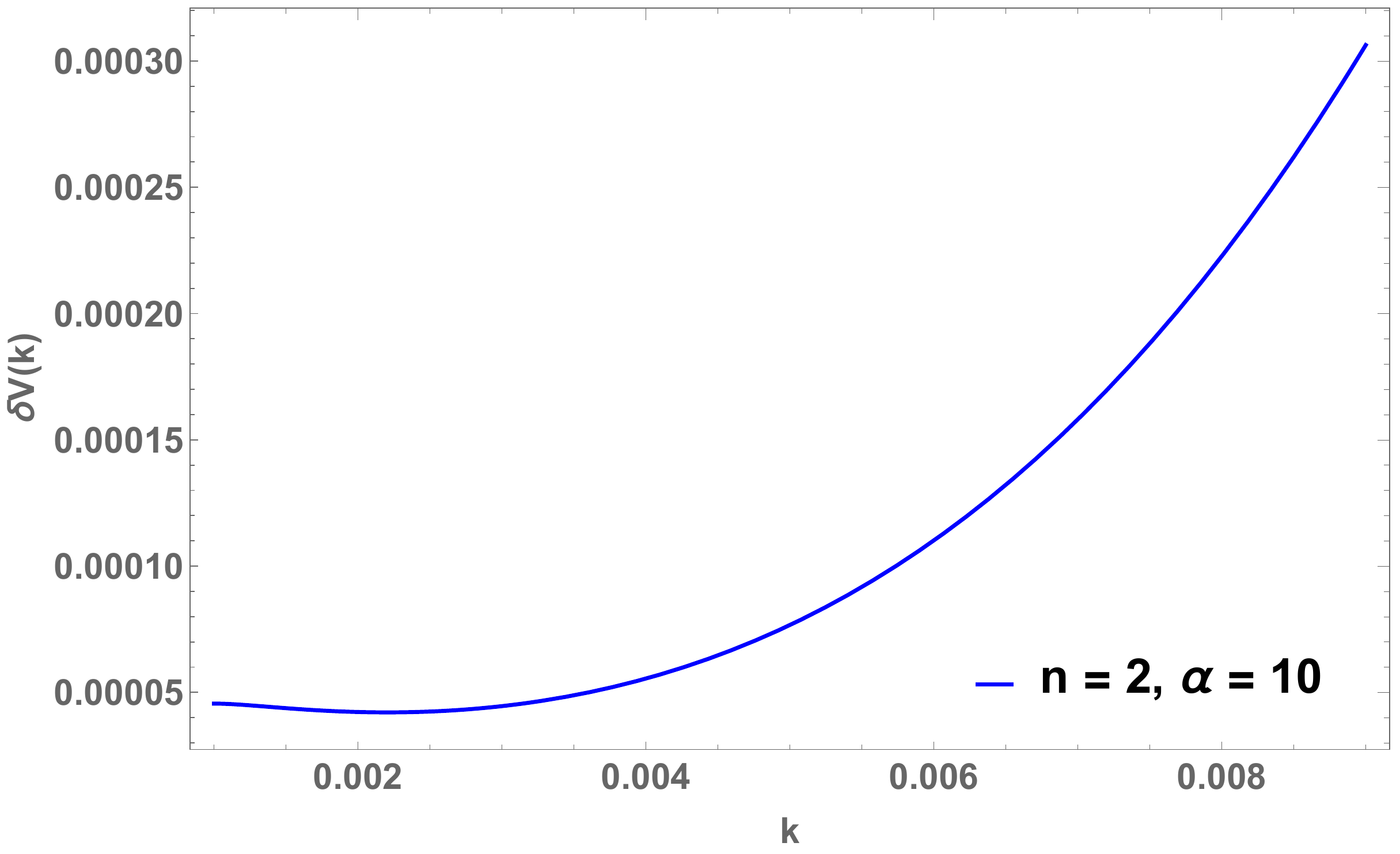}
   \subcaption{}
\end{subfigure}%

\begin{subfigure}{1\textwidth}
  \centering
   \includegraphics[width=80mm,height=80mm]{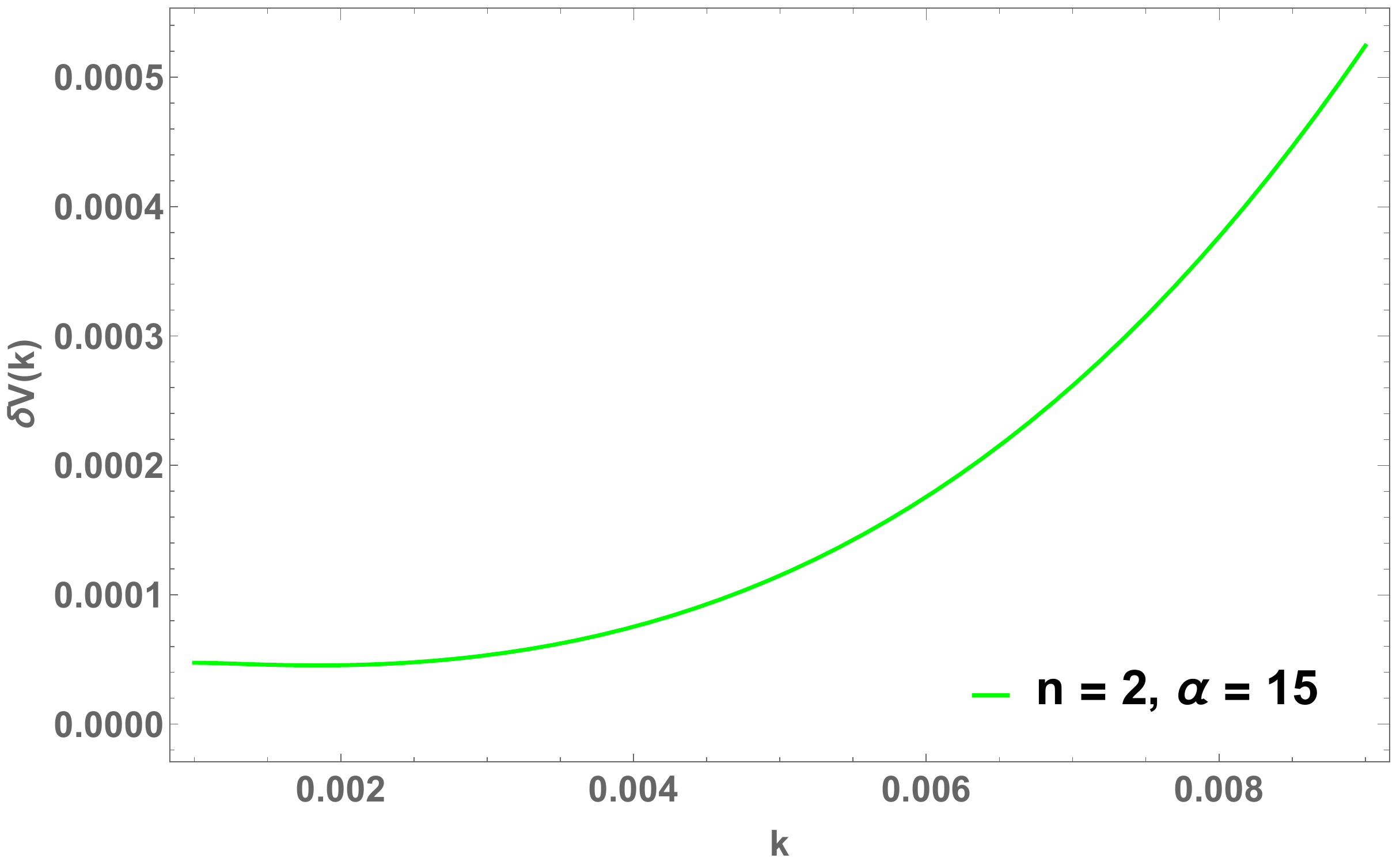}
   \subcaption{}
   \label{fig:perturbed_pot_E_2}
\end{subfigure}
 \caption{Perturbations of the $\alpha$-attractor potentials, $\delta V(k)$, for the $E$-model, for one value of $n$ and three values of $\alpha$. Perturbation increases with increase in $\alpha$ for a particular value of $k$.}
 \label{fig: perturbed part of the E_potentials}
\end{figure}

\begin{figure}[H]
\begin{subfigure}{0.5\textwidth}
  \centering
   \includegraphics[width=75mm,height=80mm]{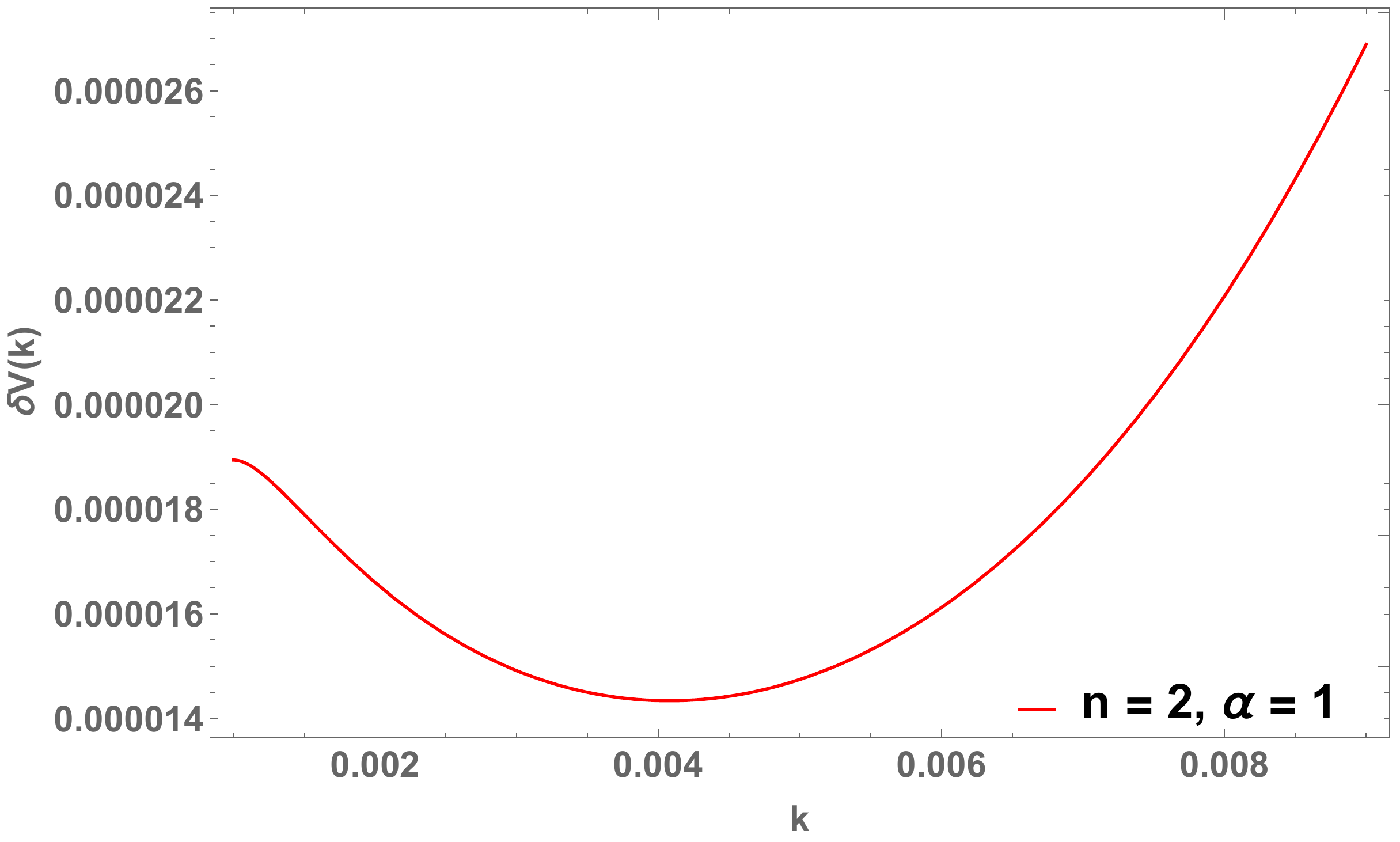}
   \subcaption{}
   \label{fig:perturbed_pot_T_1}
\end{subfigure}%
\begin{subfigure}{0.5\textwidth}
  \centering
   \includegraphics[width=75mm,height=80mm]{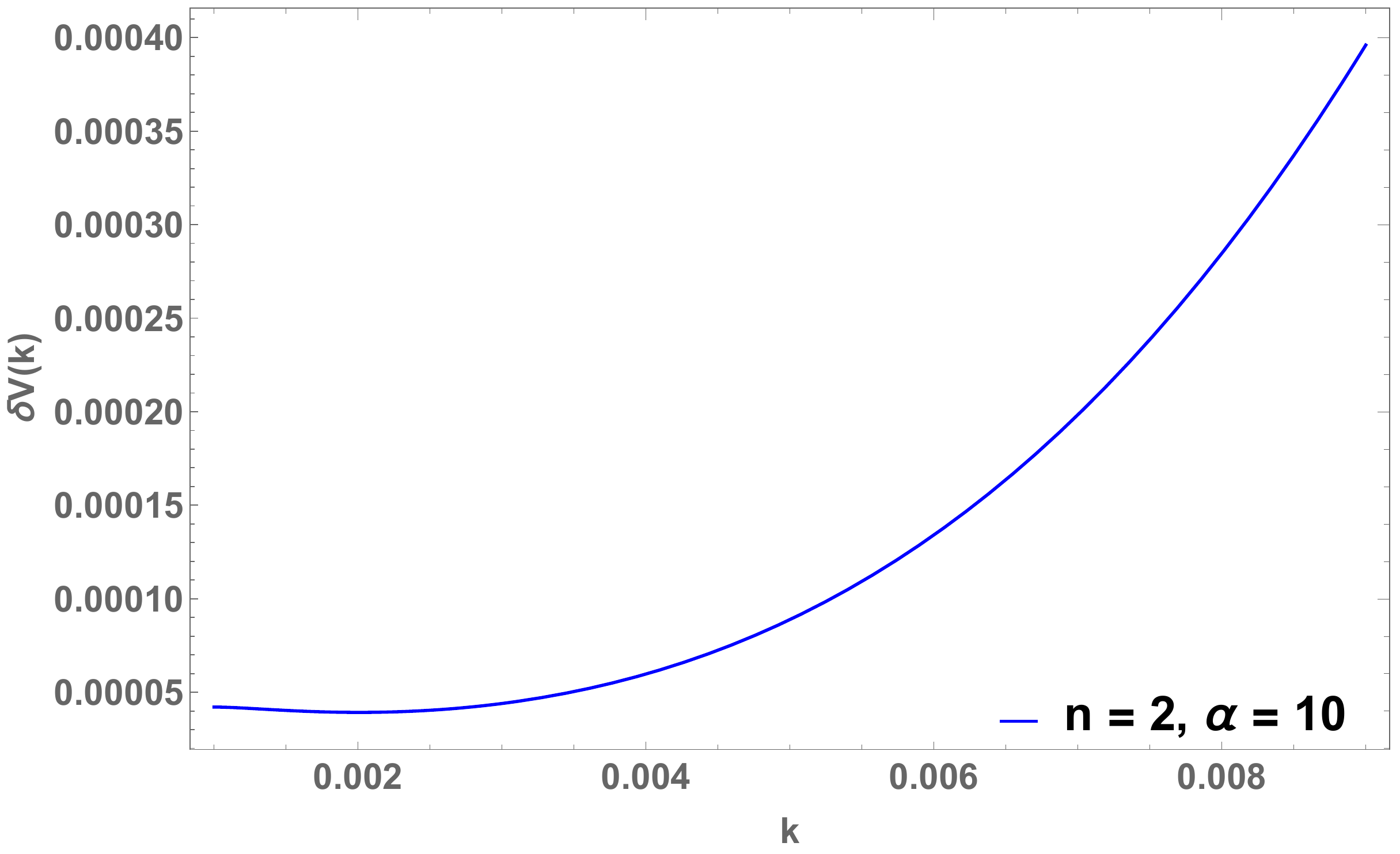}
   \subcaption{}
\end{subfigure}%

\begin{subfigure}{1\textwidth}
  \centering
   \includegraphics[width=75mm,height=80mm]{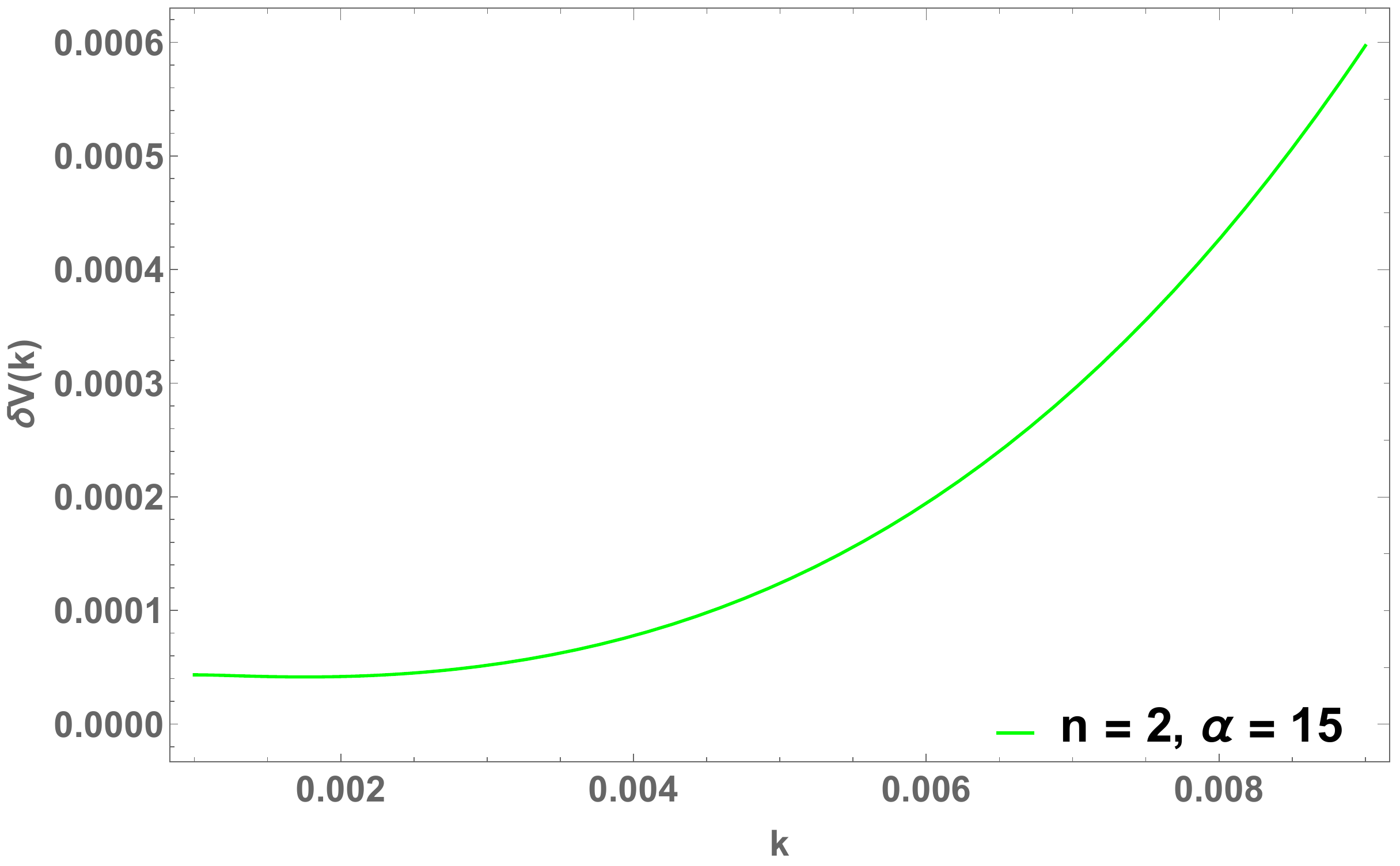}
   \subcaption{}
   \label{fig:perturbed_pot_T_2}
\end{subfigure}%
    \caption{Perturbations of the $\alpha$-attractor potentials, $\delta V(k)$, for the $T$-model, for one value of $n$ and three values of $\alpha$. Perturbation increases with increase in $\alpha$ for a particular value of $k$.}
    \label{fig: perturbed part of the T_potential}
\end{figure}

\begin{figure}[H]
\begin{subfigure}{0.5\textwidth}
  \centering
   \includegraphics[width=75mm,height=80mm]{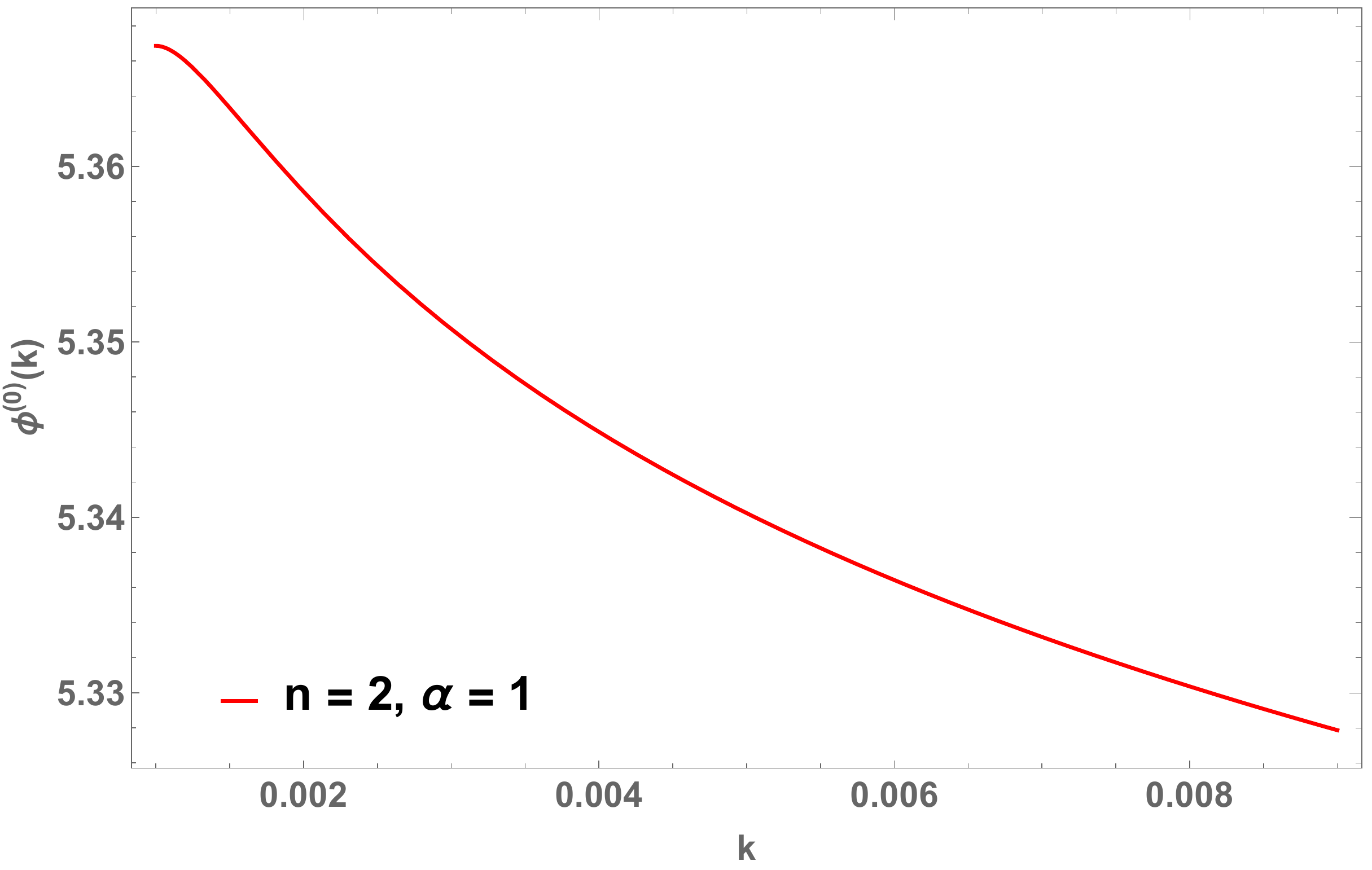}
   \subcaption{}
   \label{fig:unperturbed_inflaton_E_1}
\end{subfigure}%
\begin{subfigure}{0.5\textwidth}
  \centering
   \includegraphics[width=75mm,height=80mm]{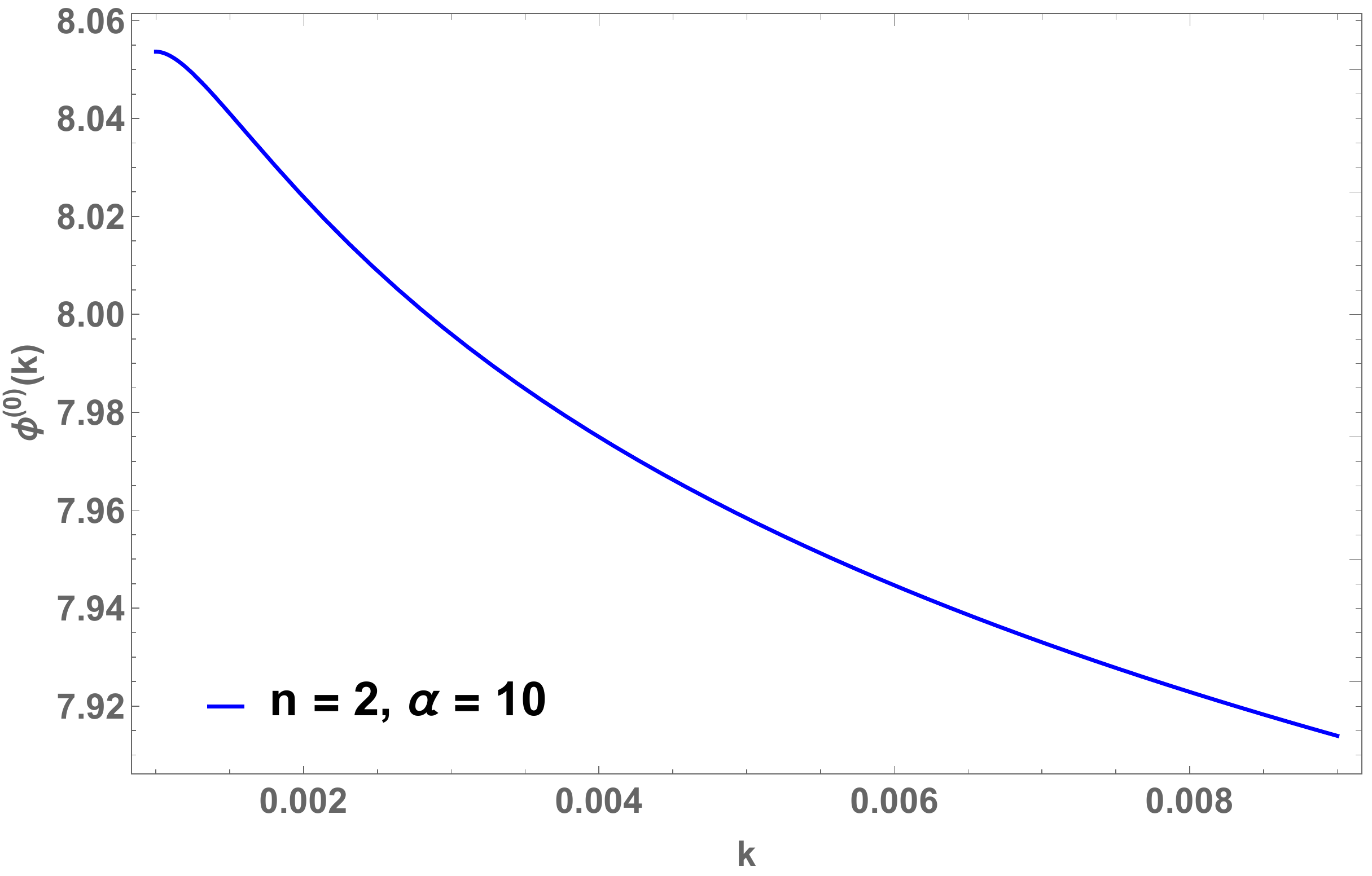}
   \subcaption{}
\end{subfigure}%

\begin{subfigure}{1\textwidth}
  \centering
   \includegraphics[width=80mm,height=80mm]{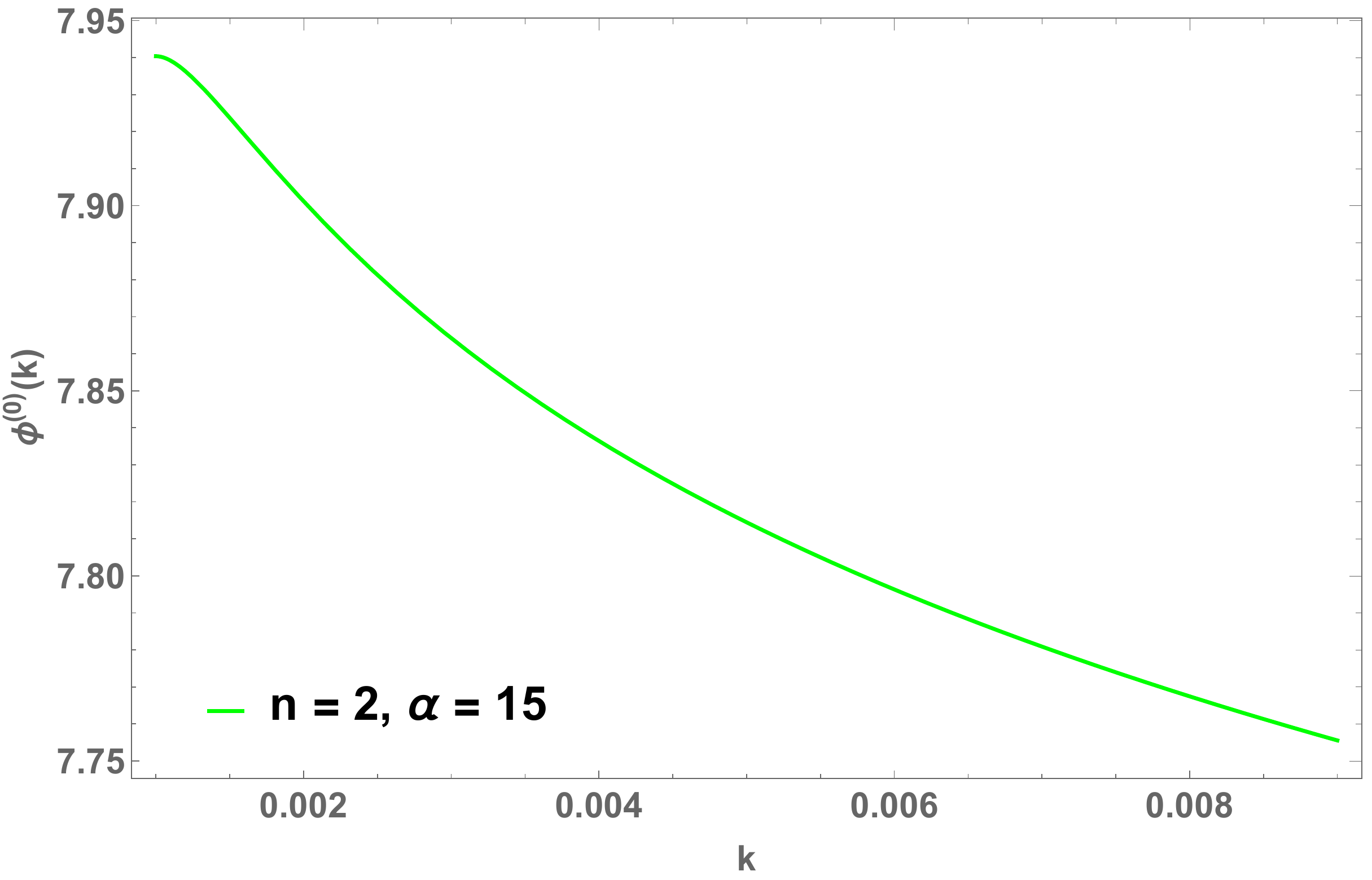}
   \subcaption{}
   \label{fig:unperturbed_inflaton_E_2}
\end{subfigure}
 \caption{Unperturbed inflaton field, $\phi^{(0)}(k)$, in the $\alpha$-attractor $E$-model potential, for one value of $n$ and three values of $\alpha$. The value of the field increases with increase in $\alpha$ for a given value of $k$, except in (Figure \ref{fig:unperturbed_inflaton_E_2}) where it slightly decreases. }
    \label{fig:unperturbed inflaton_E}
\end{figure}

\begin{figure}[H]
\begin{subfigure}{0.5\textwidth}
  \centering
   \includegraphics[width=75mm,height=80mm]{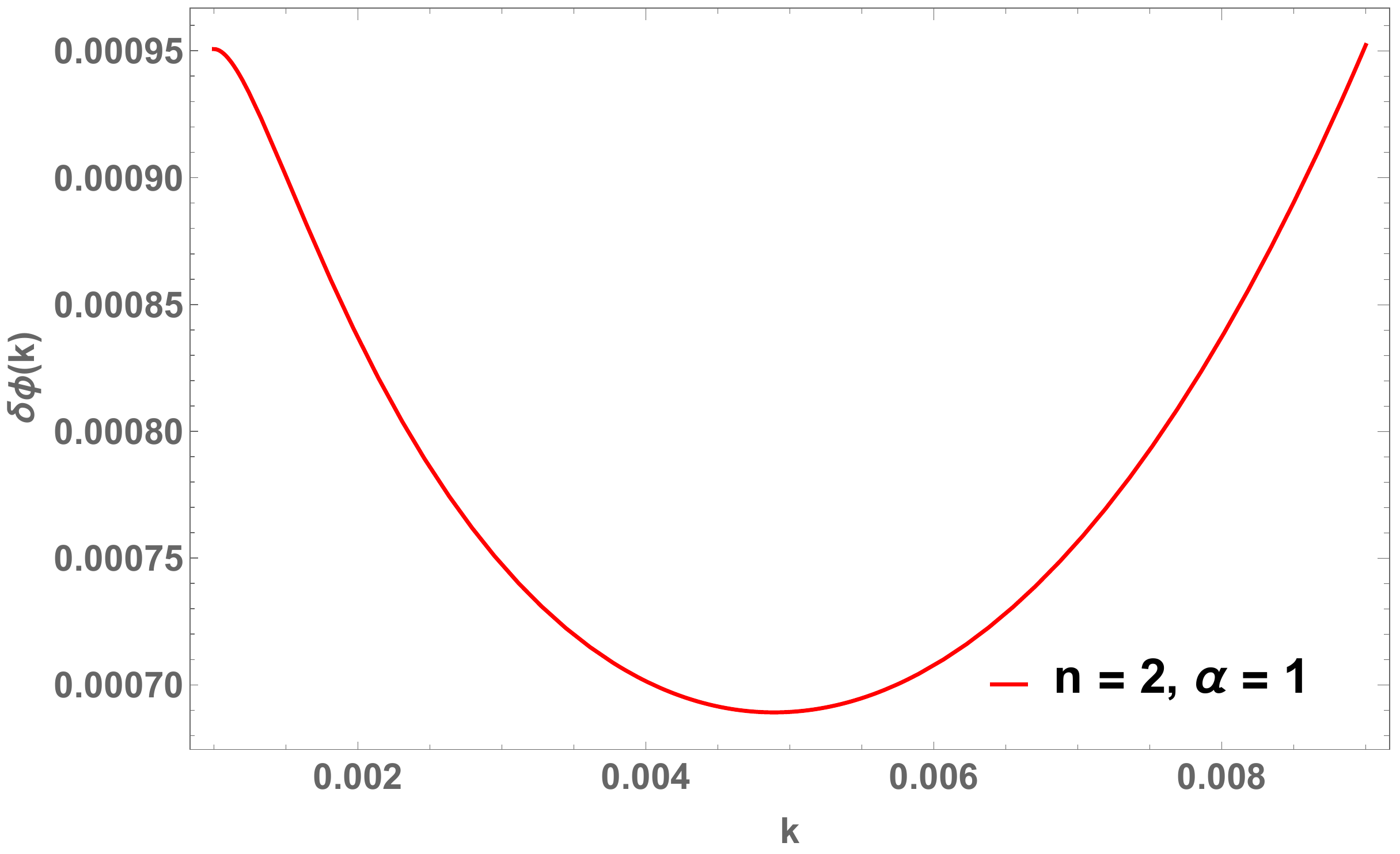}
   \subcaption{}
   \label{fig:perturbed_inflaton_E_1}
\end{subfigure}%
\begin{subfigure}{0.5\textwidth}
  \centering
   \includegraphics[width=75mm,height=80mm]{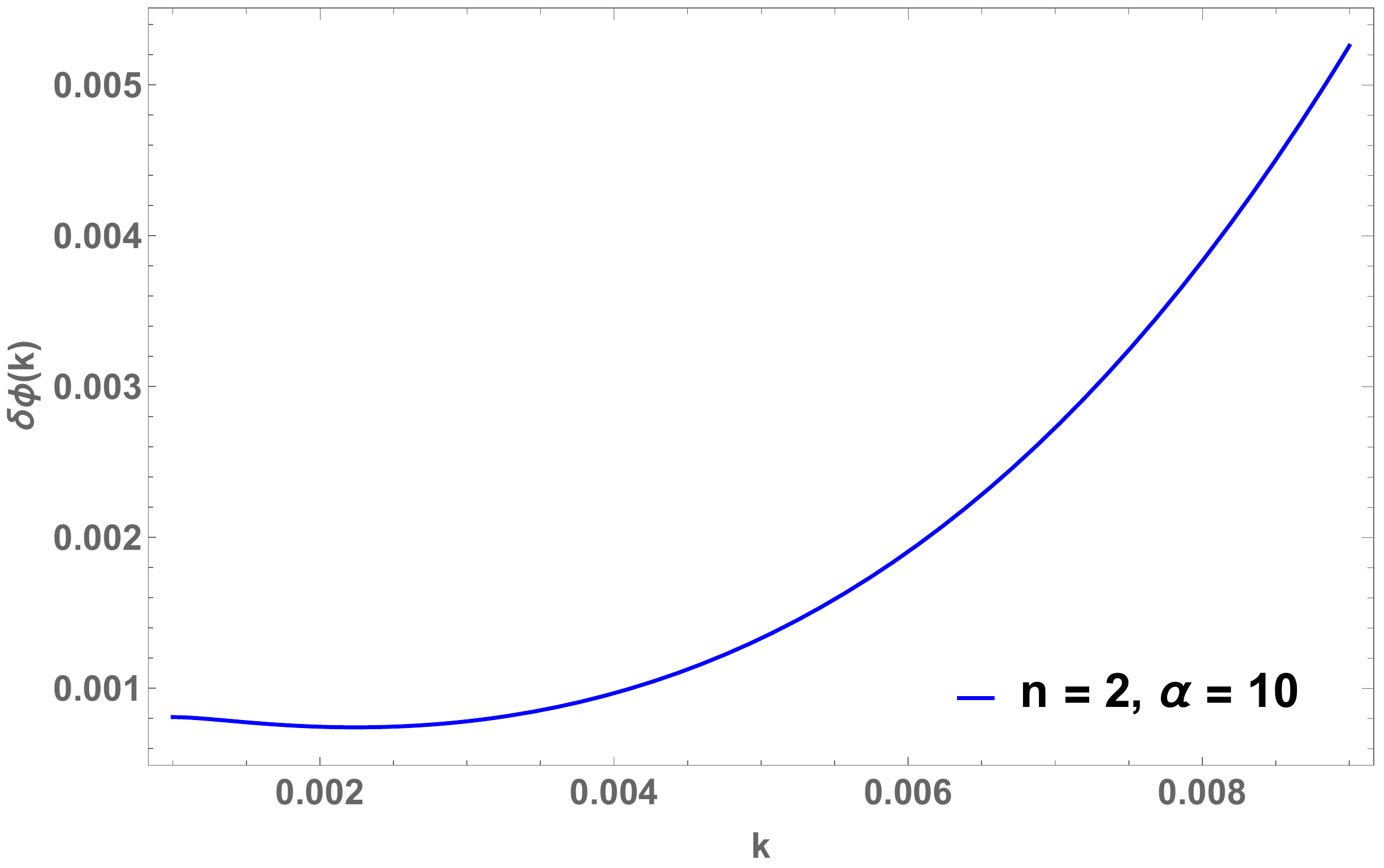}
   \subcaption{}
\end{subfigure}%

\begin{subfigure}{1\textwidth}
  \centering
   \includegraphics[width=80mm,height=80mm]{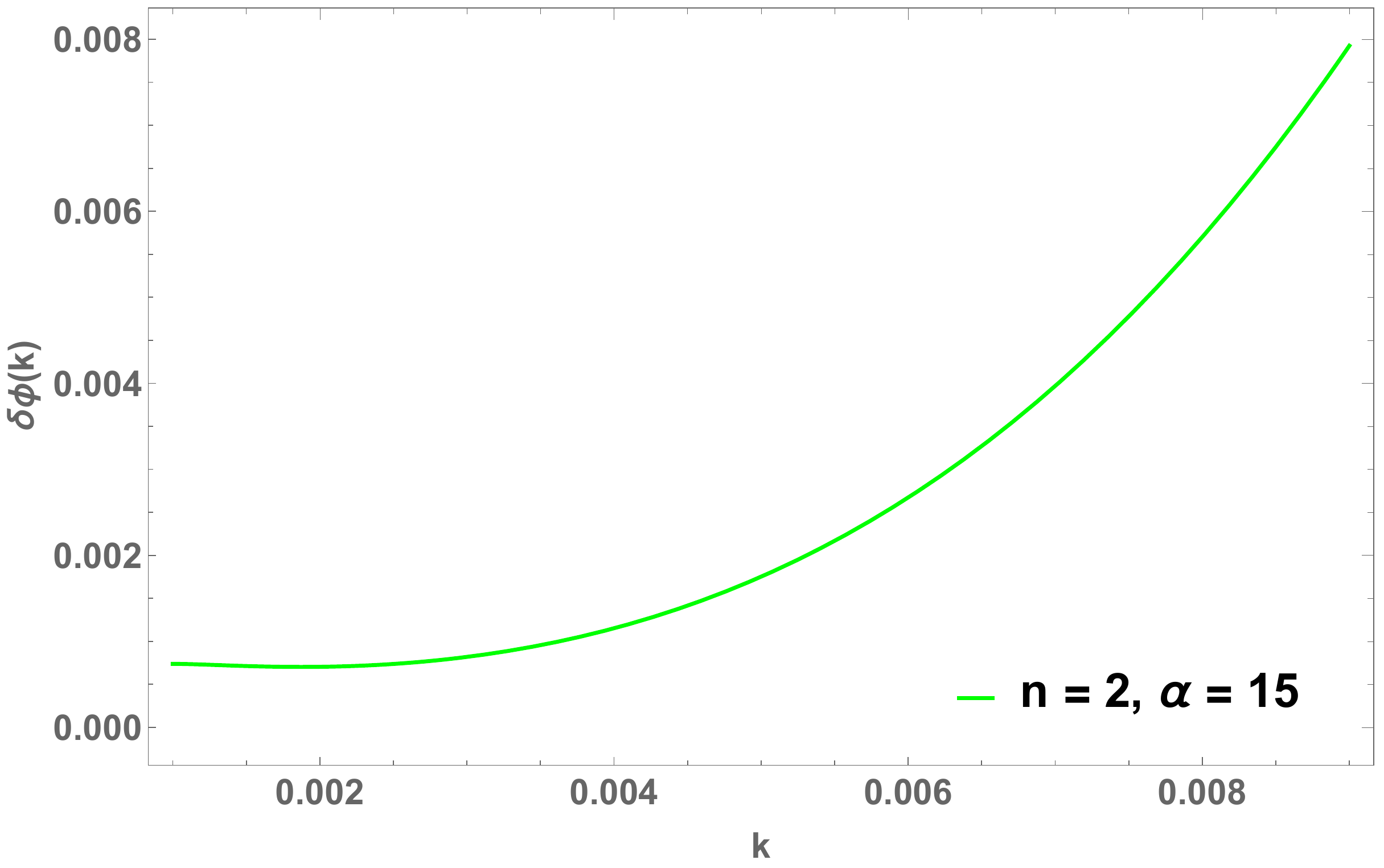}
   \subcaption{}
    \label{fig:perturbed_inflaton_E_2}
\end{subfigure}
\caption{Perturbation $\delta \phi(k)$ of the inflaton field, in the $E$-model $\alpha$-attractor potentials, for one value of $n$ and three values of $\alpha$. The perturbation increases with the value of  $\alpha$ for a particular value of $k$.  }
\label{fig: inflaton perturbation_E}
\end{figure}

\begin{figure}[H]
\begin{subfigure}{0.5\textwidth}
  \centering
   \includegraphics[width=75mm,height=80mm]{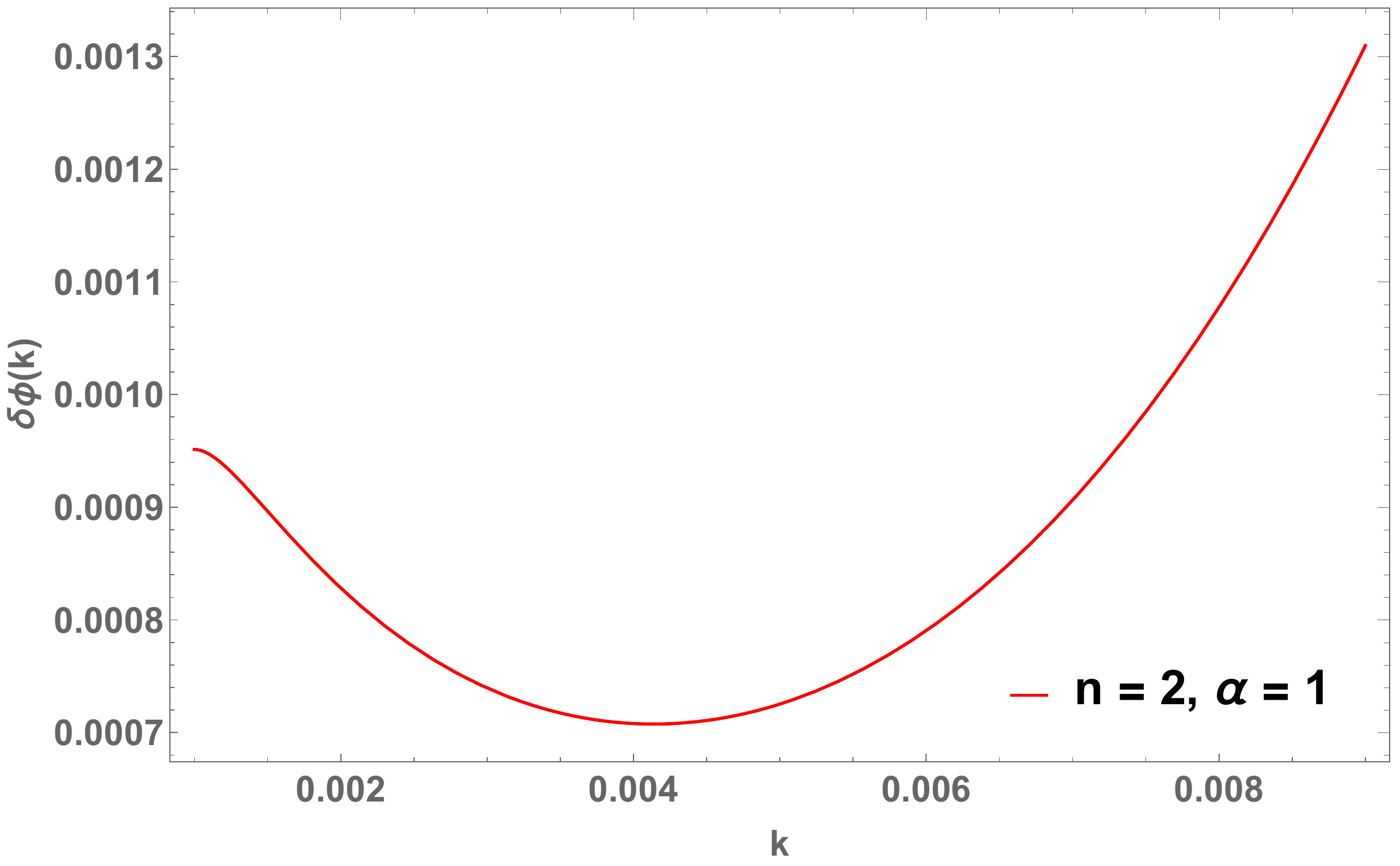}
   \subcaption{}
    \label{fig:perturbed_inflaton_T_1}
\end{subfigure}%
\begin{subfigure}{0.5\textwidth}
  \centering
   \includegraphics[width=75mm,height=80mm]{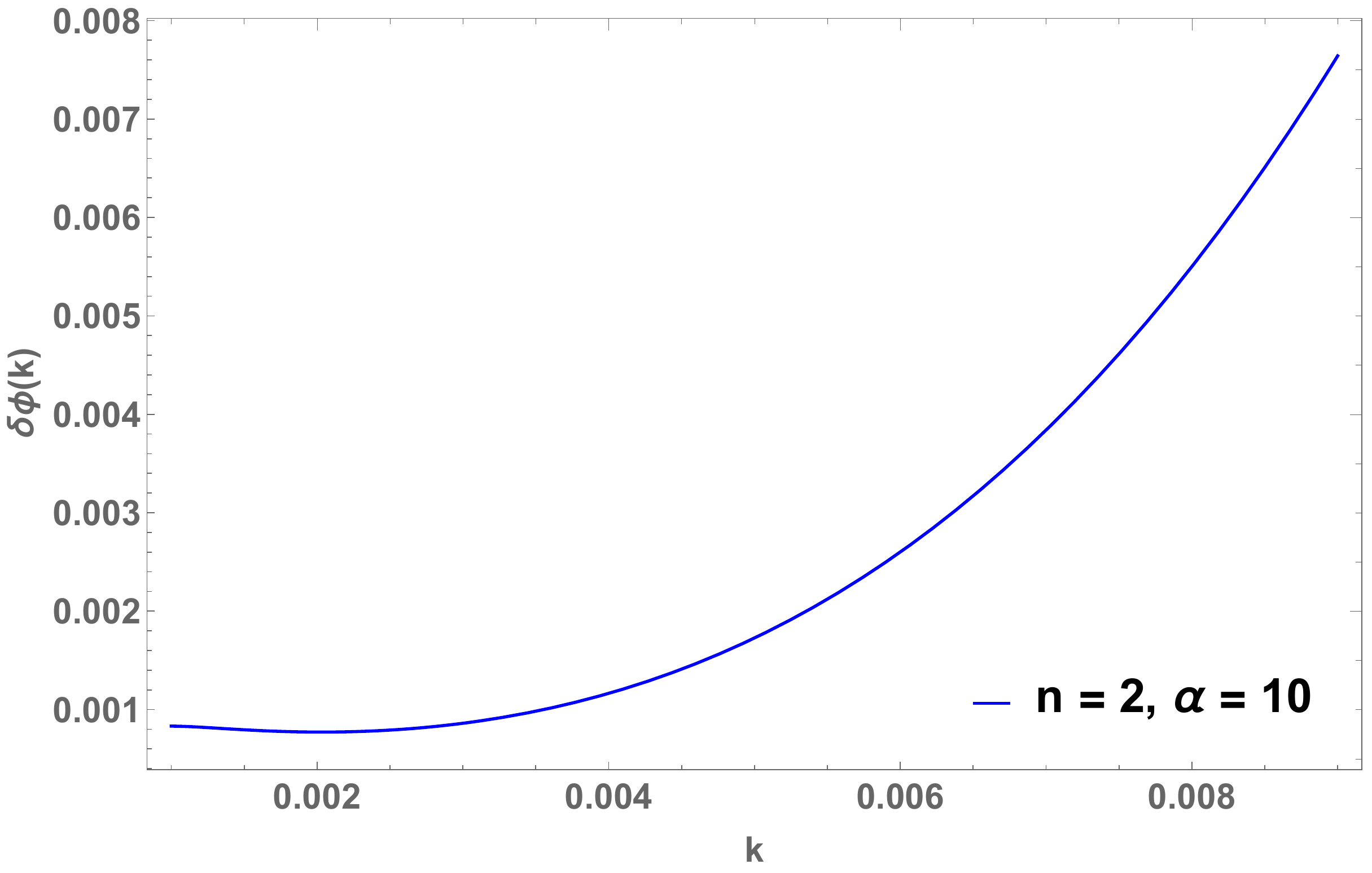}
   \subcaption{}
\end{subfigure}%

\begin{subfigure}{1.0\textwidth}
  \centering
   \includegraphics[width=80mm,height=80mm]{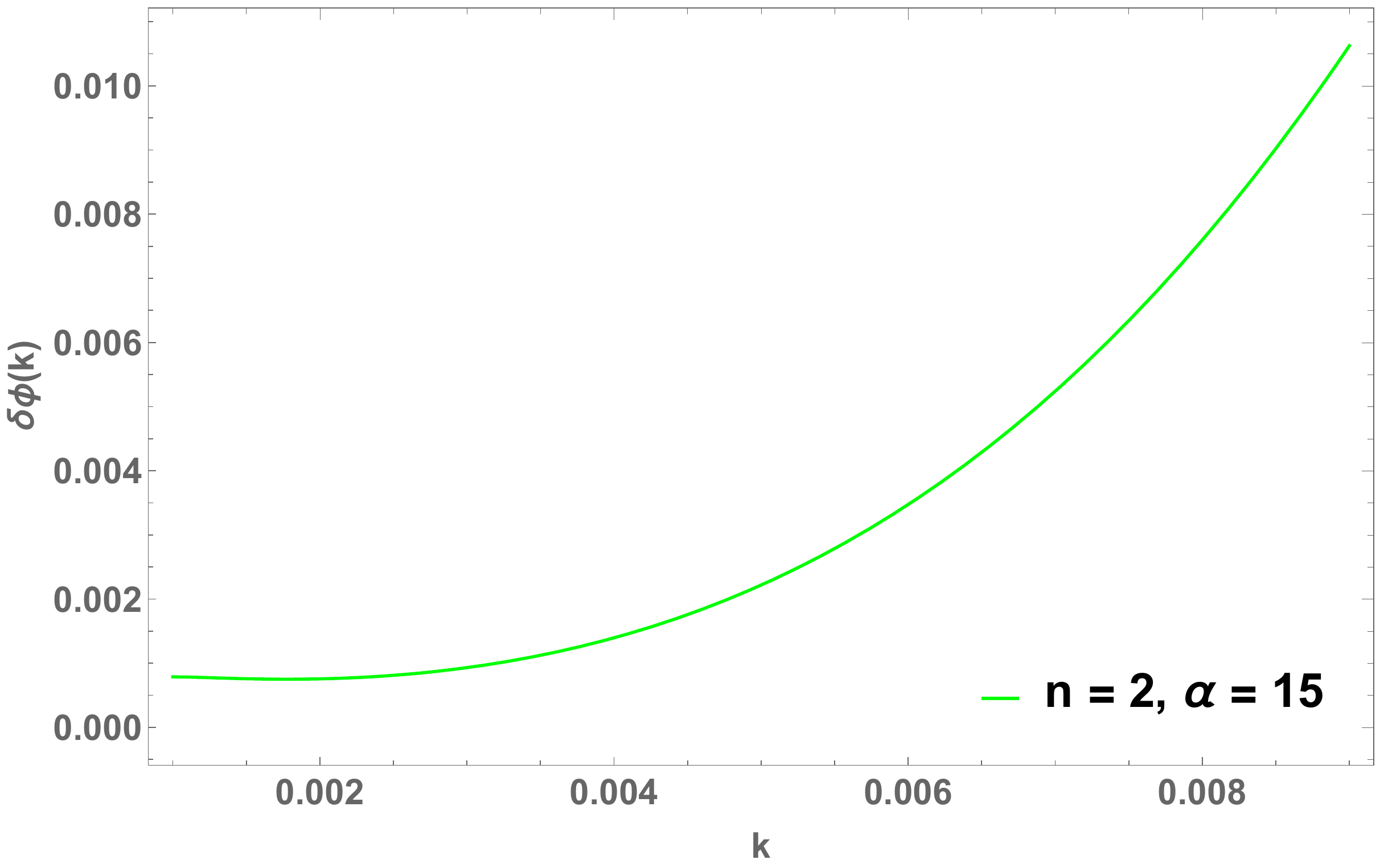}
   \subcaption{}
    \label{fig:perturbed_inflaton_T_2}
\end{subfigure}%
    \caption{Perturbation $\delta \phi(k)$ of the inflaton field, in the $T$-model $\alpha$-attractor potentials, for one value of $n$ and three values of $\alpha$. The perturbation increases with increase in $\alpha$ for a particular value of $k$. }
\label{fig:label_10}    
\end{figure}

In the following figures, we show the results of self-consistent calculations of ( \ref{eq:mode_equation1}), (\ref{eq:mode_equation2}) and (\ref{eq:mode_equation3}) for the unperturbed $\alpha$-attractor potentials (Figures \ref{fig: unperturbed part of the E_potential} for $E$-model and \ref{fig: unperturbed part of the T_potential} for $T$-model), perturbations in the potentials (Figures \ref{fig: perturbed part of the E_potentials} for $E$-model and \ref{fig: perturbed part of the T_potential} for $T$-model), the unperturbed inflaton field (Figures \ref{fig:unperturbed inflaton_E} for $E$-model and \ref{fig:  unperturbed inflaton_T} for $T$-model) and the perturbation in the inflaton field (Figure  \ref{fig: inflaton perturbation_E} for $E$-model and \ref{fig:label_10} for $T$-model) in the $k$-space. In each case, we have plotted the graphs for three values of $\alpha$, \textit{viz}., 1, 10 and 15. For $\alpha=1$ and $\alpha=10$ we have used the low-$\alpha$ limit formula  (\ref{eq:final_tensor_scalar_ratio}) and in case of $\alpha=15$ we have used the high-$\alpha$ limit formula (\ref{eq:high_alpha}), in $k$ space. To solve the nonlinear coupled differential equations, we have used the value of $\phi^{(0)} (k)$ corresponding to $N=60$, according to (\ref{eq:mode e-folds1}) and (\ref{eq:mode e-folds2}), $\phi'^{(0)} (k)=0$, $\delta\phi (k)=0.001$,  $\delta\phi' (k) =0$, $\Phi_B (k) =0.001$ and $\Phi'_B (k)=0$ at $k=0.001$, as the initial conditions. In all our calculations, we have taken the prefactor $V_0 = 1$ as $N$, $n_s$ and $r$ do not depend on its value. 
\begin{figure}[H]
\begin{subfigure}{0.5\textwidth}
  \centering
   \includegraphics[width=75mm,height=80mm]{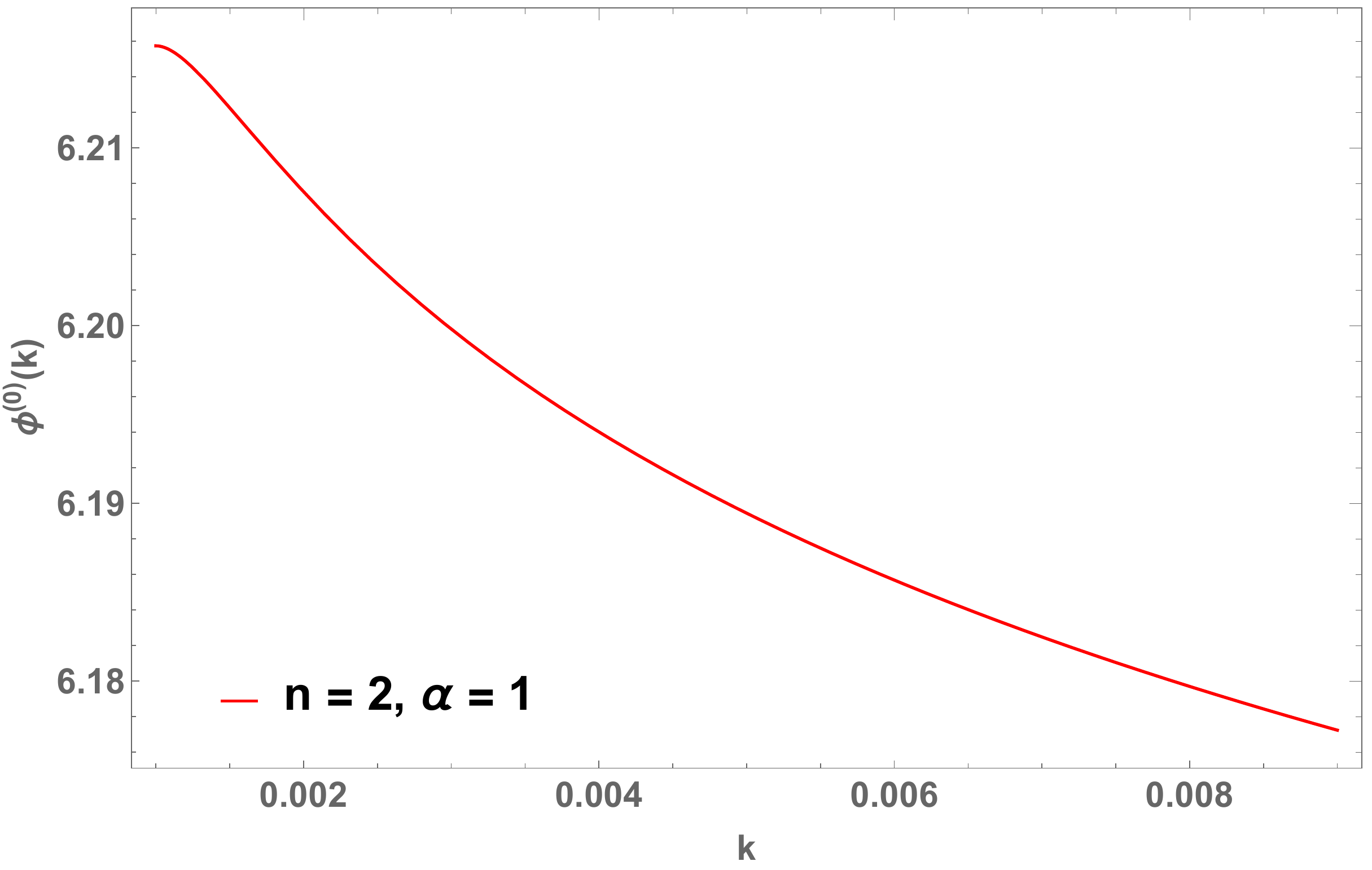}
   \subcaption{}
   \label{fig:unperturbed_inflaton_T_1}
\end{subfigure}%
\begin{subfigure}{0.5\textwidth}
  \centering
   \includegraphics[width=75mm,height=80mm]{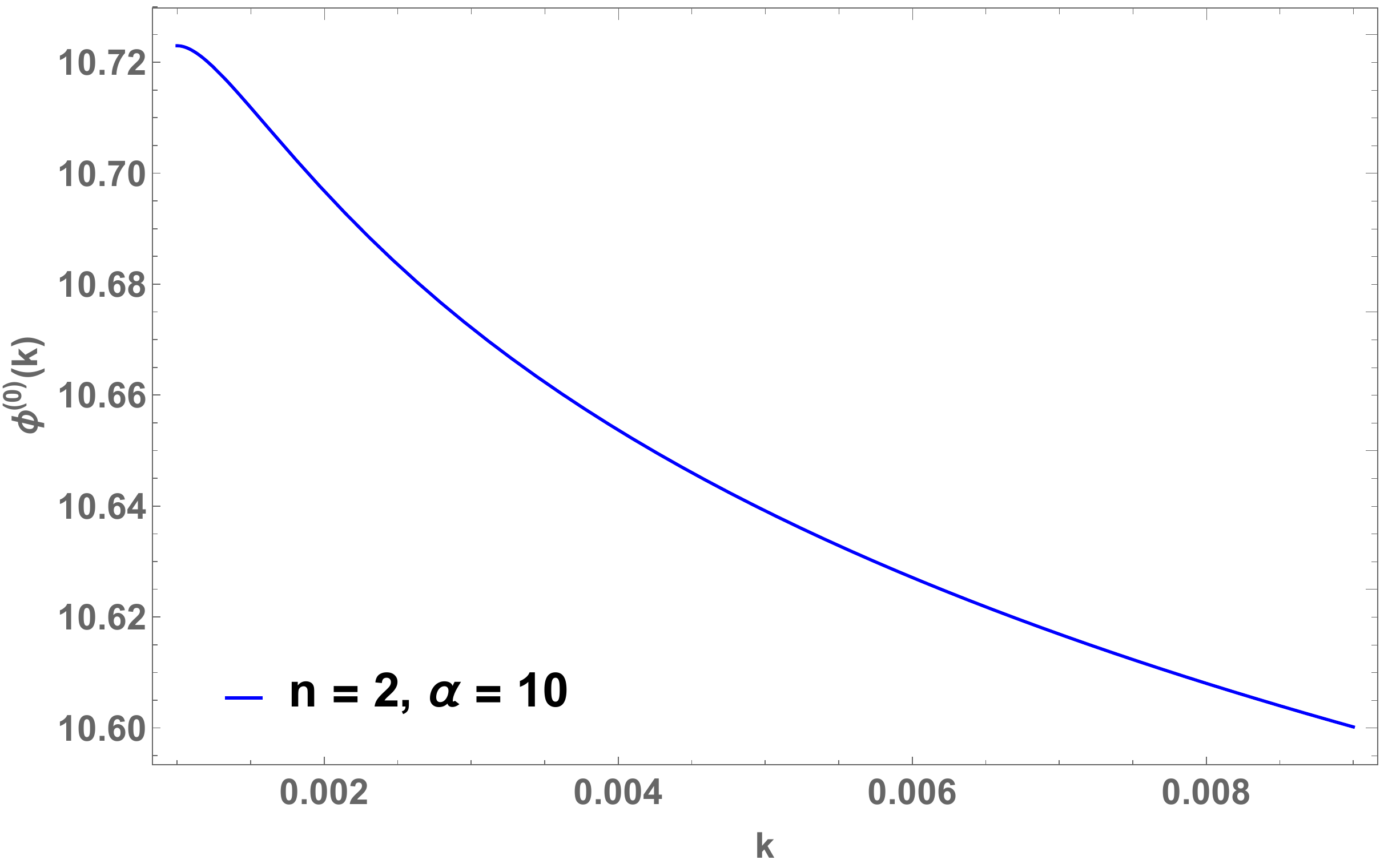}
   \subcaption{}
\end{subfigure}%

\begin{subfigure}{1\textwidth}
  \centering
   \includegraphics[width=80mm,height=80mm]{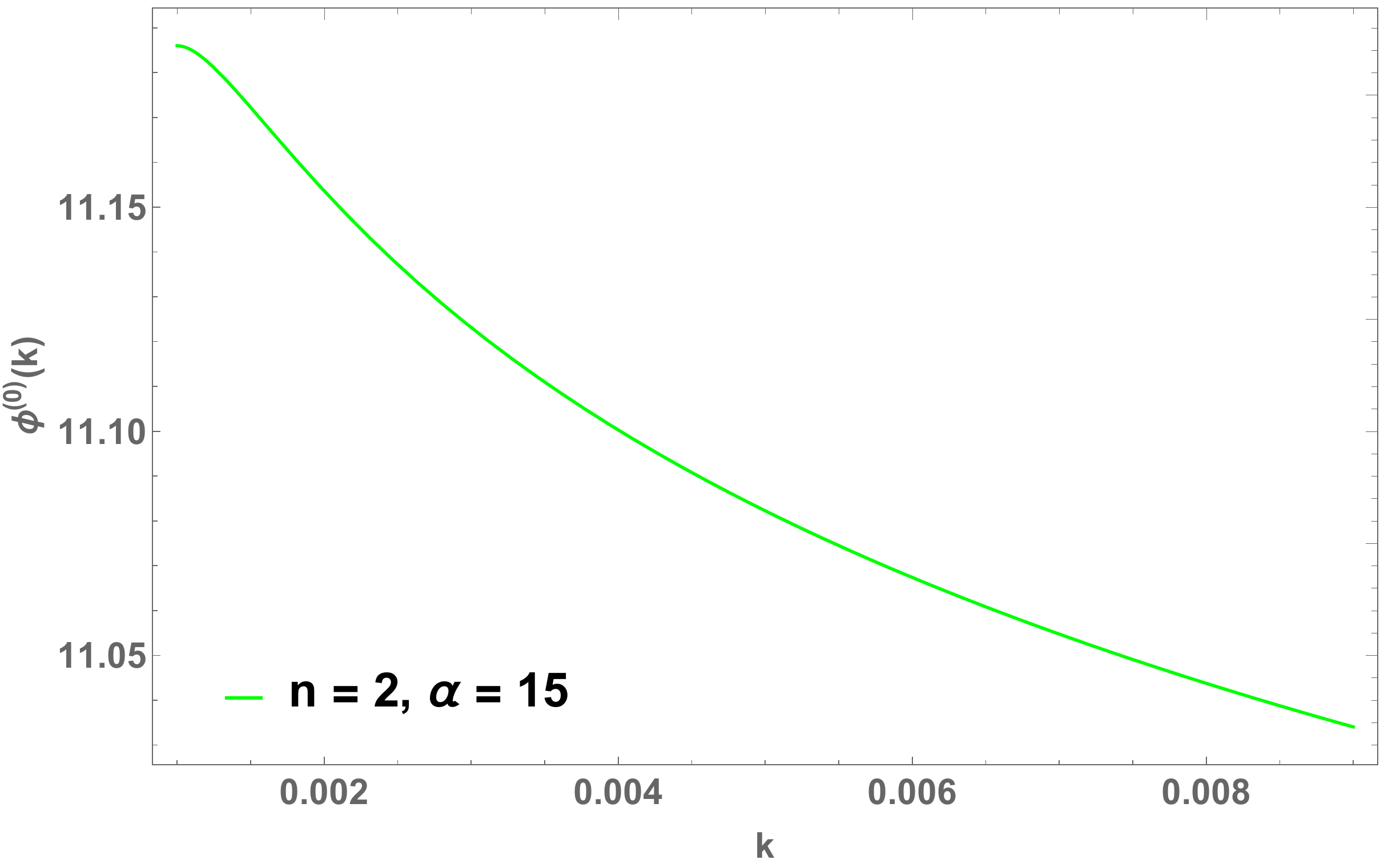}
   \subcaption{}
   \label{fig:unperturbed_inflaton_T_2}
\end{subfigure}%
    \caption{Unperturbed inflaton field, $\phi^{(0)}(k)$, in the $\alpha$-attractor $T$-model potential, for one value of $n$ and three values of $\alpha$. The value of the field increases with increase in $\alpha$ for a given value of $k$.}
    \label{fig: unperturbed inflaton_T}
\end{figure}
In Figures \ref{fig: unperturbed part of the E_potential} -\ref{fig:label_10}, the detailed $k$ space behaviour of the potentials, their perturbations and the fields and their perturbations, obtained by solving their equations self-consistently, are displayed. This figures clearly demonstrate the validity of perturbative calculations in the sub-Planckian r\'{e}gime. For instance, we observe that in Figure \ref{fig:unperturbed_pot_T_1}, $V^{(0)}(k)=0.974858$ and in Figure \ref{fig:perturbed_pot_T_1}, $\delta V(k)=1.45\times 10^{-5}$ at $k=0.004$ for the $T$-model with $\alpha=1$. Similarly in Figure \ref{fig:unperturbed_inflaton_E_1}, $\phi^{(0)}(k)=5.336$ and in Figure \ref{fig:perturbed_inflaton_E_1}, $\delta\phi(k)=7.1\times 10^{-4}$ at $k=0.006$ for the $E$-model with $\alpha=1$. Overall, it is to be noticed that, the changes in the inflaton potential corresponding to the perturbative changes in the inflaton field is also perturbative in the  background metric, $\Phi_{B}$.
(Here, $k$ is in Mpc$^{-1}$ and  all other numbers are in Planck units.)
\subsection{Power spectra and cosmological parameters in \emph{k}-space}
\begin{figure}[H]
\begin{subfigure}{0.5\textwidth}
  \centering
   \includegraphics[width=75mm,height=80mm]{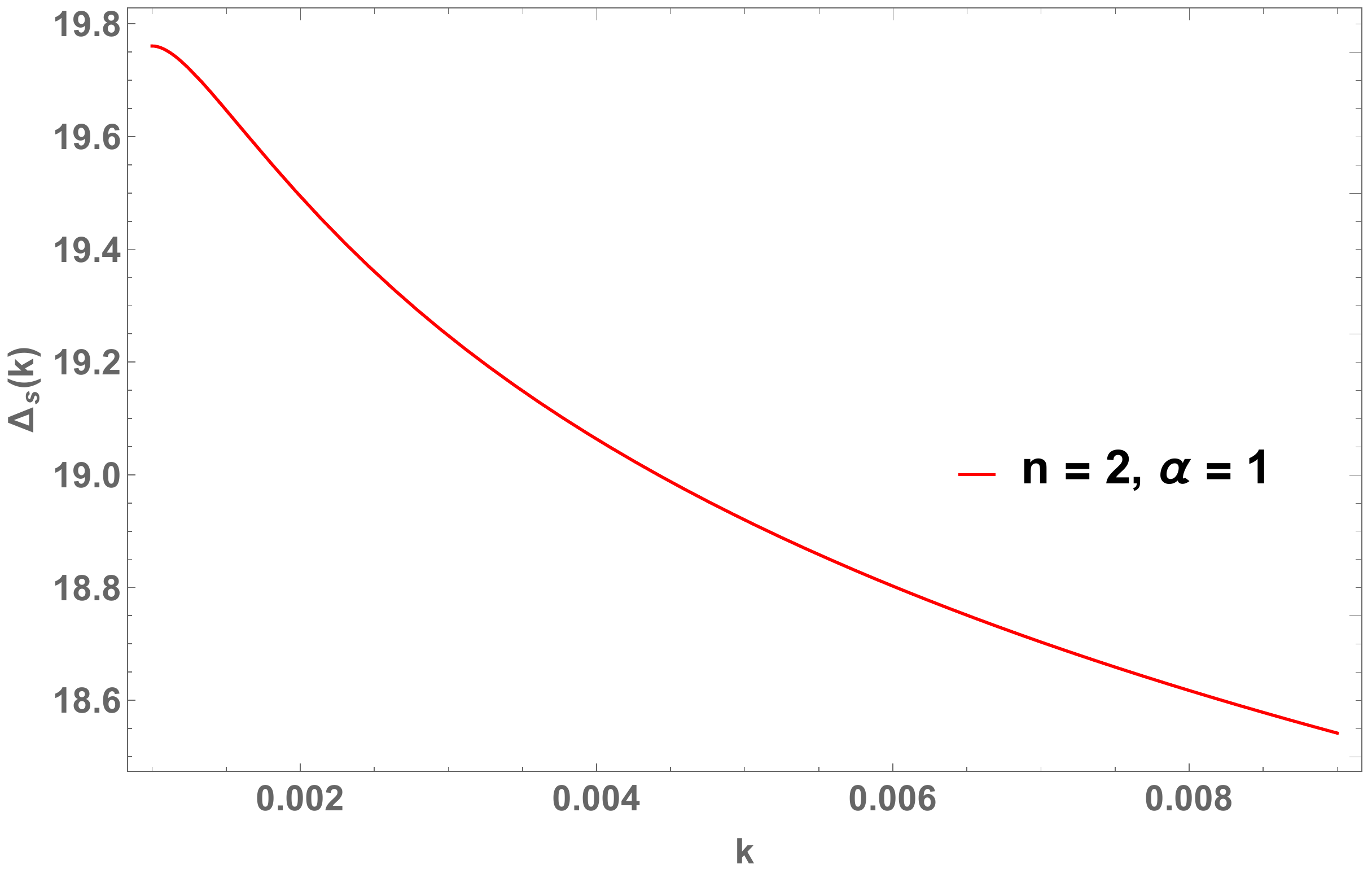}
   \subcaption{}
   \label{fig:scalar_power_spectrum_T_1}
\end{subfigure}%
\begin{subfigure}{0.5\textwidth}
  \centering
   \includegraphics[width=75mm,height=80mm]{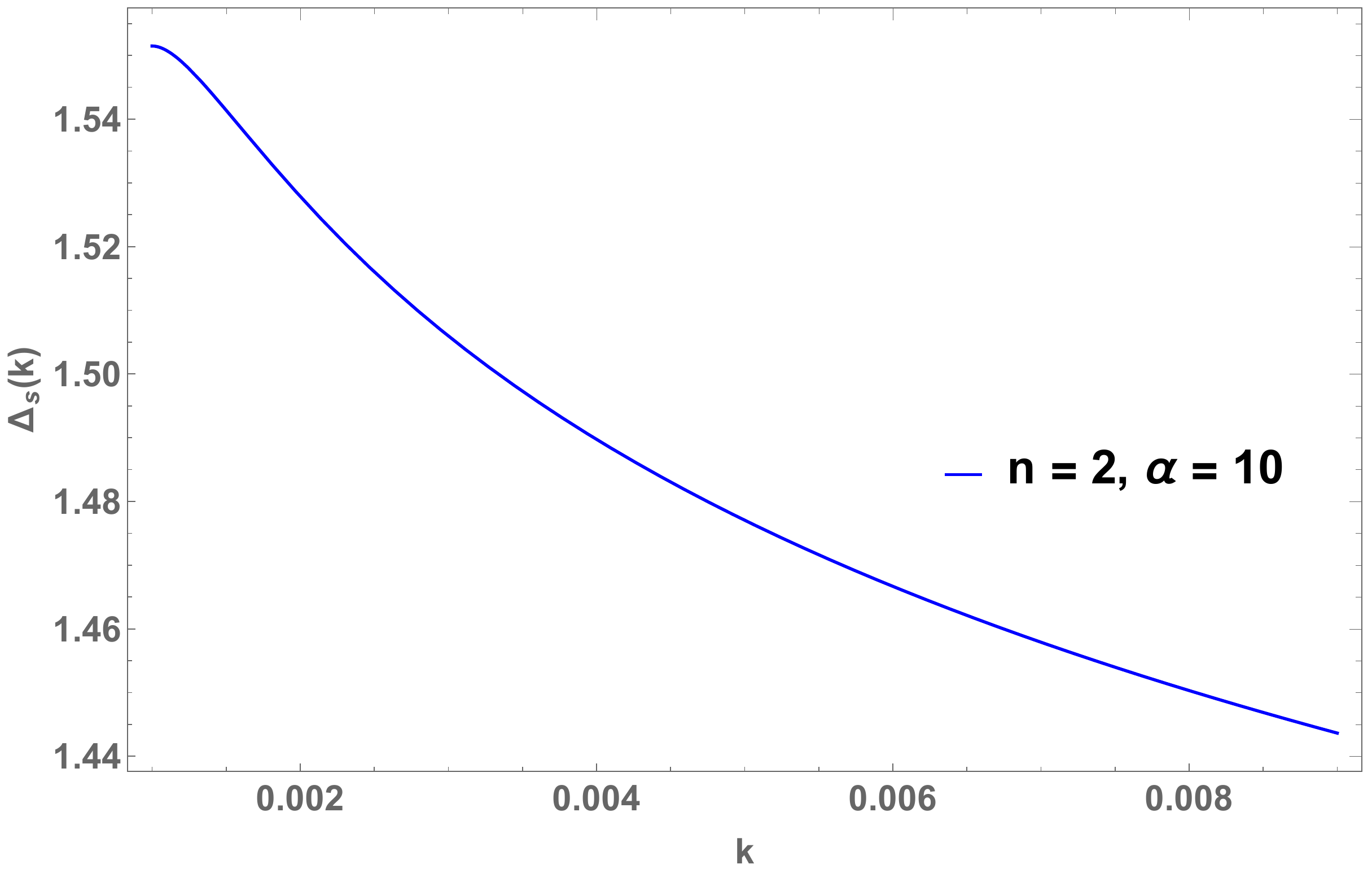}
   \subcaption{}
\end{subfigure}%

\begin{subfigure}{1\textwidth}
  \centering
   \includegraphics[width=80mm,height=80mm]{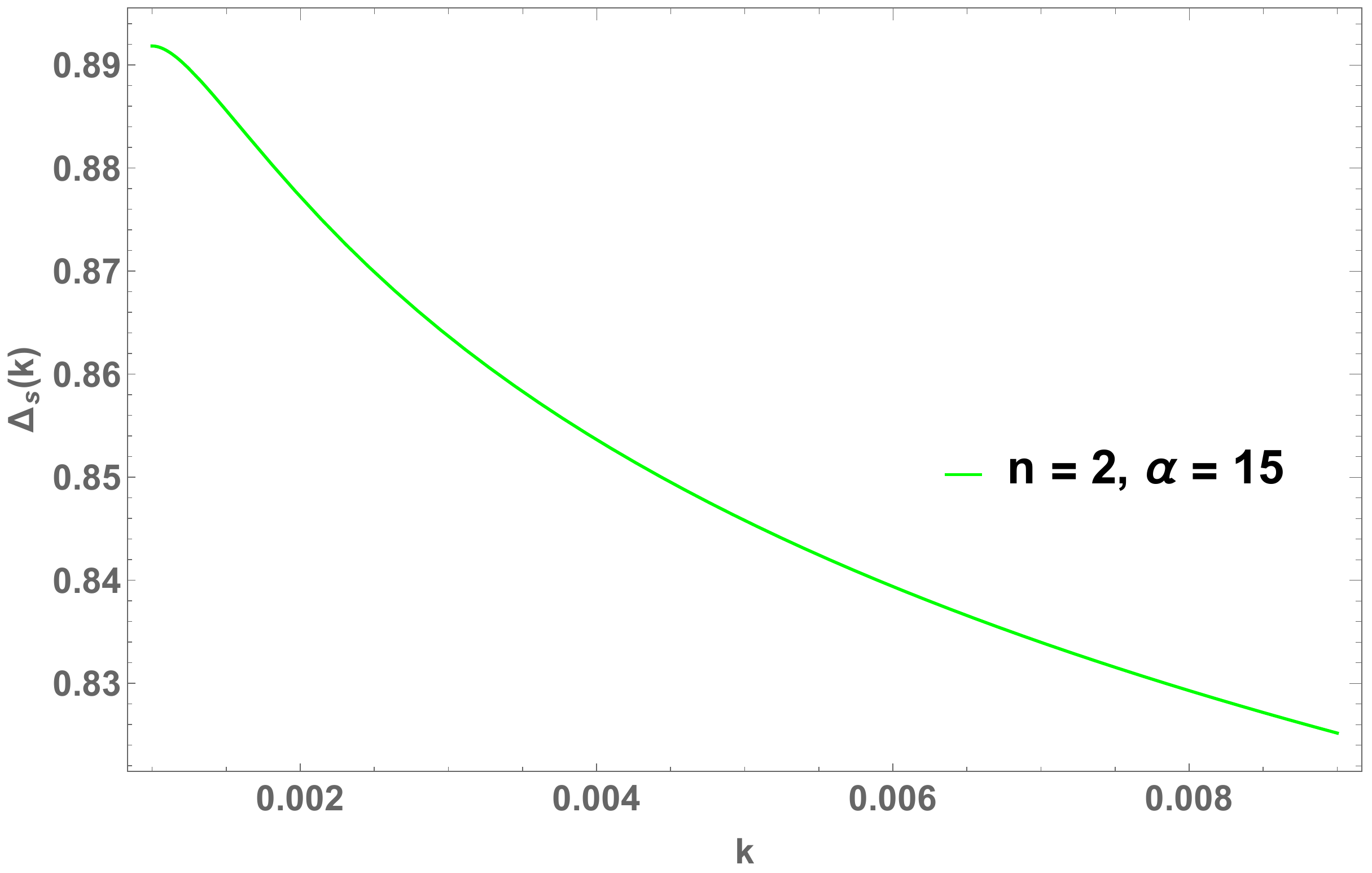}
   \subcaption{}
   \label{fig:scalar_power_spectrum_T_2}
\end{subfigure}%
    \caption{Scalar power spectrum, $\Delta_s (k)$, for the $\alpha$-attractors $T$-model potentials, for one value of $n$ and three values of $\alpha$. As $\alpha$ increases, the power spectrum decreases at a particular value of $k$, thereby, showing less correlations among the fluctuations at this value of $k$.}
    \label{fig: scalar power spectrum_T}
\end{figure}

\begin{figure}[H]
\begin{subfigure}{0.5\textwidth}
  \centering
   \includegraphics[width=75mm,height=80mm]{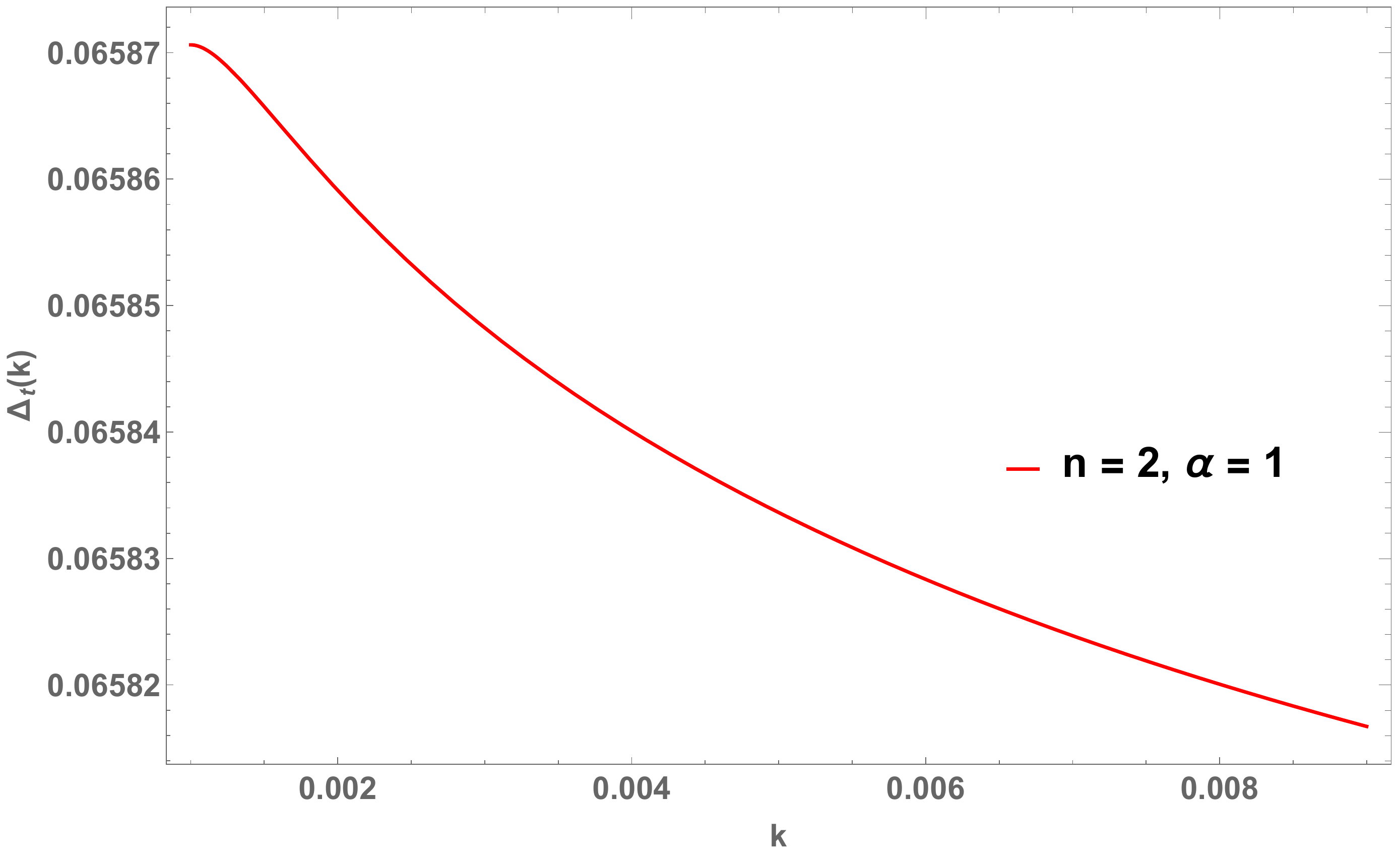} 
   \subcaption{}
   \label{fig:tensor_power_spectrum_E_1}
\end{subfigure}%
\begin{subfigure}{0.5\textwidth}
  \centering
   \includegraphics[width=75mm,height=80mm]{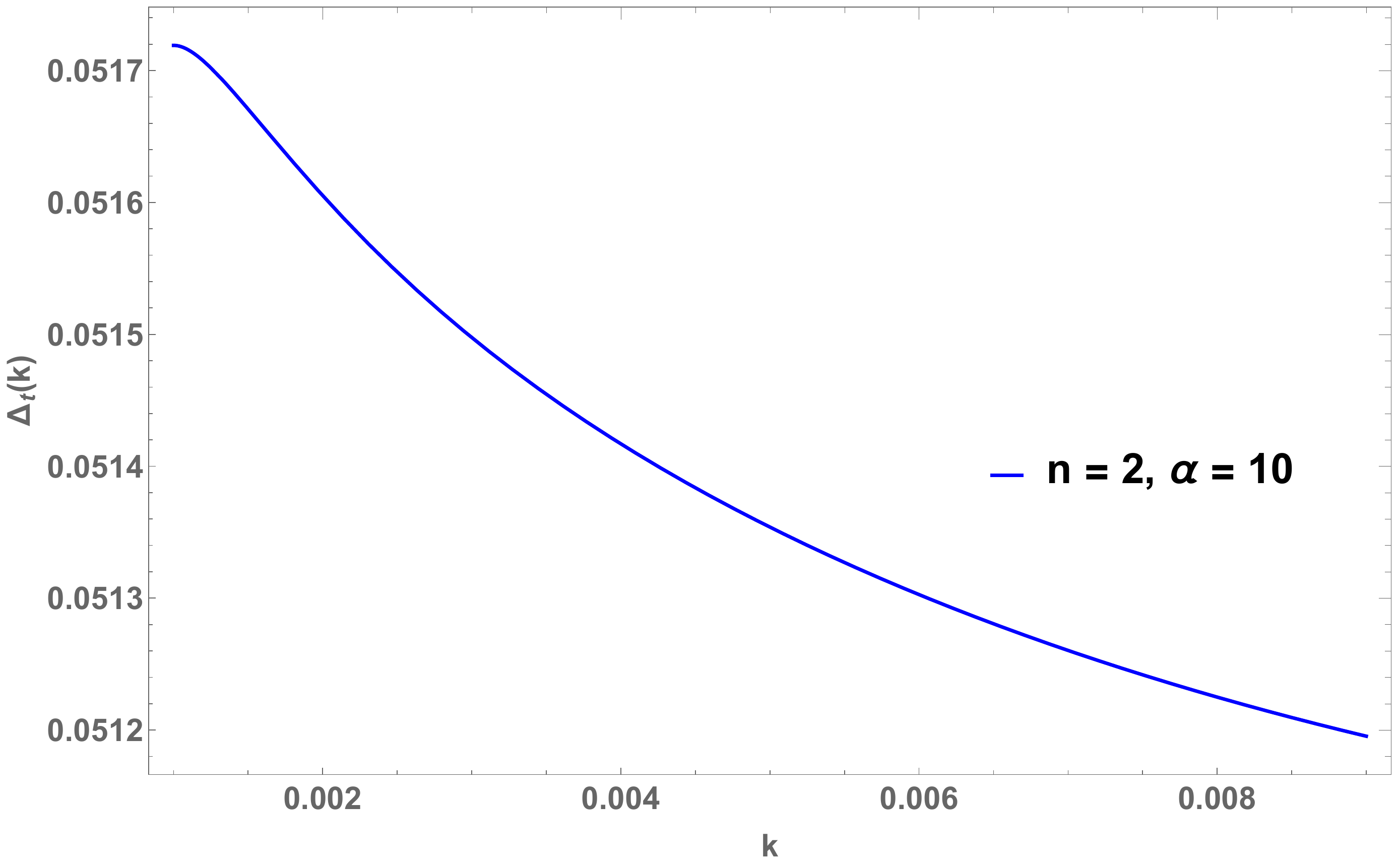}
   \subcaption{}
\end{subfigure}%

\begin{subfigure}{1\textwidth}
  \centering
   \includegraphics[width=80mm,height=80mm]{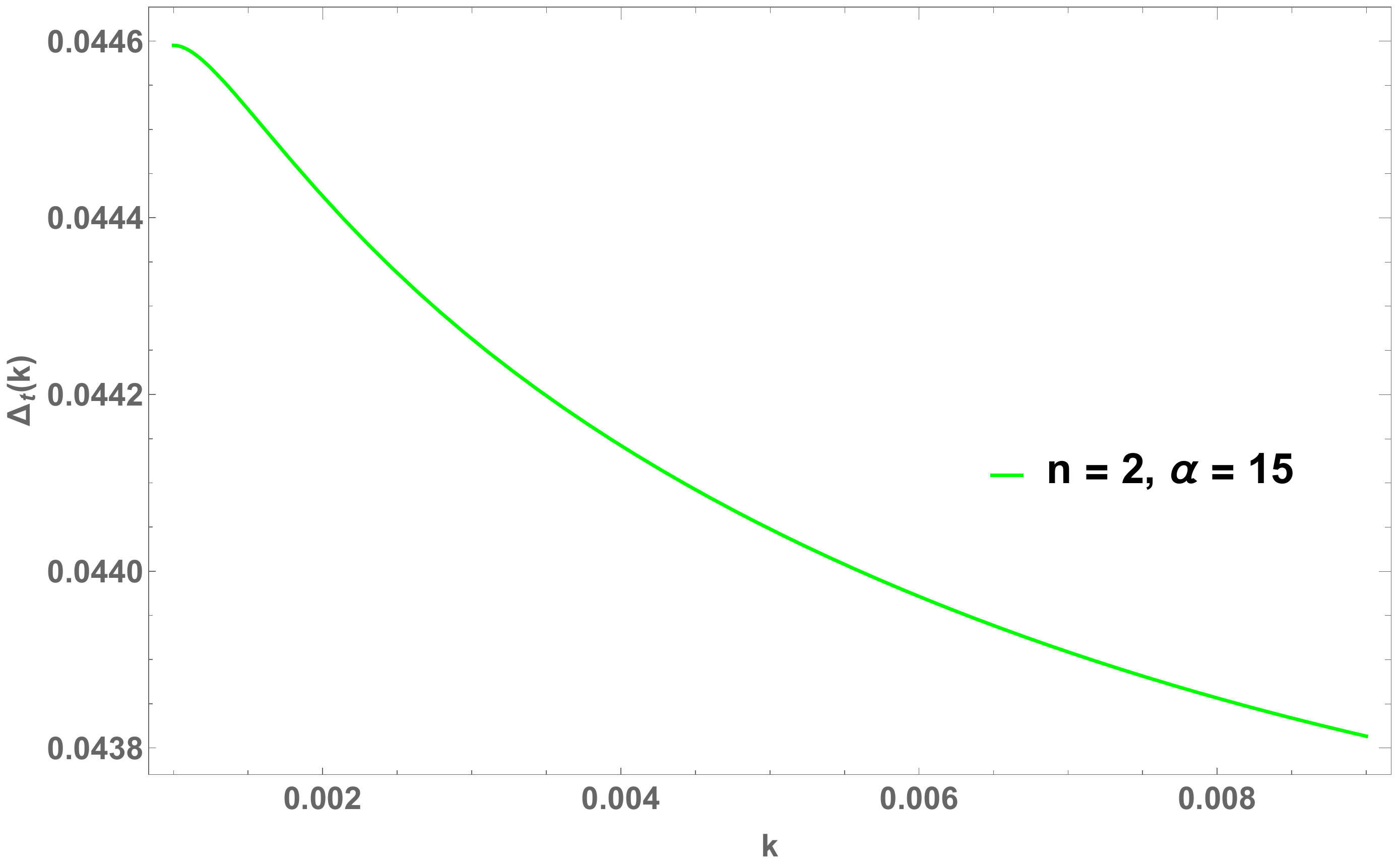}
   \subcaption{}
    \label{fig:tensor_power_spectrum_E_2}
\end{subfigure}
\caption{Tensor power spectrum, $\Delta_t (k)$, for the $\alpha$-attractor $E$-model potentials, for one value of $n$ and three values of $\alpha$. As $\alpha$ increases, the power spectrum decreases at a particular value of $k$, thereby, showing less correlations among the fluctuations at this value of $k$. However, compared to the case of the scalar power spectrum, here, the decrease in the tensor power spectrum is small. }
\label{fig: tensor power spectrum_E}
\end{figure}

\begin{figure}[H]
\begin{subfigure}{0.5\textwidth}
  \centering
   \includegraphics[width=75mm,height=80mm]{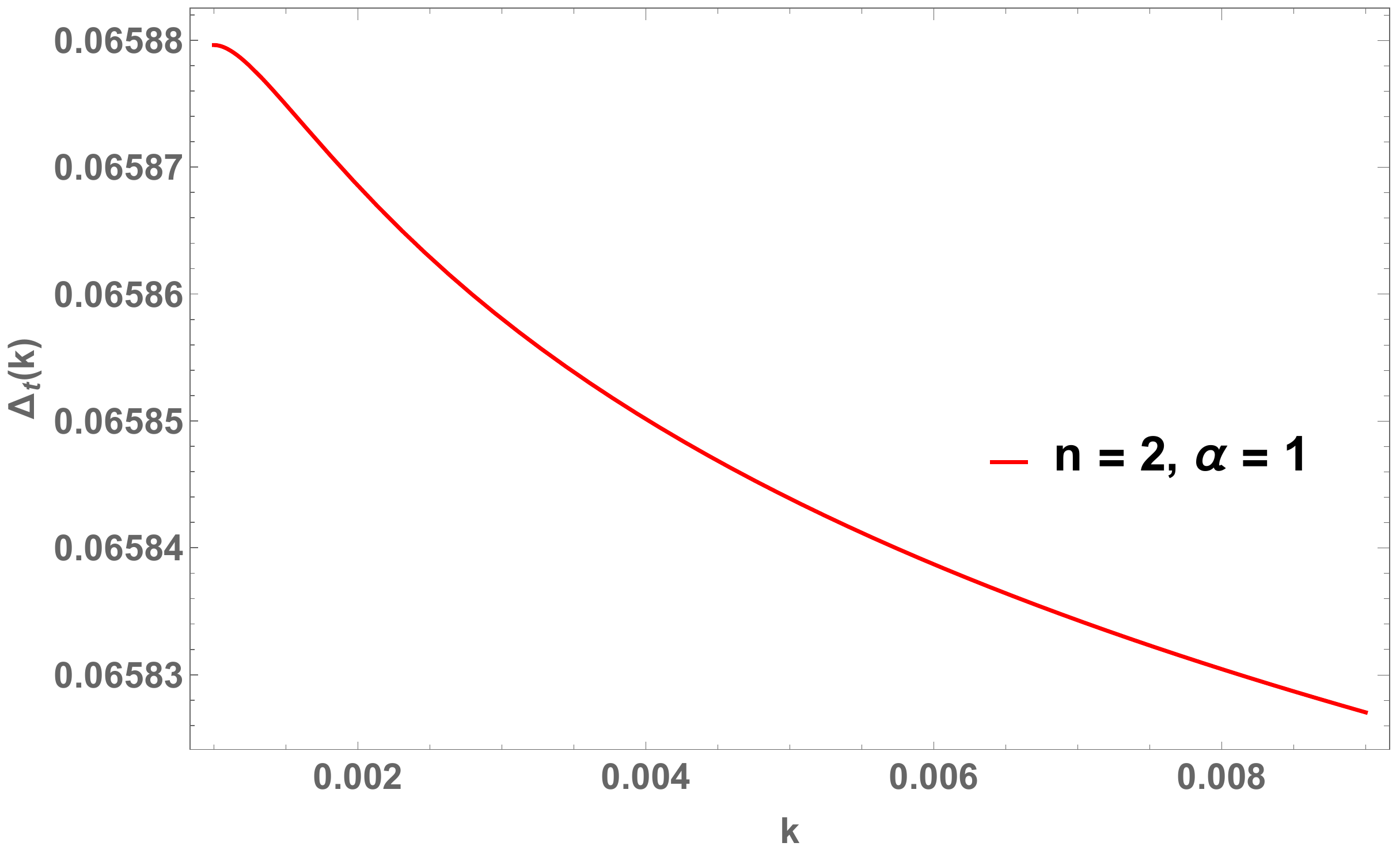}
   \subcaption{}
    \label{fig:tensor_power_spectrum_T_1}
\end{subfigure}%
\begin{subfigure}{0.5\textwidth}
  \centering
   \includegraphics[width=75mm,height=80mm]{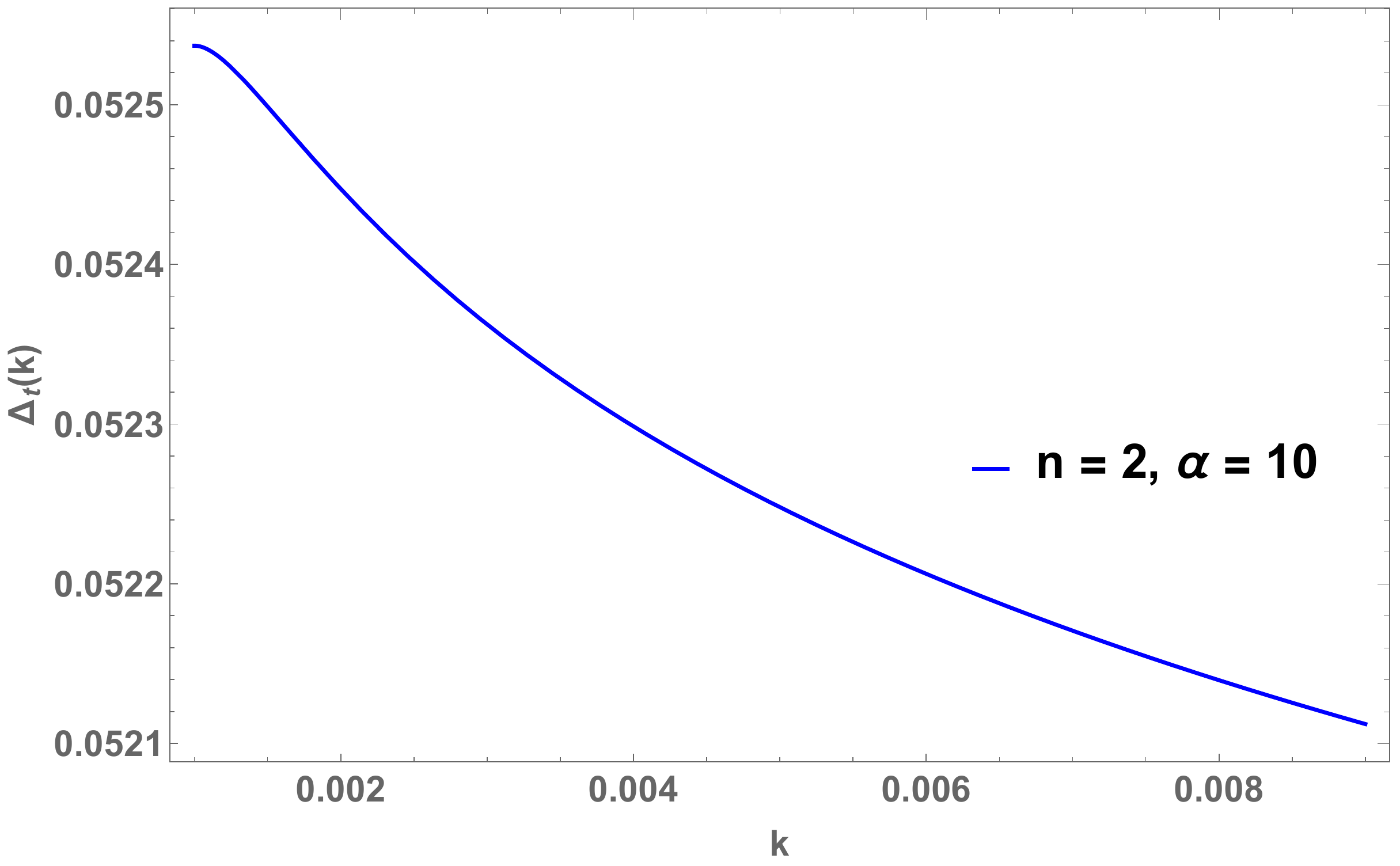}
   \subcaption{}
\end{subfigure}%

\begin{subfigure}{1\textwidth}
  \centering
   \includegraphics[width=80mm,height=80mm]{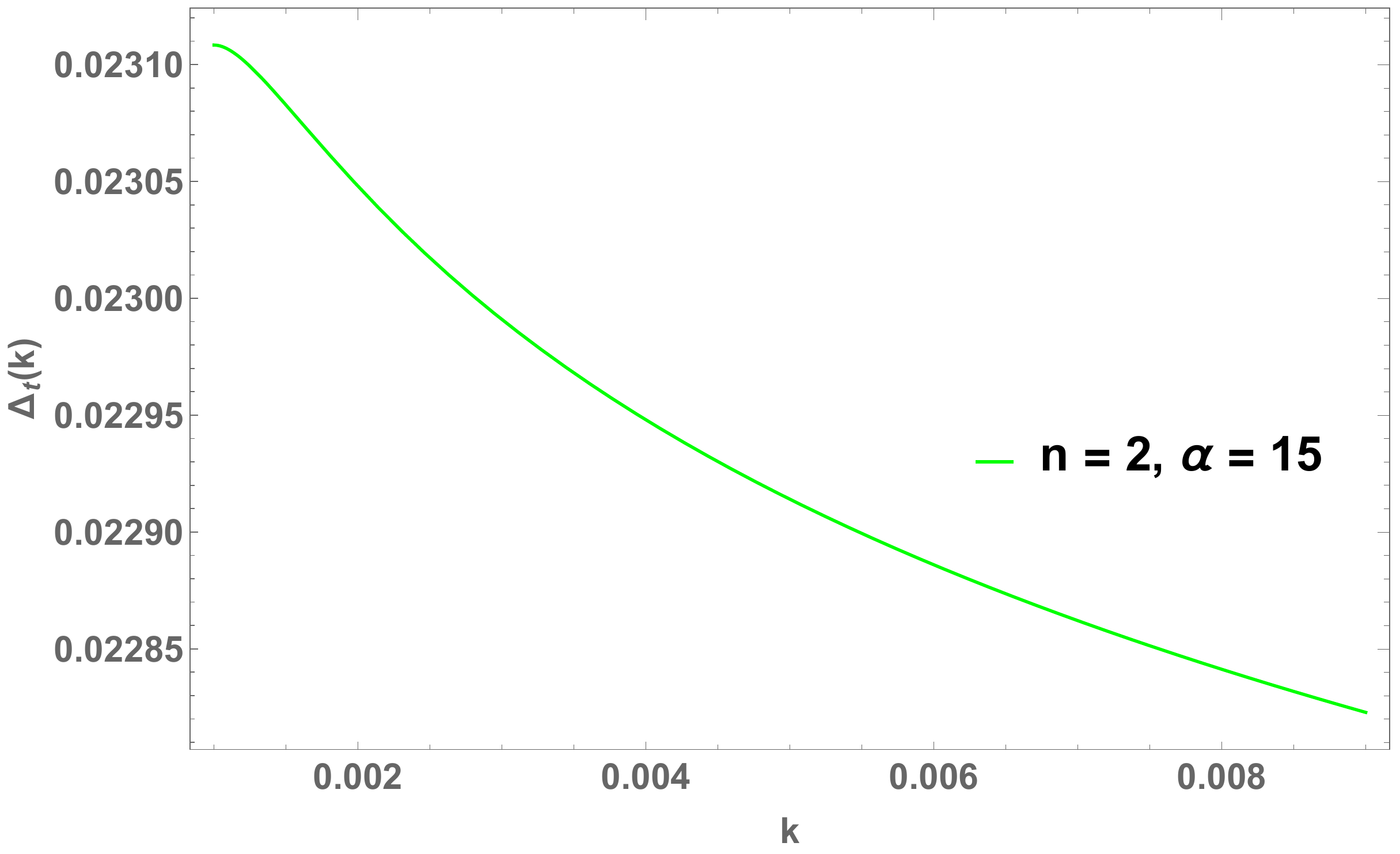}
   \subcaption{}
    \label{fig:tensor_power_spectrum_T_2}
\end{subfigure}%
    \caption{Tensor power spectrum, $\Delta_t (k)$, for the $\alpha$-attractor $T$-model potentials, for one value of $n$ and three values of $\alpha$. As $\alpha$ increases, the power spectrum decreases at a particular value of $k$, thereby, showing less correlations among the fluctuations at this value of $k$. However, compared to the case of the scalar power spectrum, here, the decrease in the tensor power spectrum is small.}
    \label{fig:fig_14}
\end{figure}

We have plotted the scalar (Figures \ref{fig:  scalar power spectrum_E} for the $E$-model and \ref{fig: scalar power spectrum_T} for the $T$-model) and the tensor (Figures \ref{fig: tensor power spectrum_E} for the $E$-model and \ref{fig:fig_14} for the $T$-model) power spectra corresponding to various values of $\alpha$ against the sub-Planckian mode momenta $k$. These figures clearly show that the magnitudes of both the scalar and the tensor power spectra decrease as $\alpha$ increases. However, the decrease in the former is more than that in the latter. Thus, $r$ has the tendency to increase as $\alpha$ increases. Interestingly, the analysis of Planck data \cite{Akrami:2018odb} shows a correlation between the low values of $\alpha$ $(\alpha_{E}< 19.95$ and $\alpha_{T}< 10)$ and the small values of $r$ ($r_{0.002} <0.056)$. Table $5$ in Ref.\cite{Akrami:2018odb} shows that the  models with these values of $\alpha$ are favored, whereas the power-law potentials (which correspond to large values of $\alpha$) are strongly disfavored. It is also shown (Figures \ref{fig: tensor power spectrum_E}  and \ref{fig:fig_14}) that the magnitudes of the tensor power spectra are quite small, which corresponds to the fact the primordial gravitational waves originated from tiny tensor fluctuations. 
\begin{figure}[H]
\begin{subfigure}{0.5\textwidth}
  \centering
   \includegraphics[width=75mm,height=55mm]{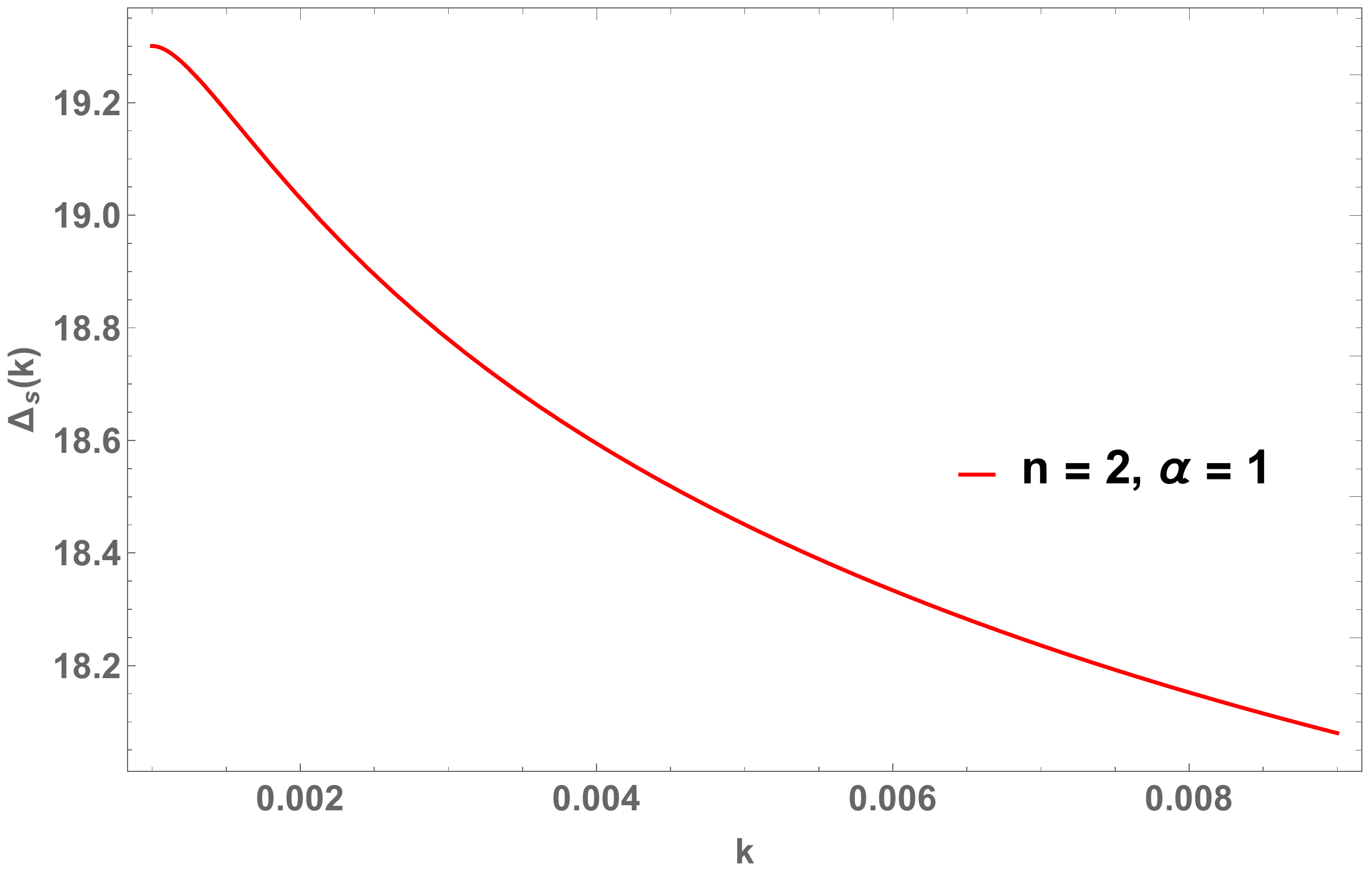}
   \subcaption{}
   \label{fig:scalar_power_spectrum_E_1}
\end{subfigure}%
\begin{subfigure}{0.5\textwidth}
  \centering
   \includegraphics[width=75mm,height=55mm]{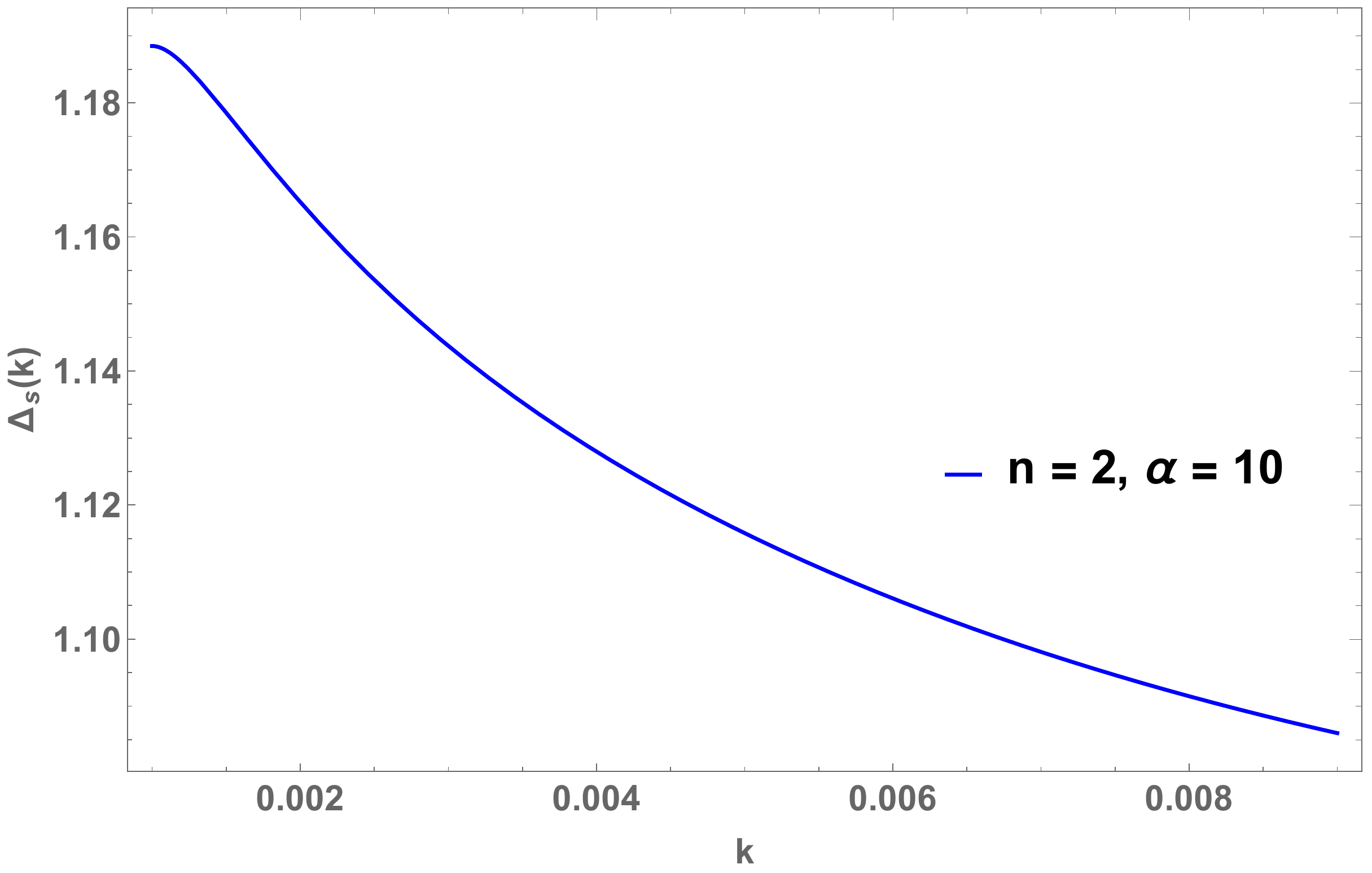}
   \subcaption{}
\end{subfigure}%

\begin{subfigure}{1.0\textwidth}
  \centering
   \includegraphics[width=75mm,height=55mm]{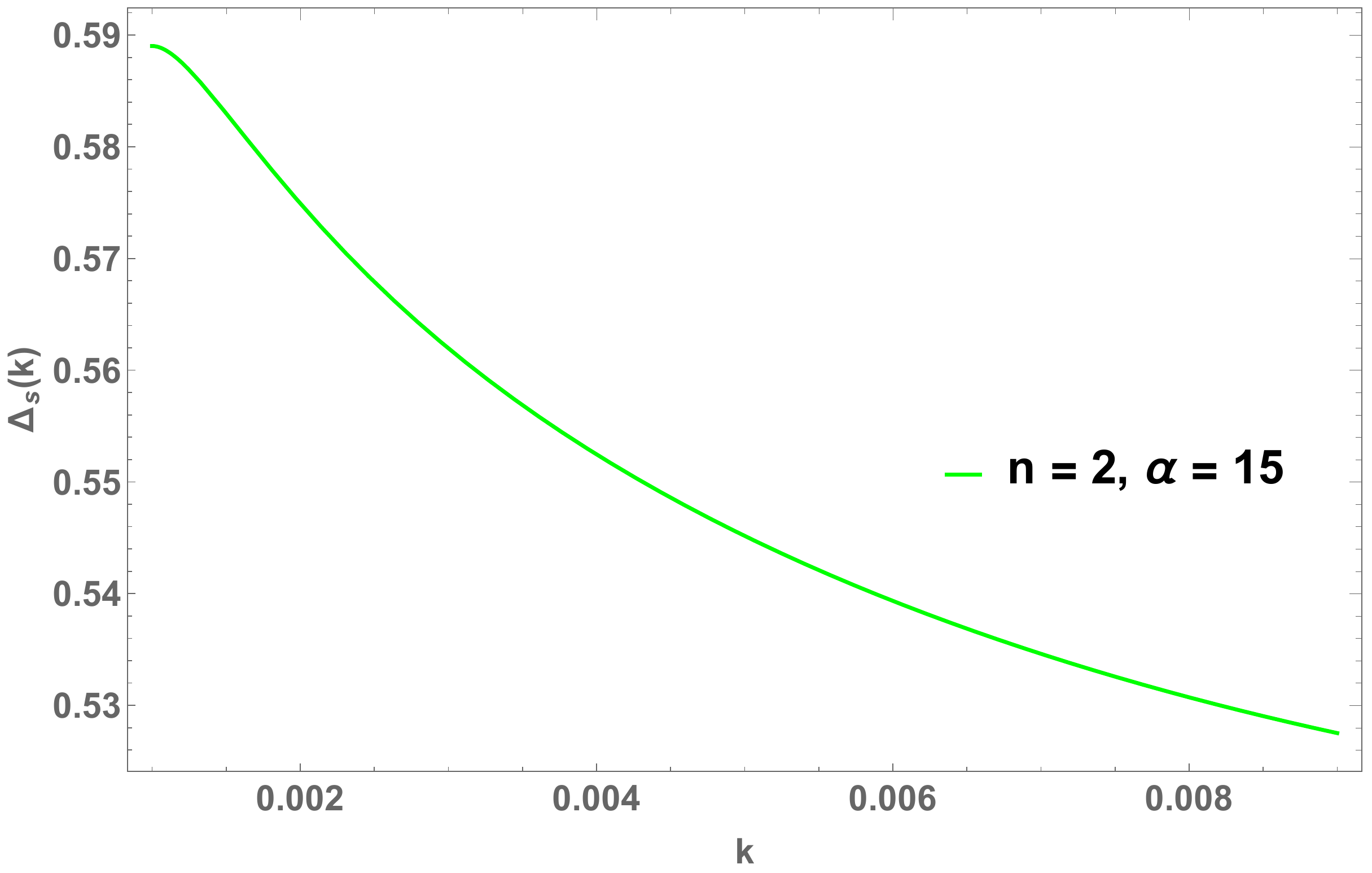}
   \subcaption{}
   \label{fig:scalar_power_spectrum_E_2}
\end{subfigure}
\caption{Scalar power spectrum, $\Delta_s (k)$, for the $\alpha$-attractor $E$-model potentials, for one value of $n$ and three values of $\alpha$. As $\alpha$ increases, the power spectrum decreases at a particular value of $k$, thereby, showing less correlations among the fluctuations at this value of $k$. } 
\label{fig: scalar power spectrum_E}
\end{figure}

\begin{figure}[H]
\begin{subfigure}{0.5\textwidth}
  \centering
   \includegraphics[width=75mm,height=80mm]{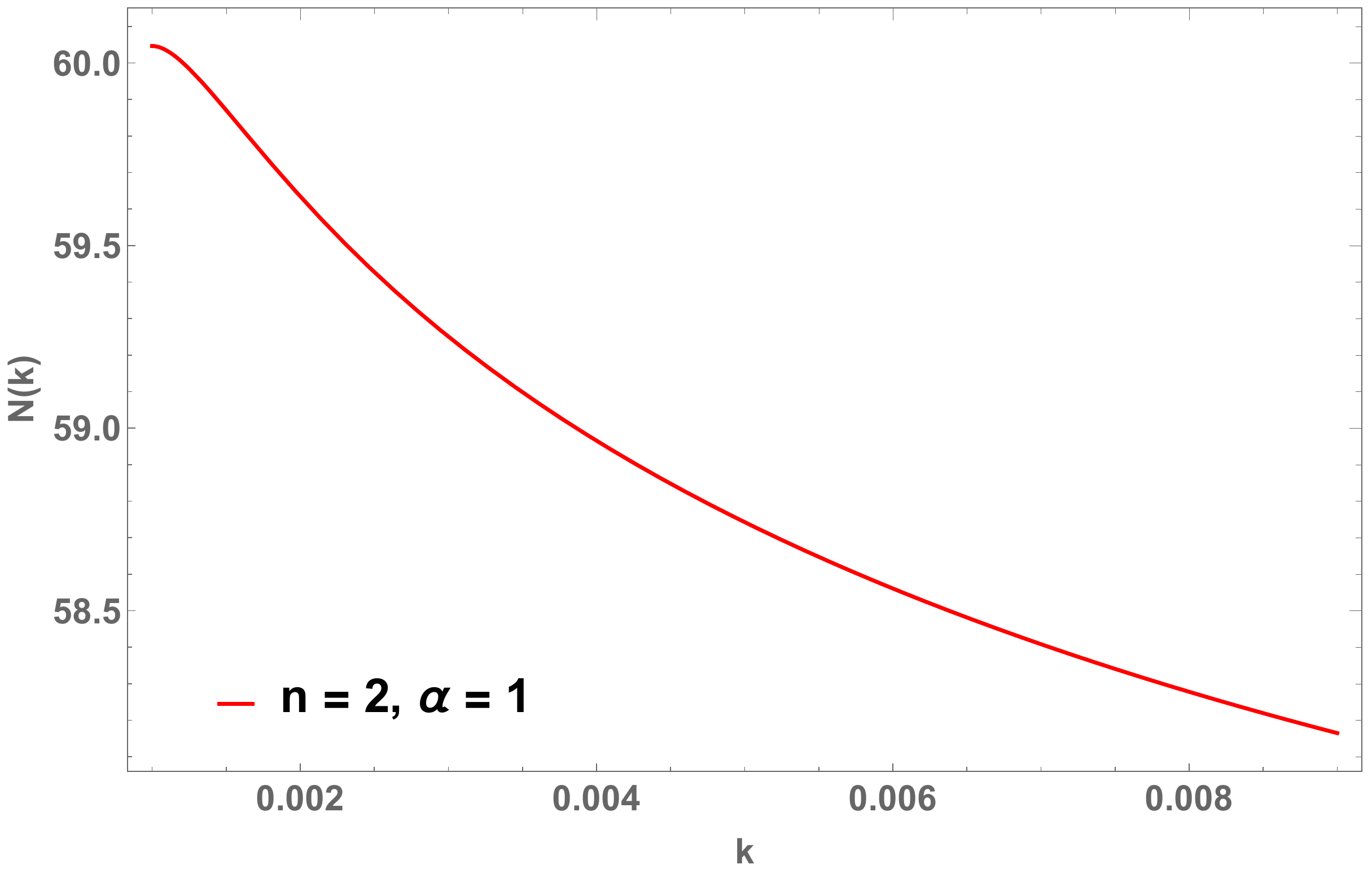}
   \subcaption{}
   \label{fig:efolds_E_1}
\end{subfigure}%
\begin{subfigure}{0.5\textwidth}
  \centering
   \includegraphics[width=75mm,height=80mm]{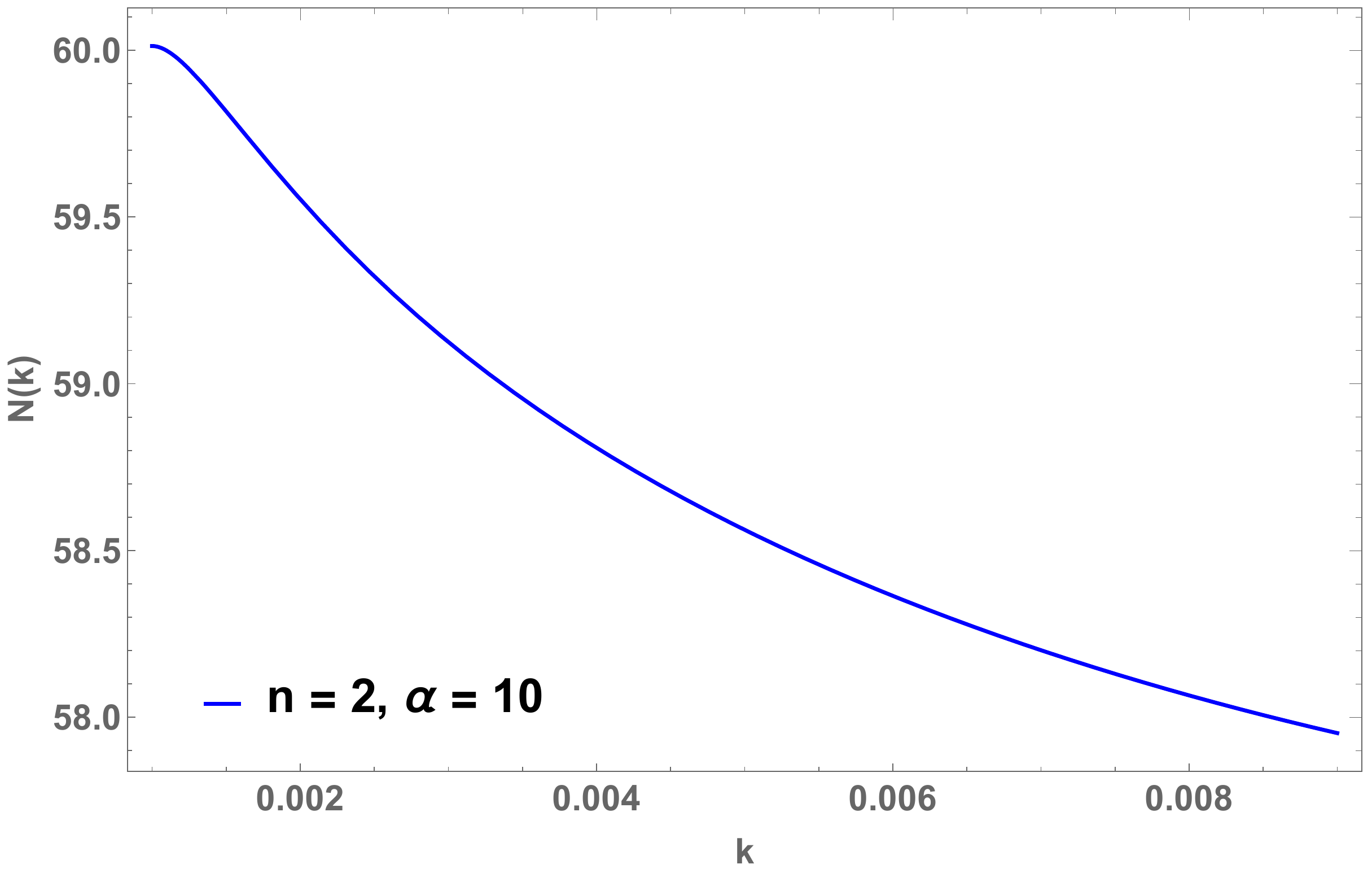}
   \subcaption{}
\end{subfigure}%

\begin{subfigure}{1\textwidth}
  \centering
   \includegraphics[width=80mm,height=80mm]{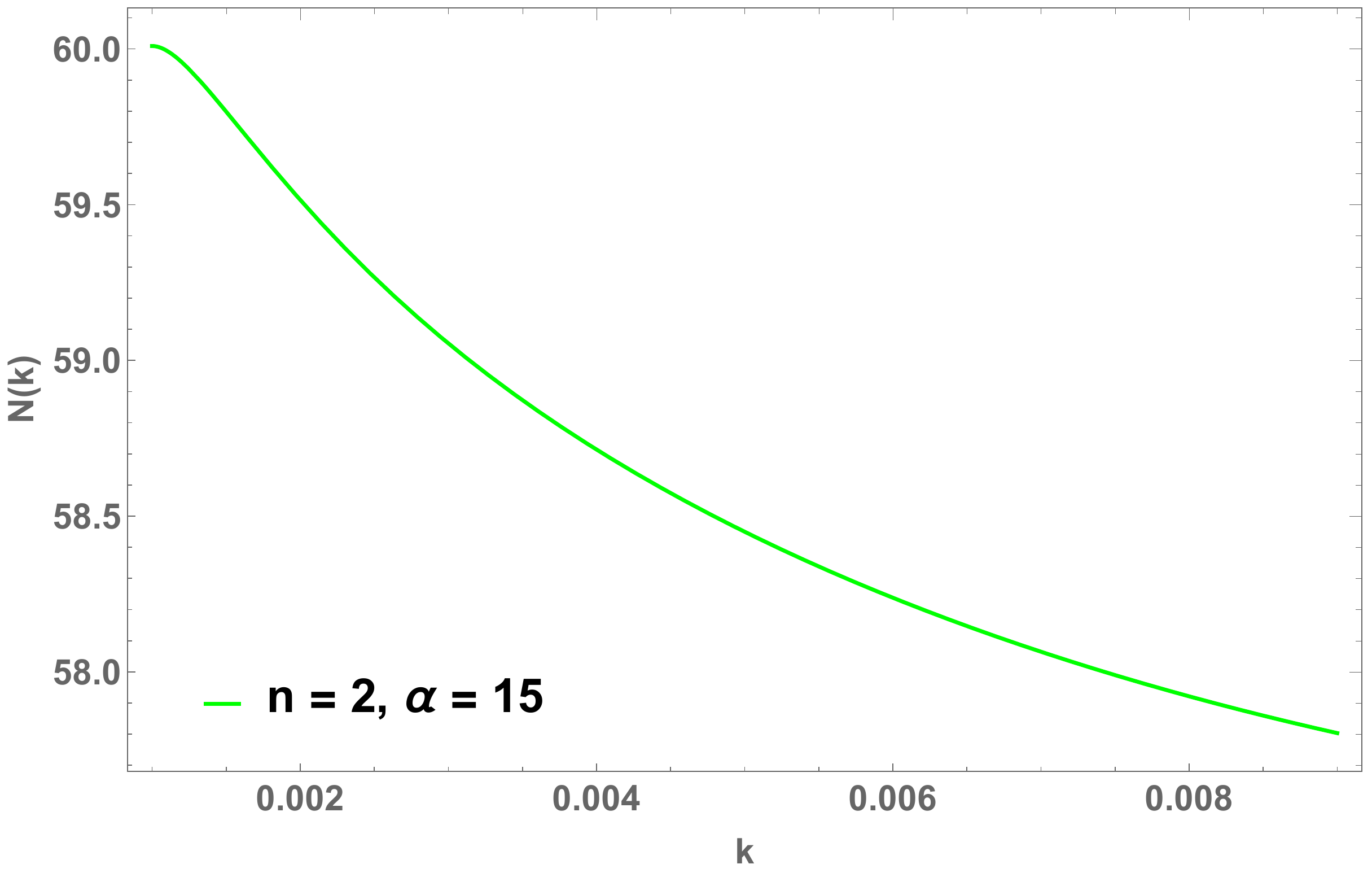}
   \subcaption{}
   \label{fig:efolds_E_2}
\end{subfigure}
\caption{Number of e-folds, $N(k)$, for the  $\alpha$ attractor $E$-model potentials,  for one value of $n$ and three values of $\alpha$. Very small variations in the $e$-folds is observed, due to the variation in $\alpha$. }
\label{fig:number of e-folds_E}
\end{figure}

\begin{figure}[H]
\begin{subfigure}{0.5\textwidth}
  \centering
   \includegraphics[width=75mm,height=80mm]{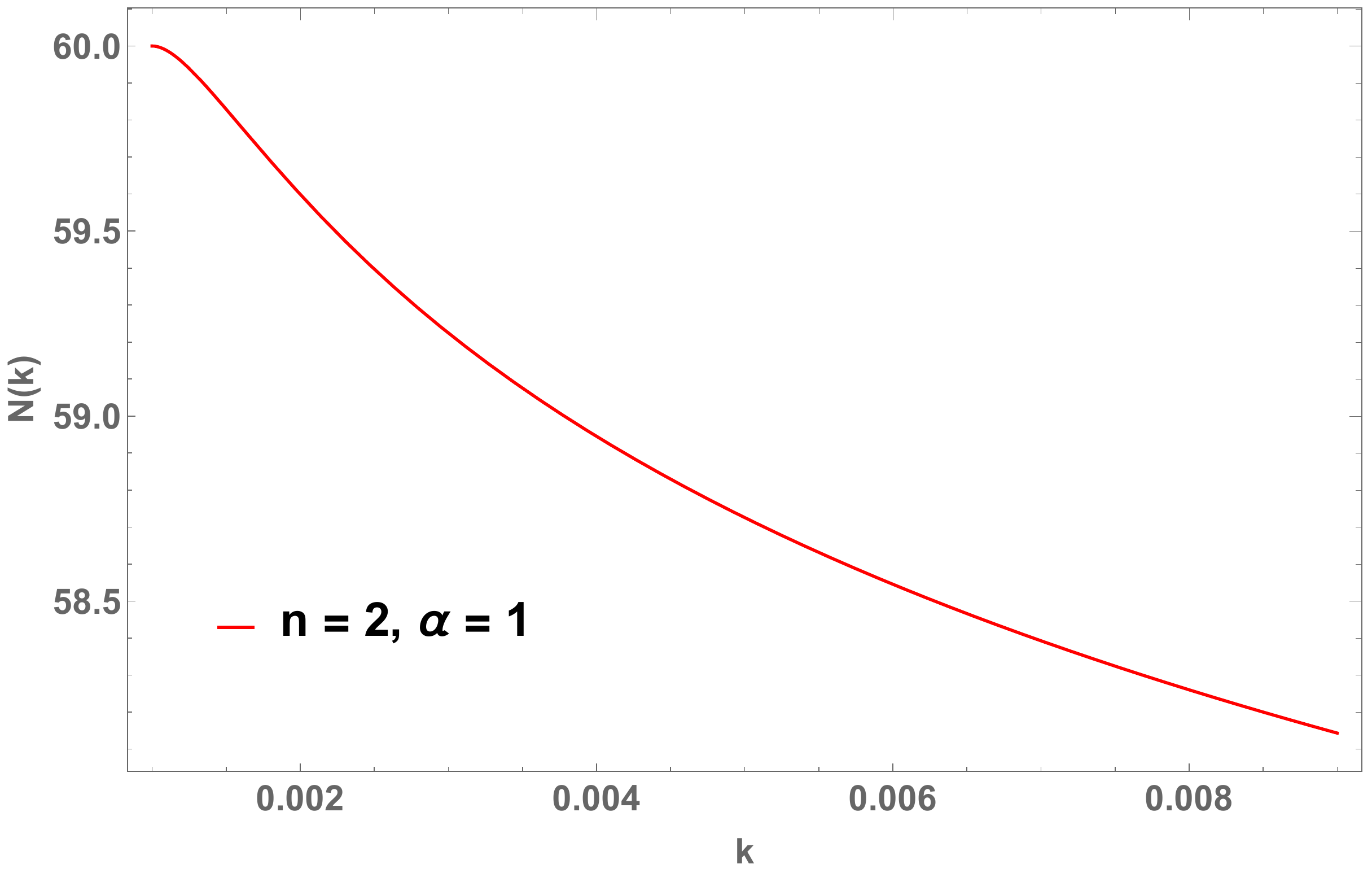}
   \subcaption{}
   \label{fig:efolds_T_1}
\end{subfigure}%
\begin{subfigure}{0.5\textwidth}
  \centering
   \includegraphics[width=75mm,height=80mm]{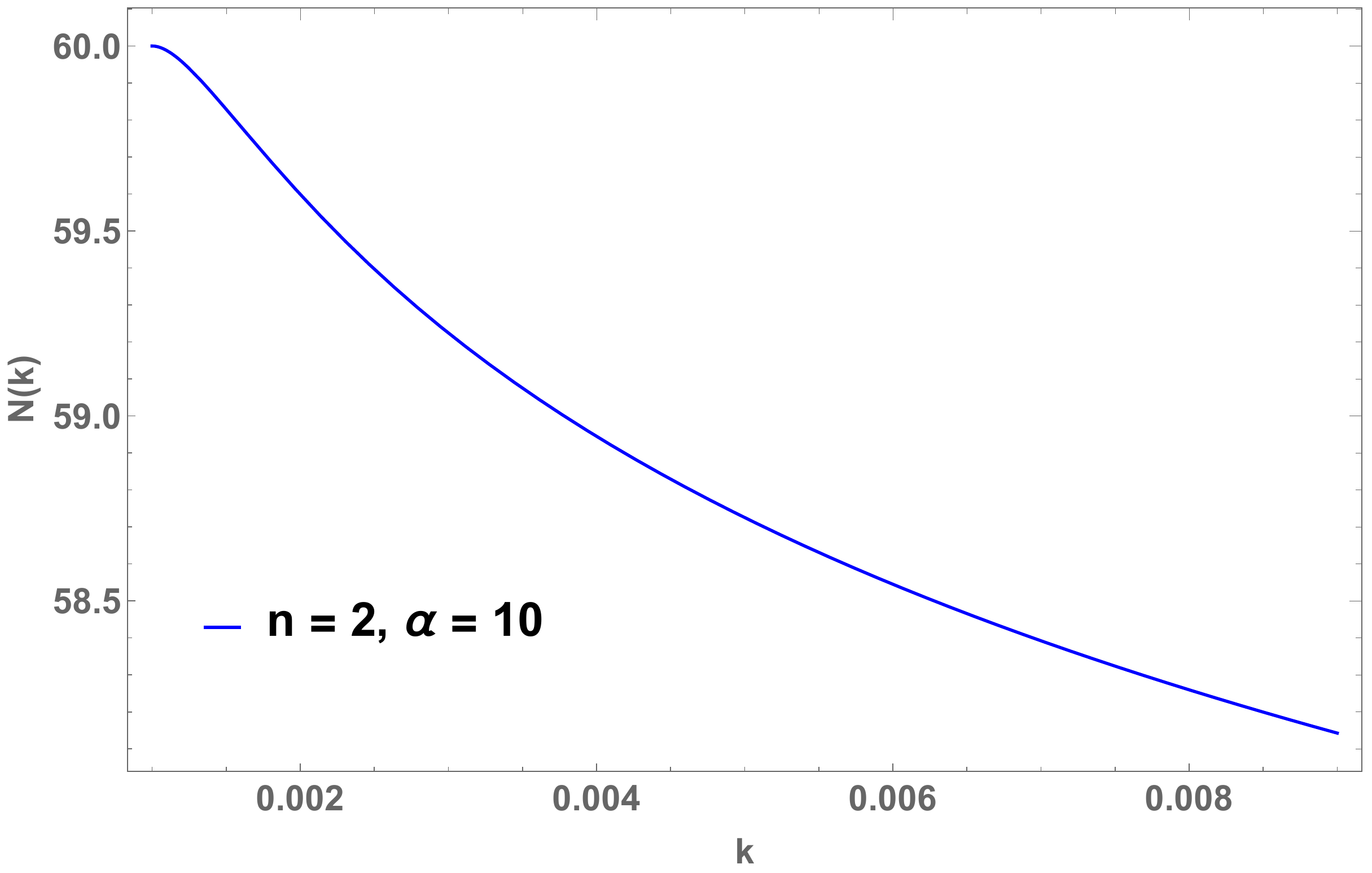}
   \subcaption{}
\end{subfigure}%

\begin{subfigure}{1\textwidth}
  \centering
   \includegraphics[width=75mm,height=80mm]{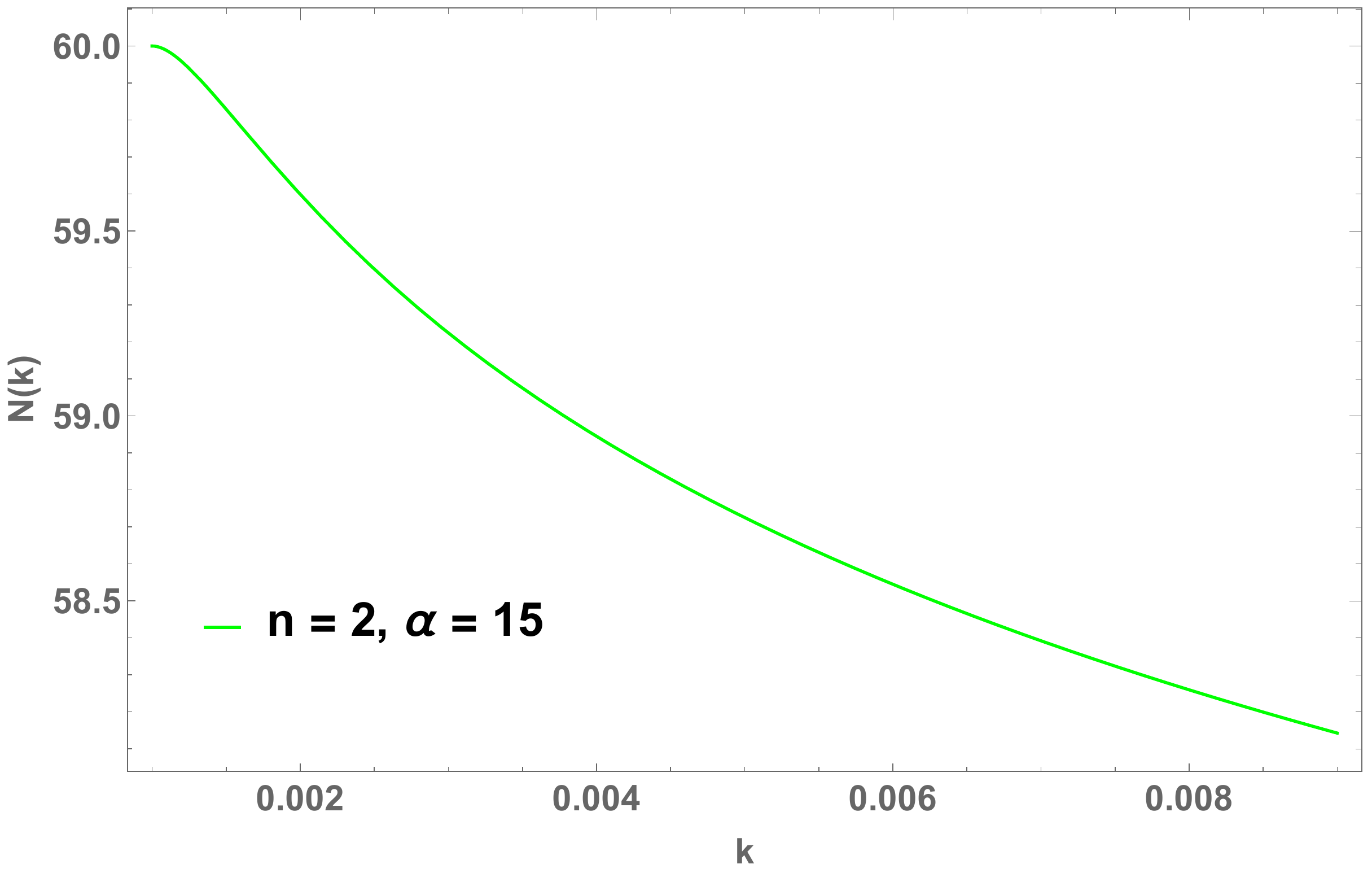}
   \subcaption{}
   \label{fig:efolds_T_2}
\end{subfigure}%
    \caption{Number of e-folds, $N(k)$, for the  $\alpha$ attractor $T$-model potentials,  for one value of $n$ and three values of $\alpha$. Almost no variation in $e$-folds is observed, due to the variation in $\alpha$.}
    \label{fig: number of e-folds_T}
\end{figure}

\begin{figure}[H]
\begin{subfigure}{0.5\textwidth}
  \centering
   \includegraphics[width=75mm,height=80mm]{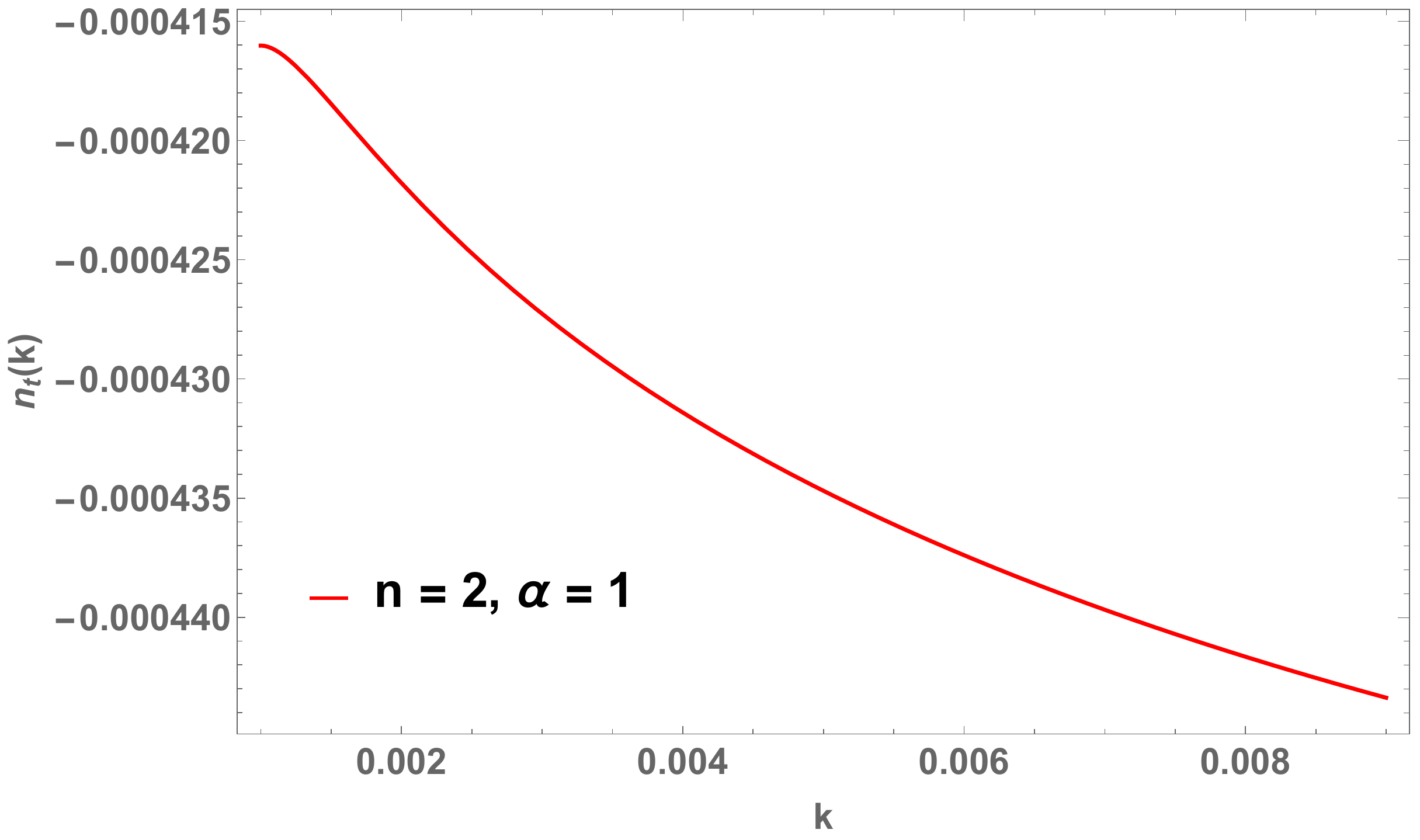}
   \subcaption{}
   \label{fig:tensor_spectral_index_E_1}
\end{subfigure}%
\begin{subfigure}{0.5\textwidth}
  \centering
   \includegraphics[width=75mm,height=80mm]{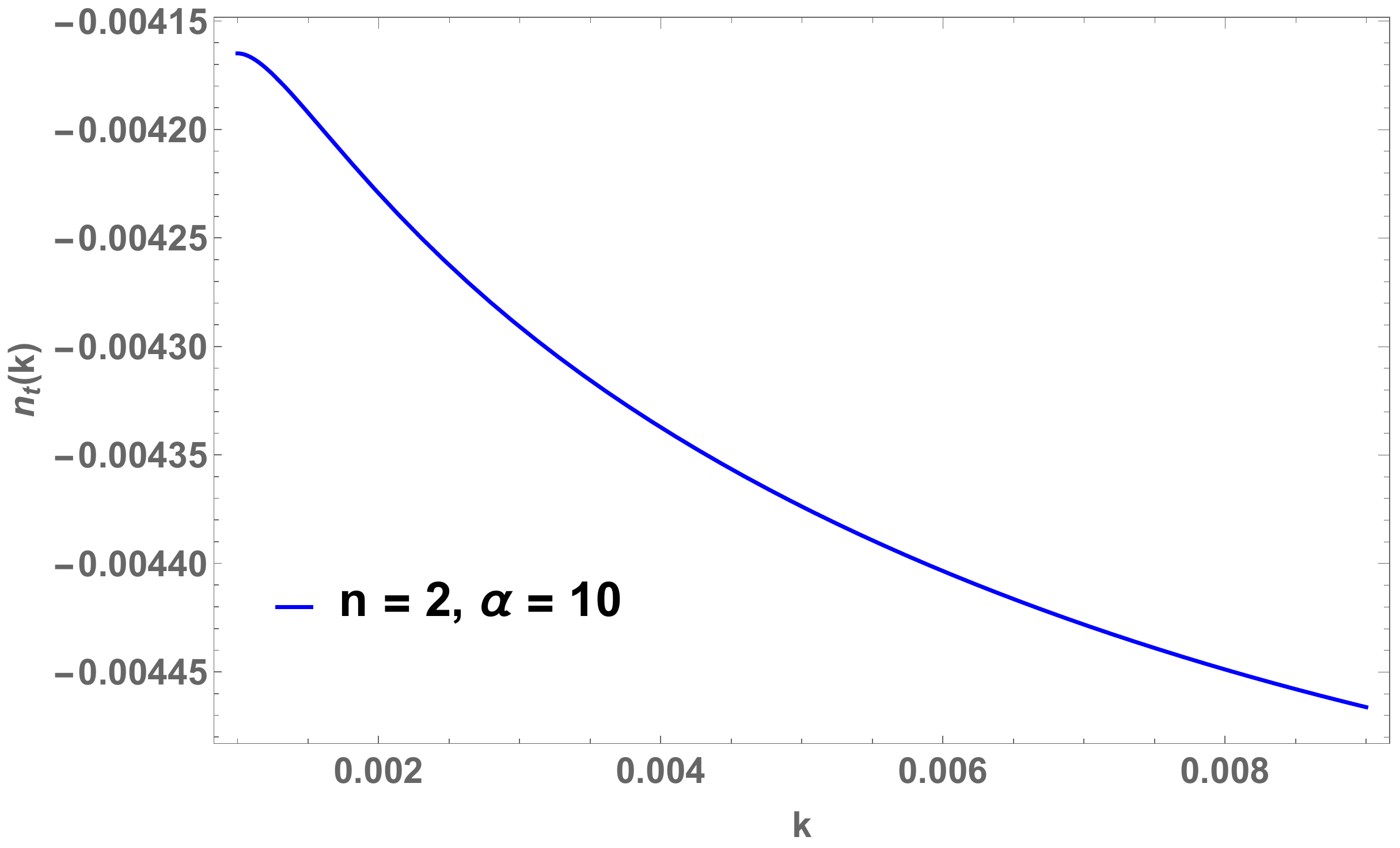}
   \subcaption{}
\end{subfigure}%

\begin{subfigure}{1\textwidth}
  \centering
   \includegraphics[width=80mm,height=80mm]{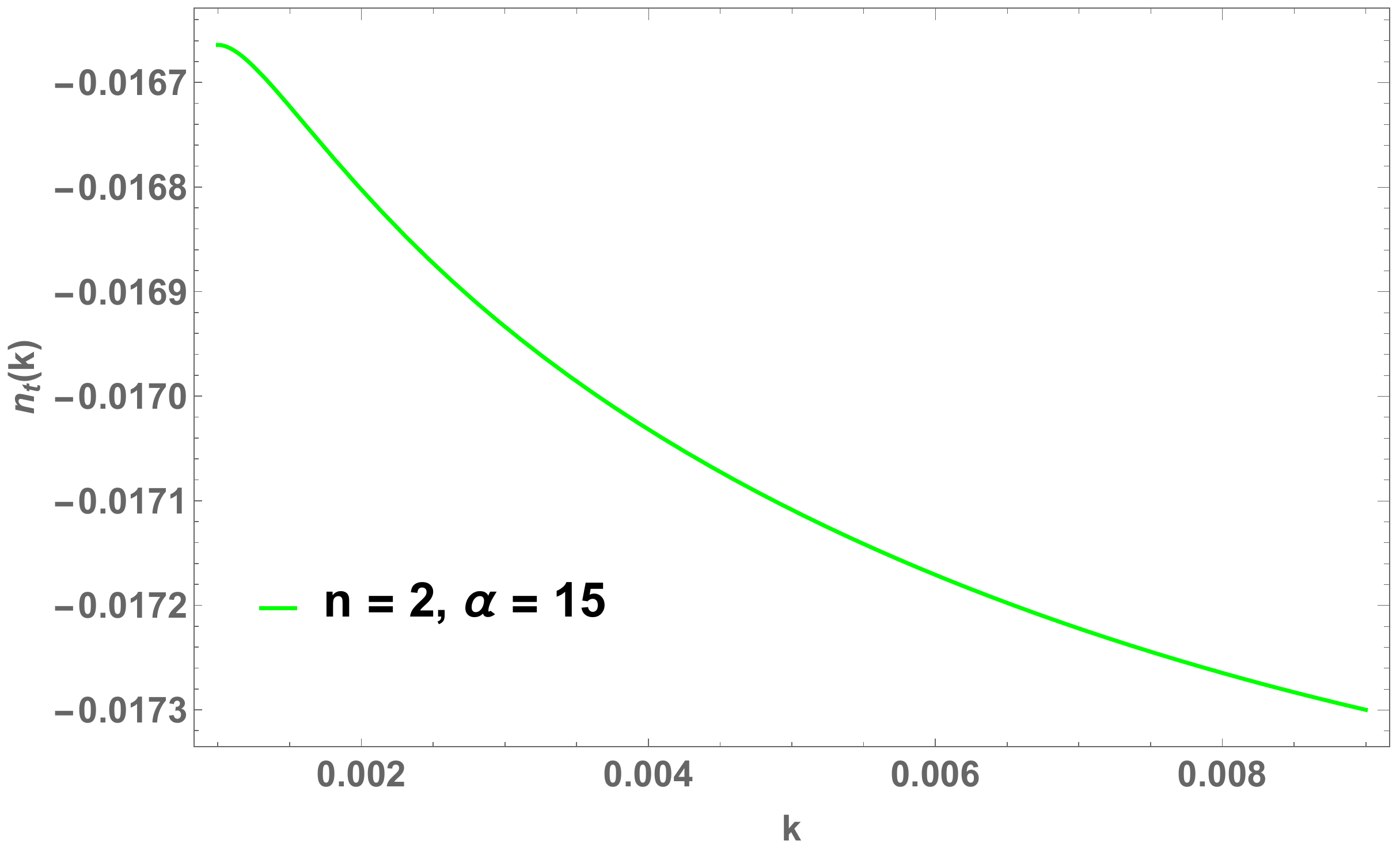}
   \subcaption{}
   \label{fig:tensor_spectral_index_E_2}
\end{subfigure}
\caption{Tensor spectral indices, $n_t (k)$, for the $\alpha$ attractor $E$-model potentials, for one value of $n$ and three values of $\alpha$.  Variations in the magnitude  $n_t(k)$,  with respect to $\alpha$, are found.}
\label{fig: tensor spectral index_E}
\end{figure}

\begin{figure}[H]
\begin{subfigure}{0.5\textwidth}
  \centering
   \includegraphics[width=75mm,height=80mm]{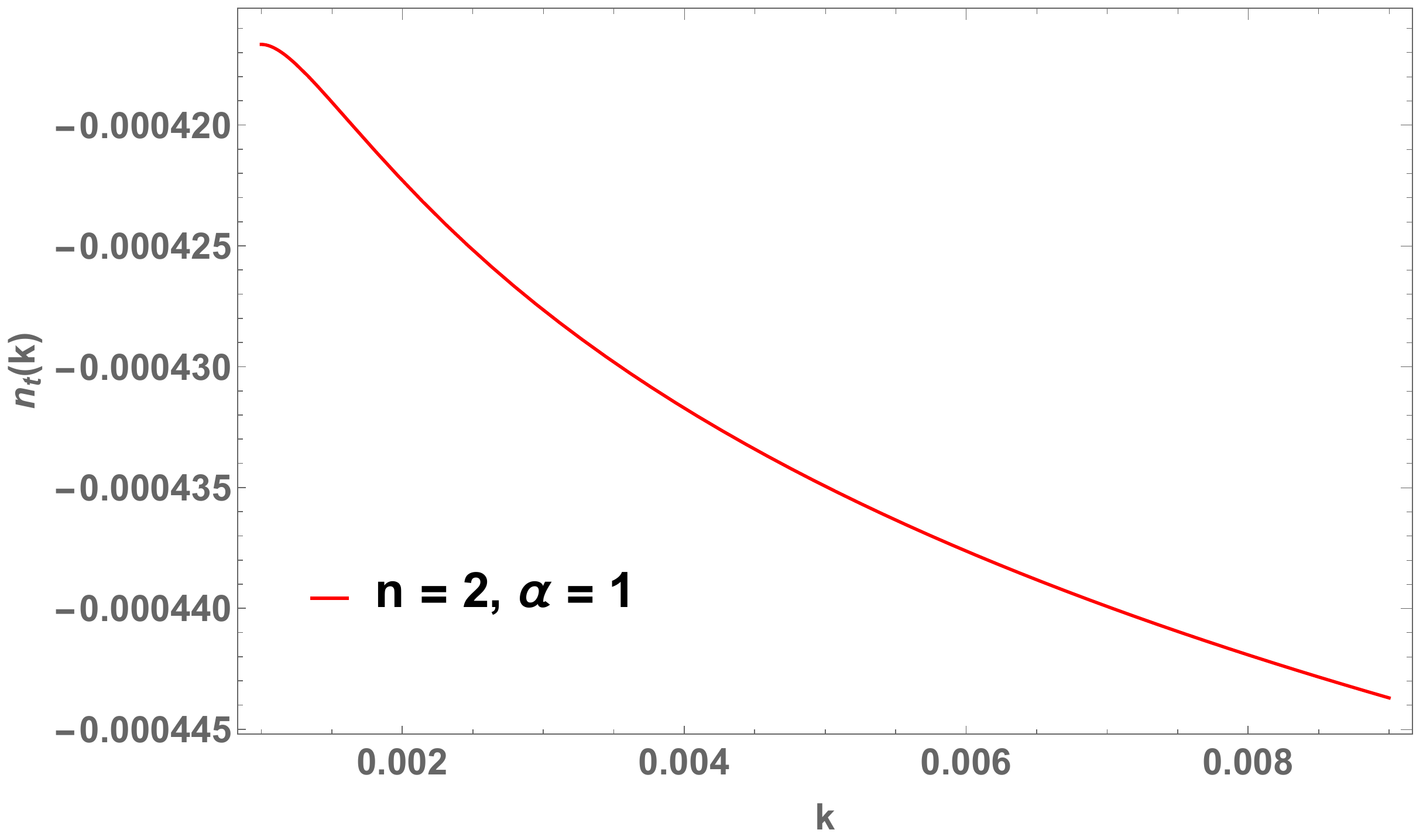}
   \subcaption{}
   \label{fig:tensor_spectral_index_T_1}
\end{subfigure}%
\begin{subfigure}{0.5\textwidth}
  \centering
   \includegraphics[width=75mm,height=80mm]{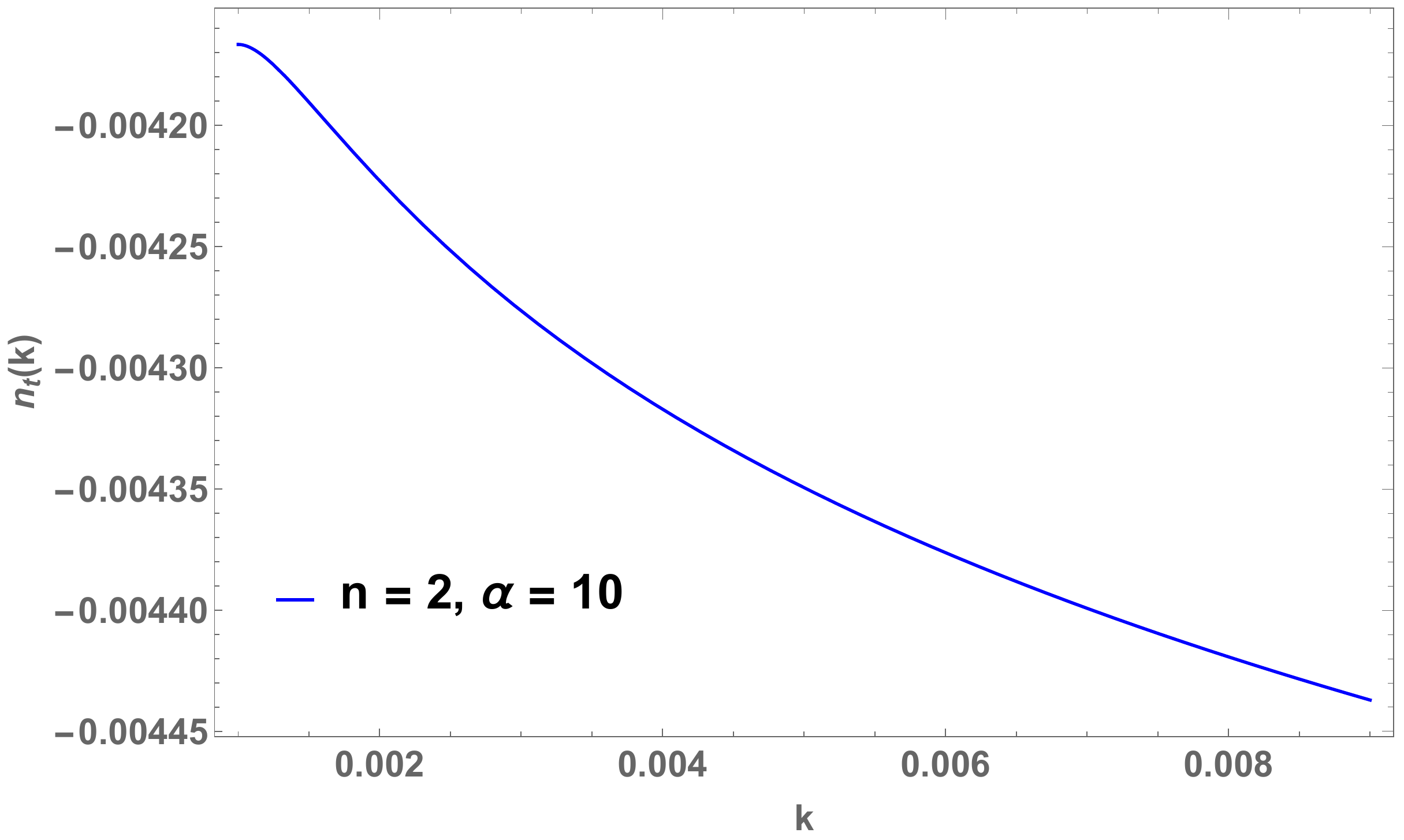}
   \subcaption{}
\end{subfigure}%

\begin{subfigure}{1\textwidth}
  \centering
   \includegraphics[width=80mm,height=80mm]{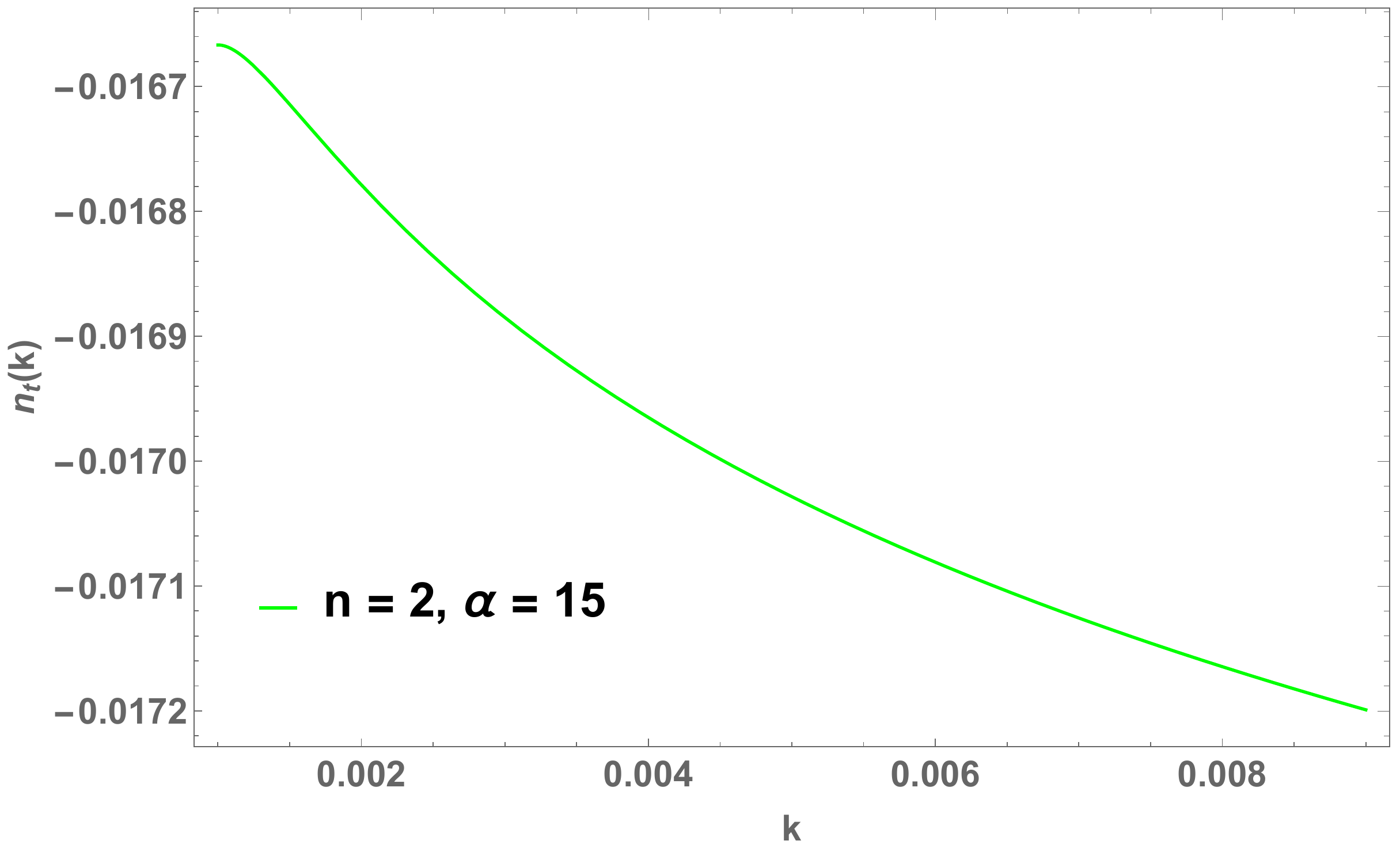}
   \subcaption{}
   \label{fig:tensor_spectral_index_T_2}
\end{subfigure}%
    \caption{Tensor spectral indices, $n_t (k)$, for the $\alpha$ attractor $T$-model potentials, for one value of $n$ and three values of $\alpha$. Variations in the magnitude  $n_t(k)$,  with respect to $\alpha$, are found.}
    \label{fig: tensor spectral index_T}
\end{figure}

In Figures \ref{fig:number of e-folds_E} for the $E$-model and \ref{fig: number of e-folds_T} for the  $T$-model, variations of number of e-folds,  against mode momenta have been shown. In the calculation of the e-folds, we have considered $N=60$ at the lowest mode, $k= 0.001$. We observe that during inflation, the number of e-folds shows little variation across the quantum fluctuations, which is to be expected. So far  as the mode dependence of the number of e-folds is concerned, we find that $N(k)$ decreases as $k$ increases, signifying that the the high-momentum modes undergo less number of e-folds than the low-momentum modes. 
\begin{figure}[H]
\begin{subfigure}{0.5\textwidth}
  \centering
   \includegraphics[width=75mm,height=80mm]{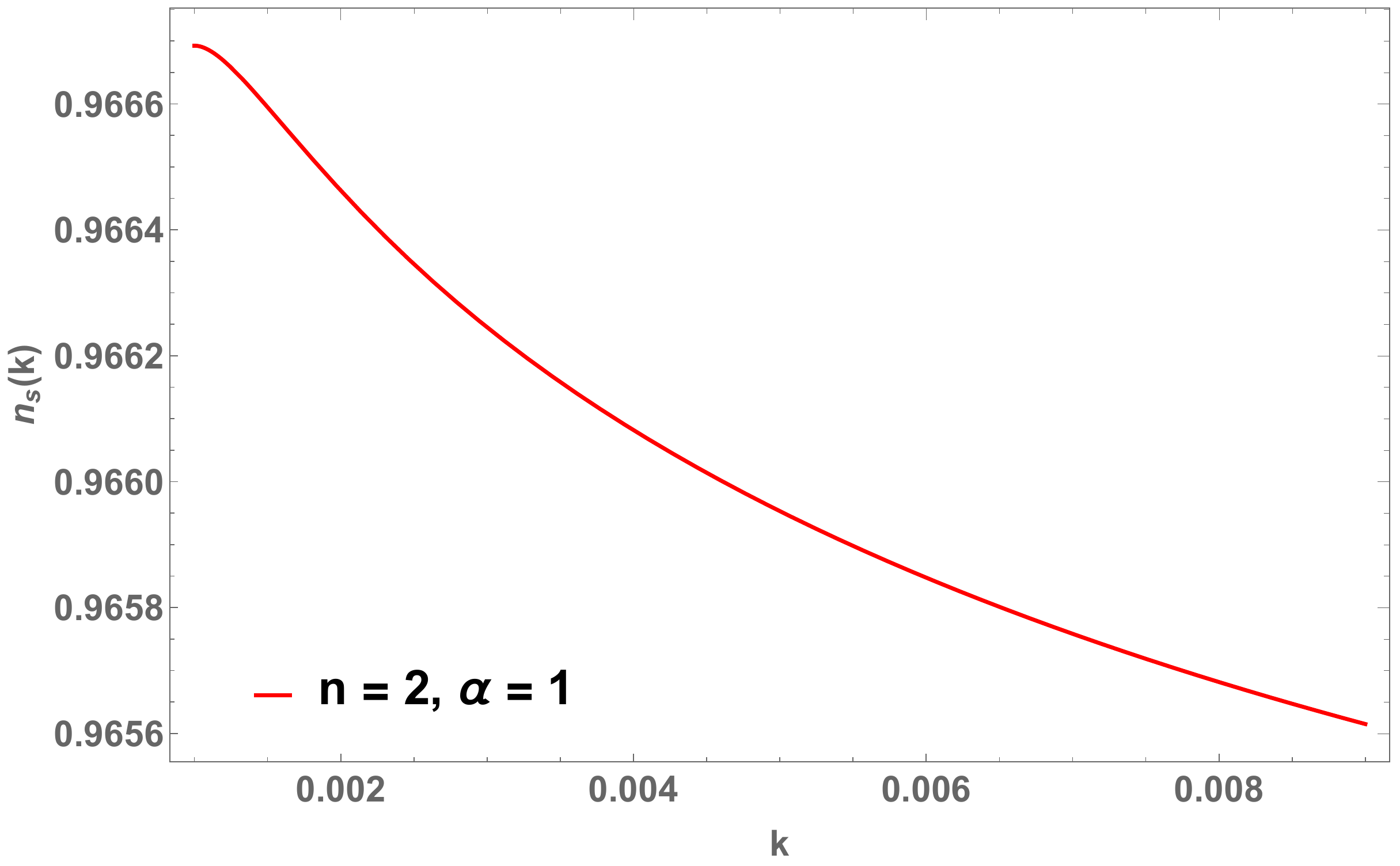}
   \subcaption{}
   \label{fig:scalar_spectral_index_E_1}
\end{subfigure}%
\begin{subfigure}{0.5\textwidth}
  \centering
   \includegraphics[width=80mm,height=80mm]{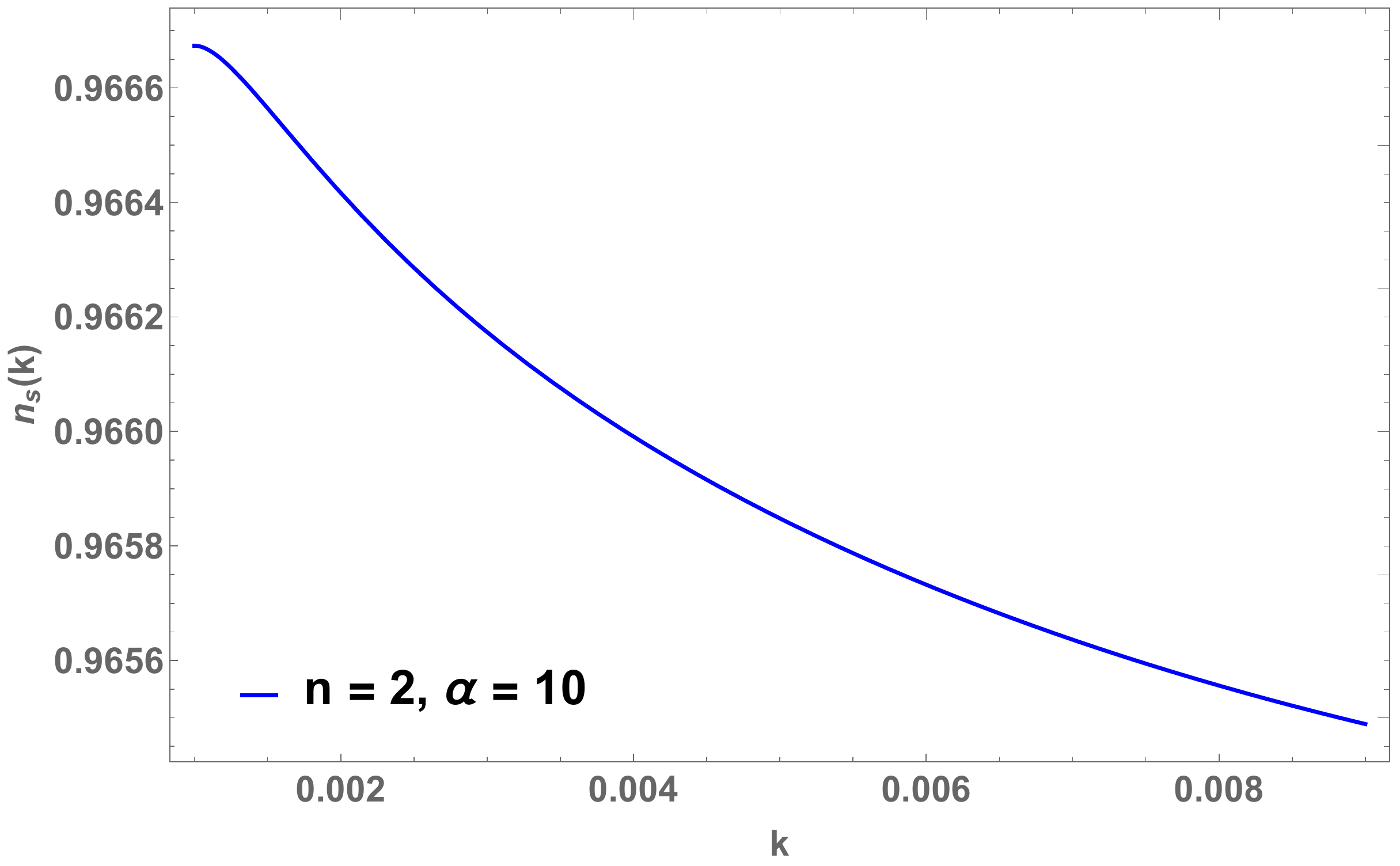}
   \subcaption{}
\end{subfigure}%

\begin{subfigure}{1\textwidth}
  \centering
   \includegraphics[width=80mm,height=80mm]{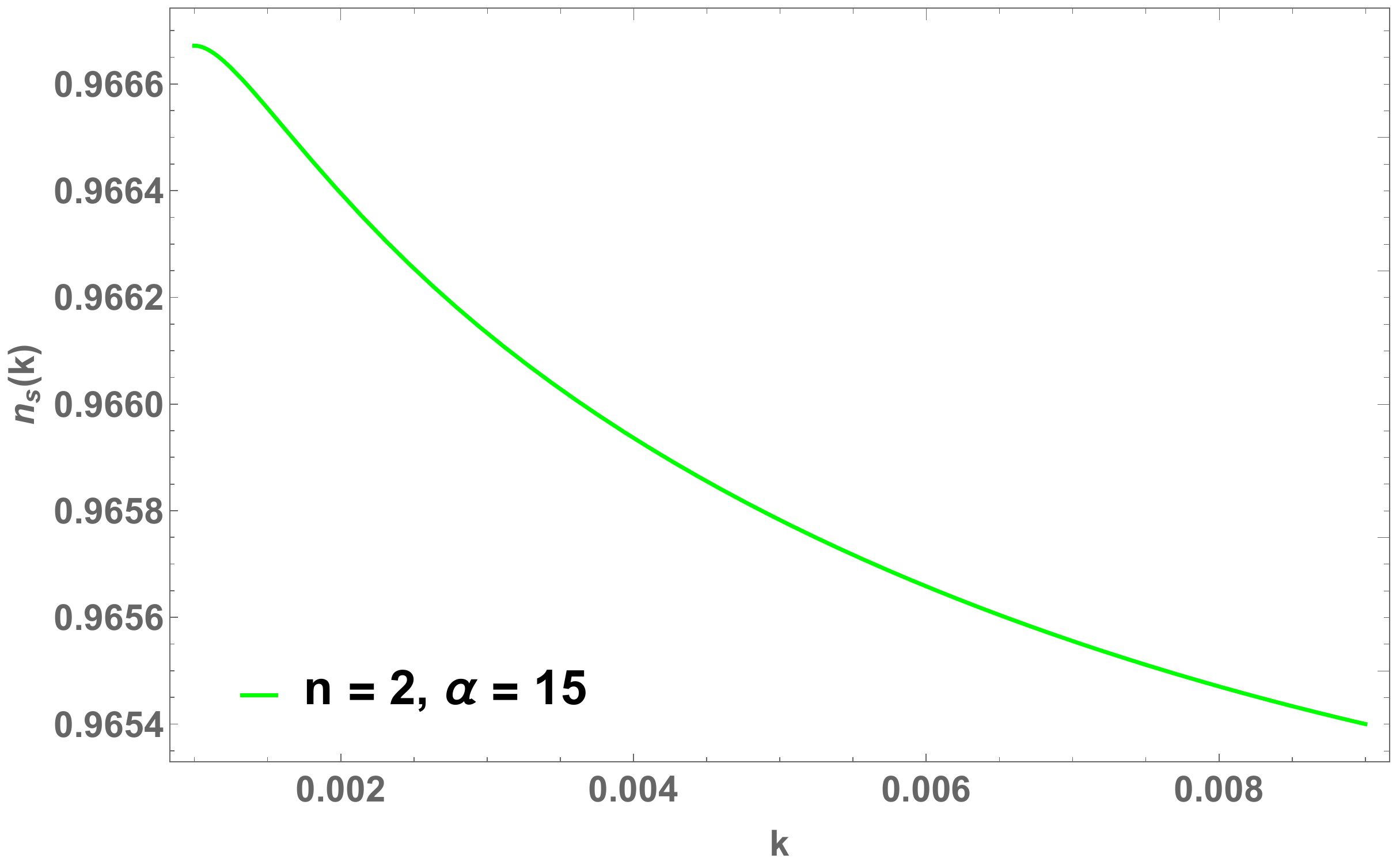}
   \subcaption{}
    \label{fig:scalar_spectral_index_E_2}
\end{subfigure}
\caption{Scalar spectral indices, $n_s (k)$, for the $\alpha$ attractor $E$-model potentials, for one value of $n$ and three values of $\alpha$.  No variation of $n_{s}(k)$, with respect to $\alpha$, is found.}
\label{fig: scalar spectral index_E}
\end{figure}
\begin{figure}[H]
\begin{subfigure}{0.5\textwidth}
  \centering
   \includegraphics[width=75mm,height=80mm]{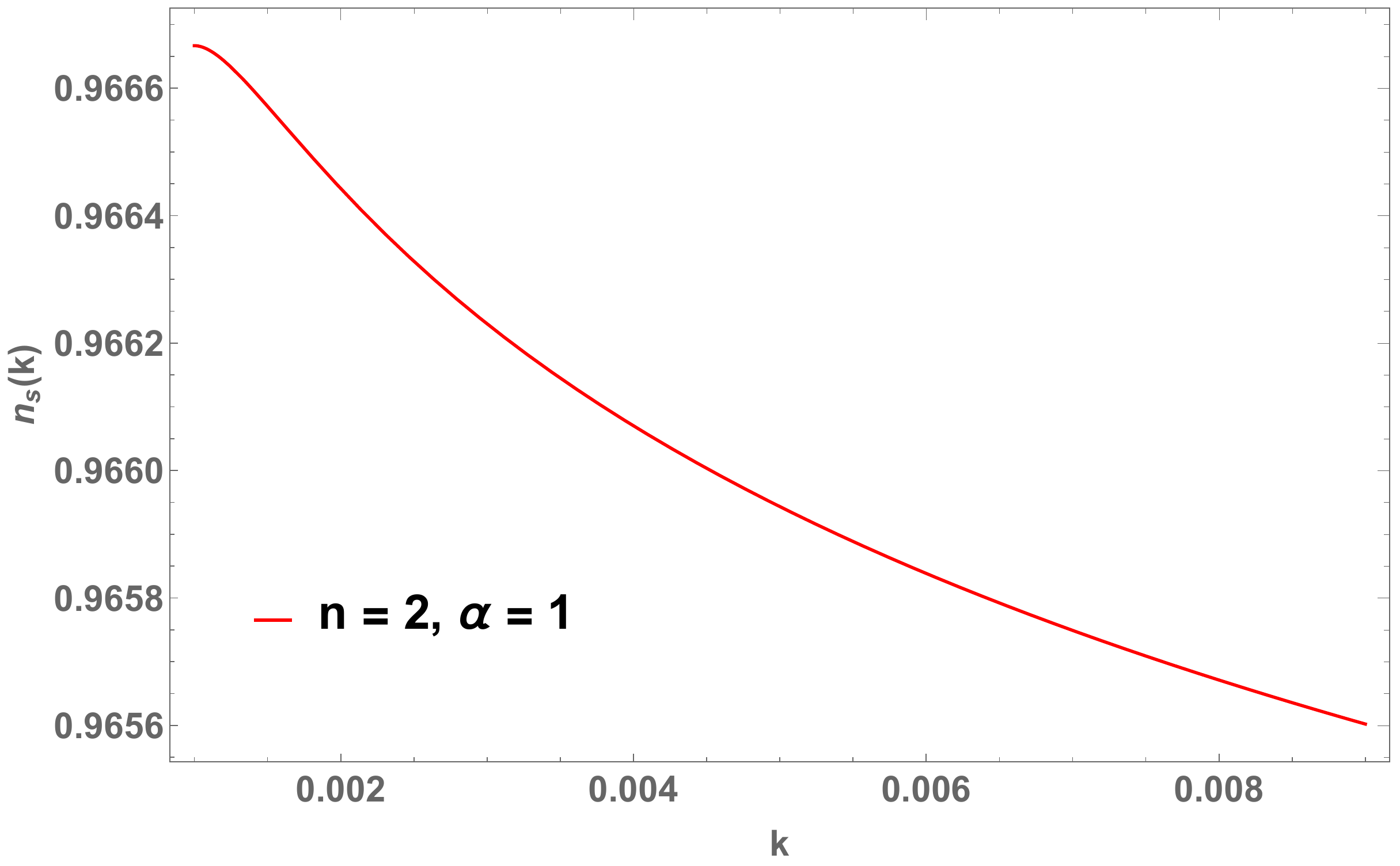}
   \subcaption{}
    \label{fig:scalar_spectral_index_T_1}
\end{subfigure}%
\begin{subfigure}{0.5\textwidth}
  \centering
   \includegraphics[width=75mm,height=80mm]{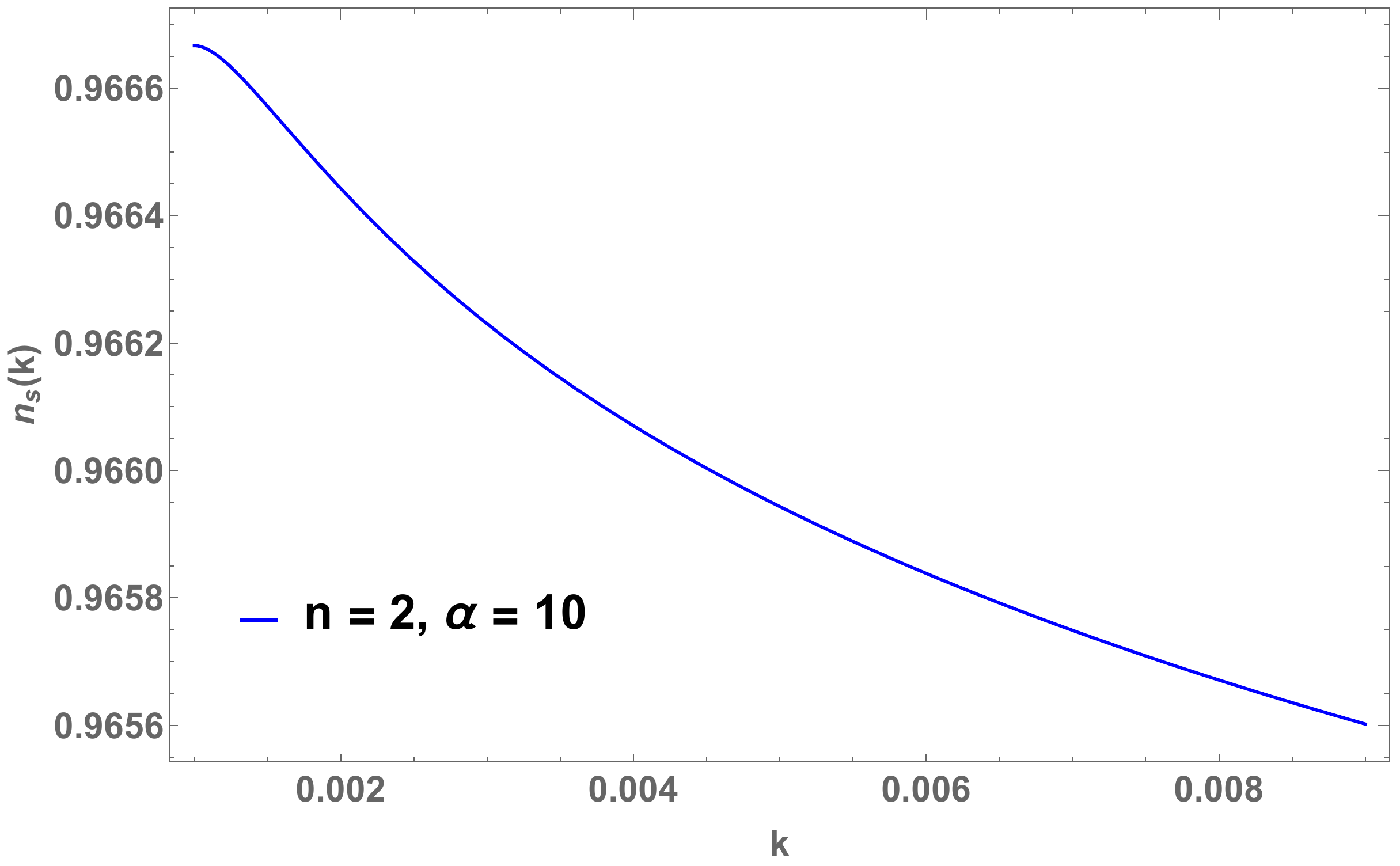}
   \subcaption{}
\end{subfigure}%

\begin{subfigure}{1\textwidth}
  \centering
   \includegraphics[width=80mm,height=80mm]{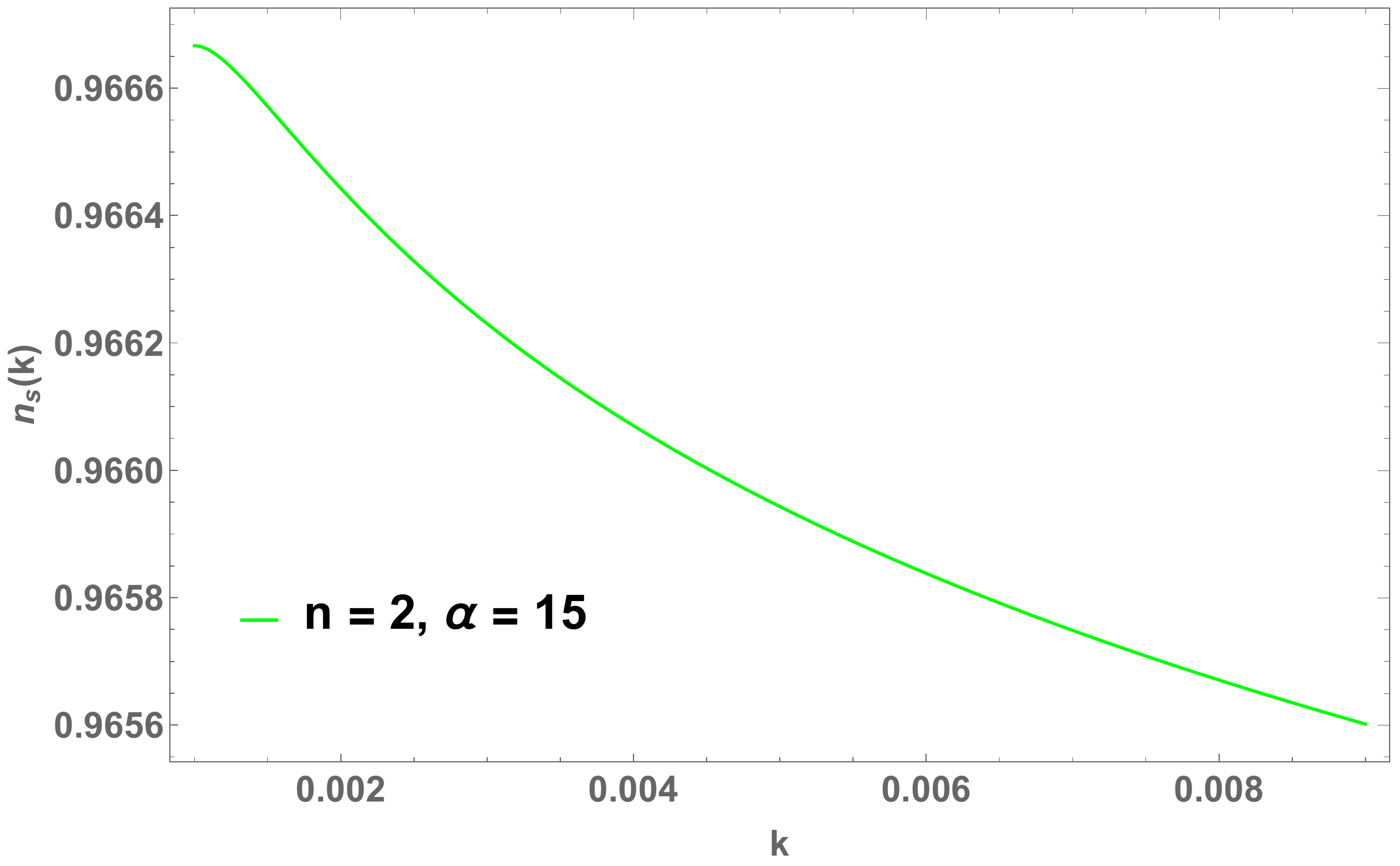}
   \subcaption{}
    \label{fig: scalar_spectral_index_T_2}
\end{subfigure}%
    \caption{Scalar spectral indices, $n_s (k)$, for the $\alpha$ attractor $T$-model potentials, for one value of $n$ and three values of $\alpha$. No variation of $n_{s}(k)$, with respect to $\alpha$, is found.}
    \label{fig: scalar spectral index_T}
\end{figure}
In Figures \ref{fig: scalar spectral index_E} - \ref{fig: tensor spectral index_T}, the mode variations of the scalar and the tensor spectral indices are displayed for  the $E$ and the $T$ models,  for a number of values of $\alpha$. No variations in  the scalar spectral indices are observed for variation in $\alpha$, as the former does not depend on the latter (see equation (\ref{eq:final_scalar_index})). On the other hand, $n_{t}(k)$ varies with $\alpha$ (see equation (\ref{eq:final_tensor_index})).\par
\begin{figure}[H]
\begin{subfigure}{0.5\textwidth}
  \centering
   \includegraphics[width=75mm,height=80mm]{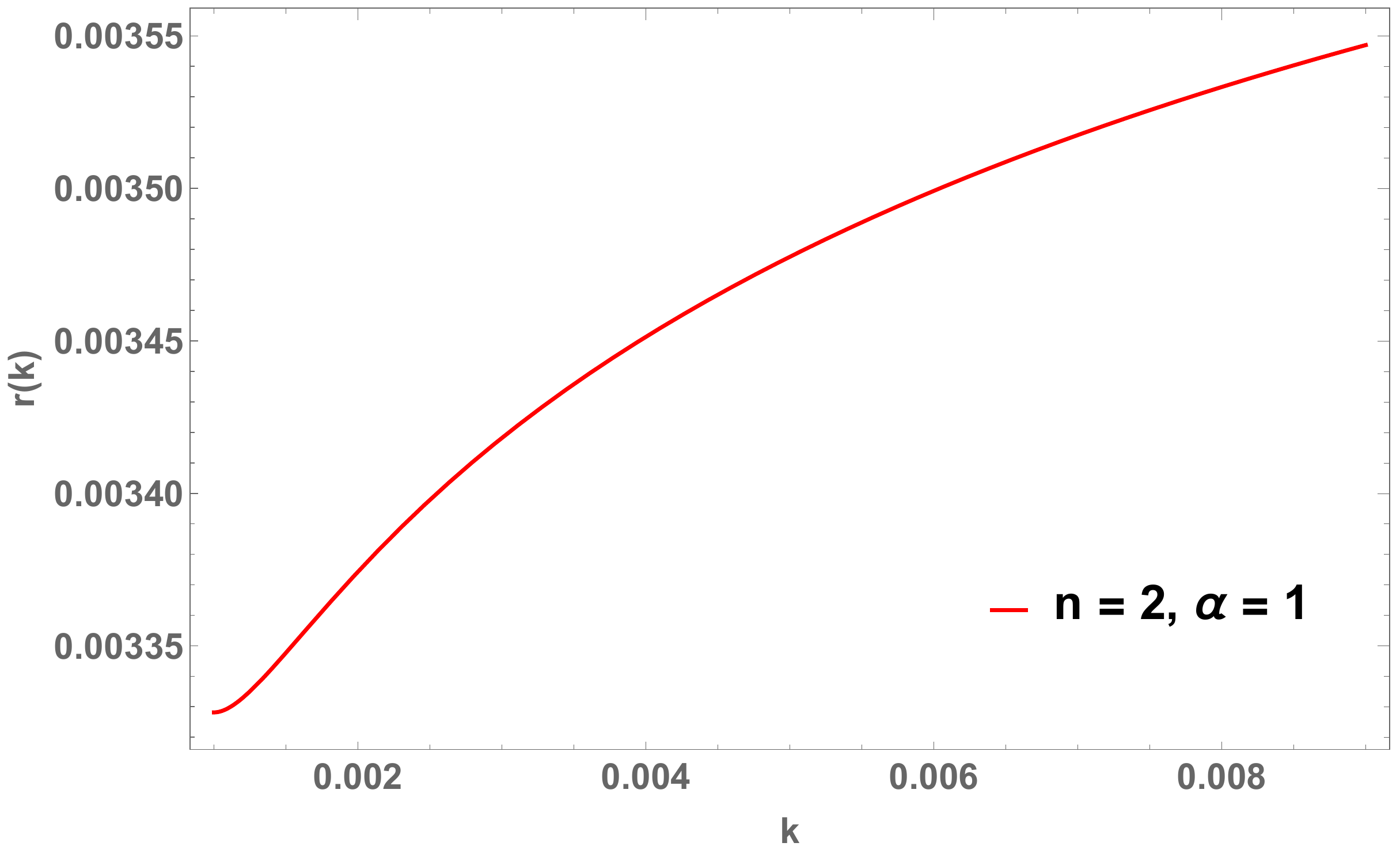}
   \subcaption{}
   \label{fig:ratio_E_1}
\end{subfigure}%
\begin{subfigure}{0.5\textwidth}
  \centering
   \includegraphics[width=75mm,height=80mm]{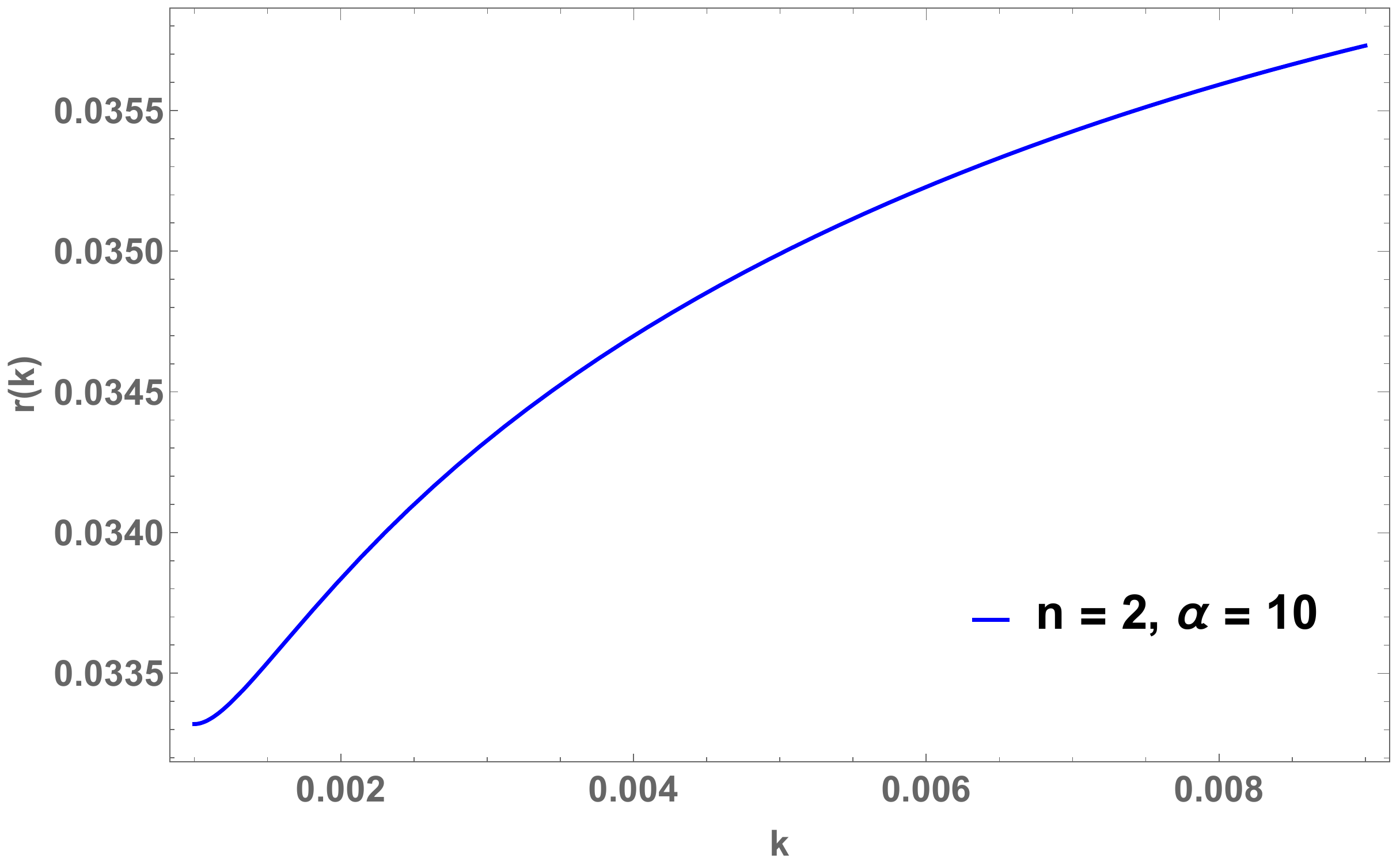}
   \subcaption{}
\end{subfigure}%

\begin{subfigure}{1\textwidth}
  \centering
   \includegraphics[width=80mm,height=80mm]{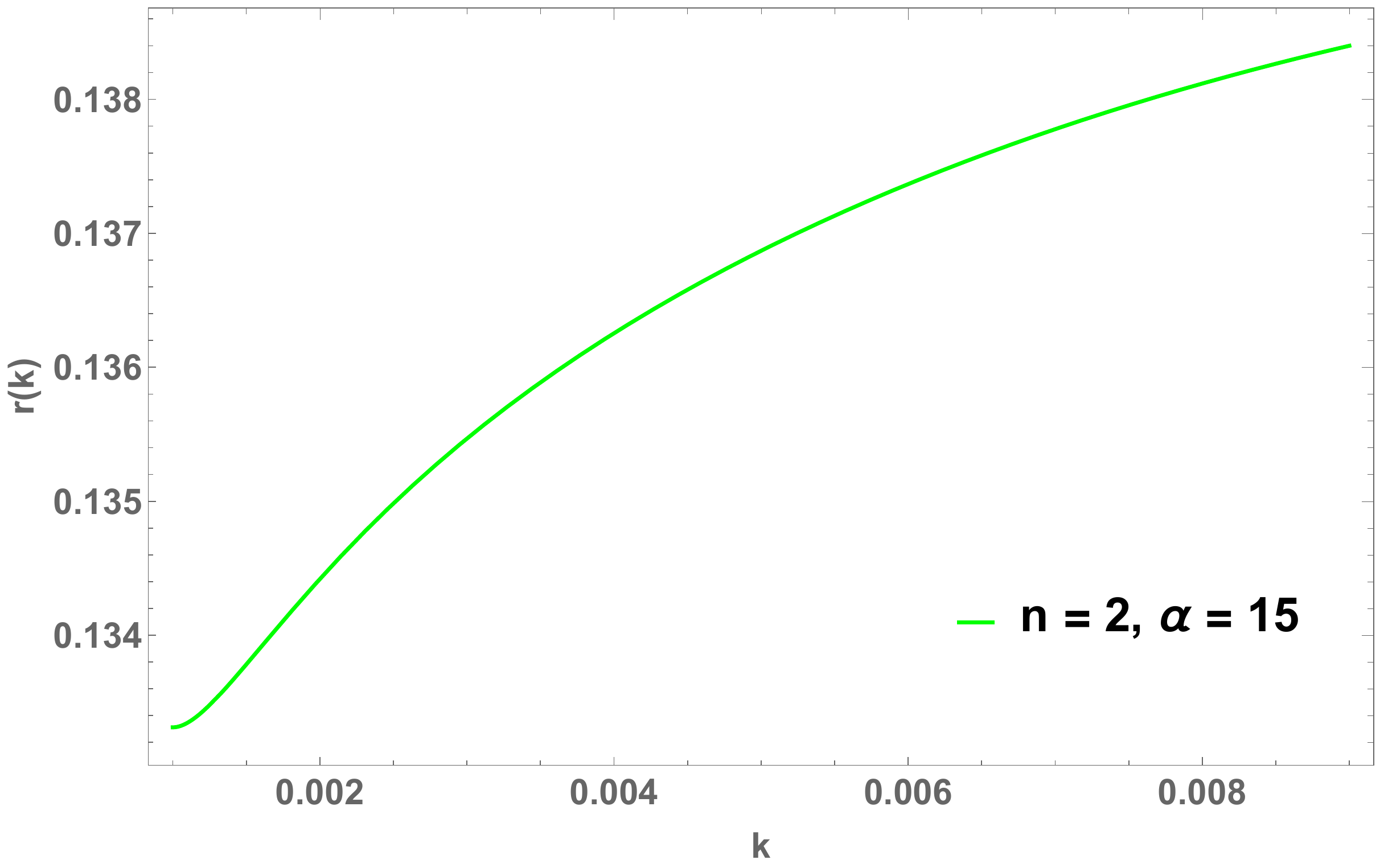}
   \subcaption{}
   \label{fig:ratio_E_2}
\end{subfigure}
\caption{Tensor to scalar ratio, $r(k)$, for the $\alpha$ attractor $E$-model potentials, for one value of $n$ and three values of $\alpha$. The values of $r(k)$ increase with the increase in $\alpha$.}
\label{fig: tensor to scalar ratio_E}
\end{figure}
\begin{figure}[H]
\begin{subfigure}{0.5\textwidth}
  \centering
   \includegraphics[width=75mm,height=80mm]{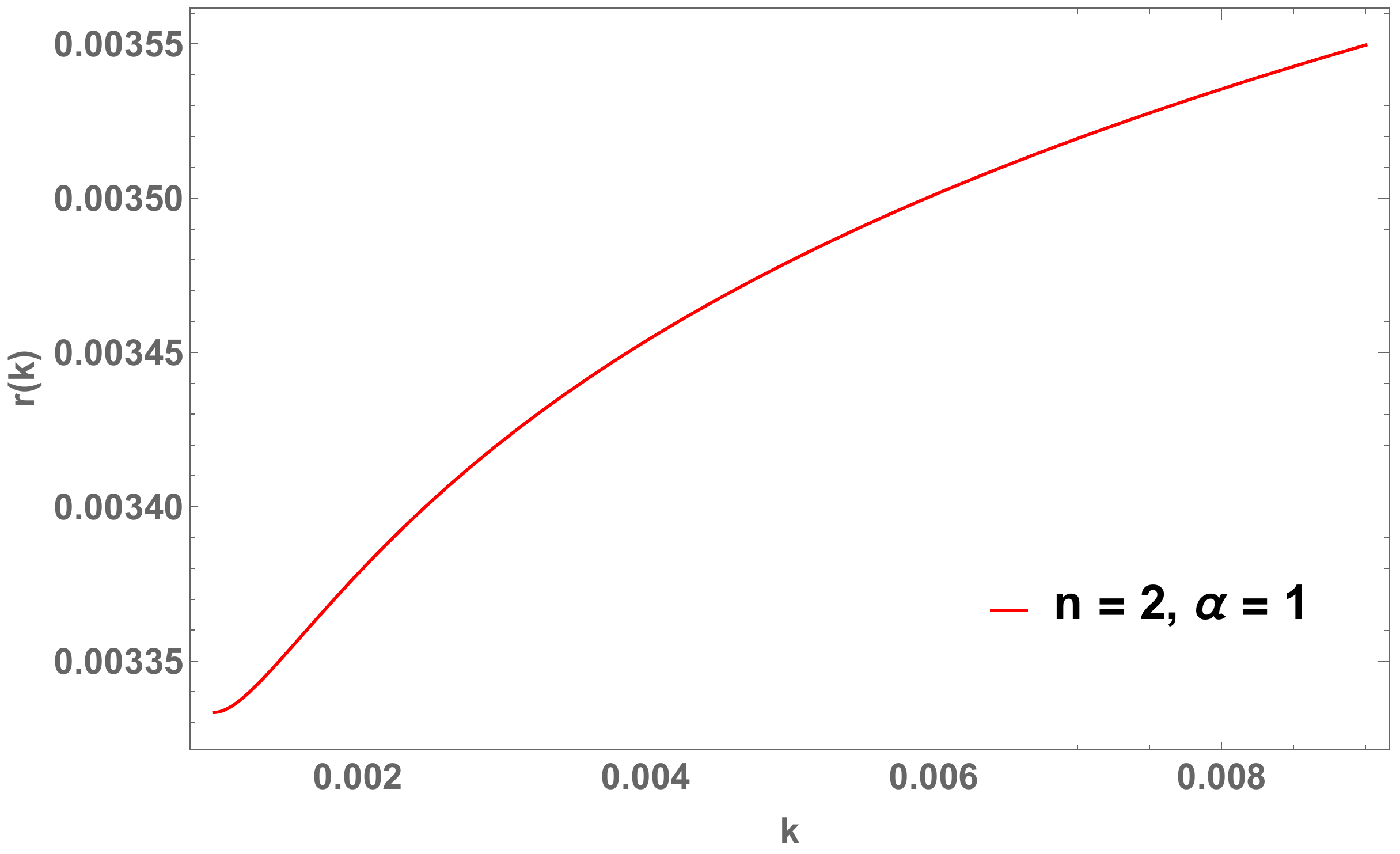}
   \subcaption{}
   \label{fig:ratio_T_1}
\end{subfigure}%
\begin{subfigure}{0.5\textwidth}
  \centering
   \includegraphics[width=75mm,height=80mm]{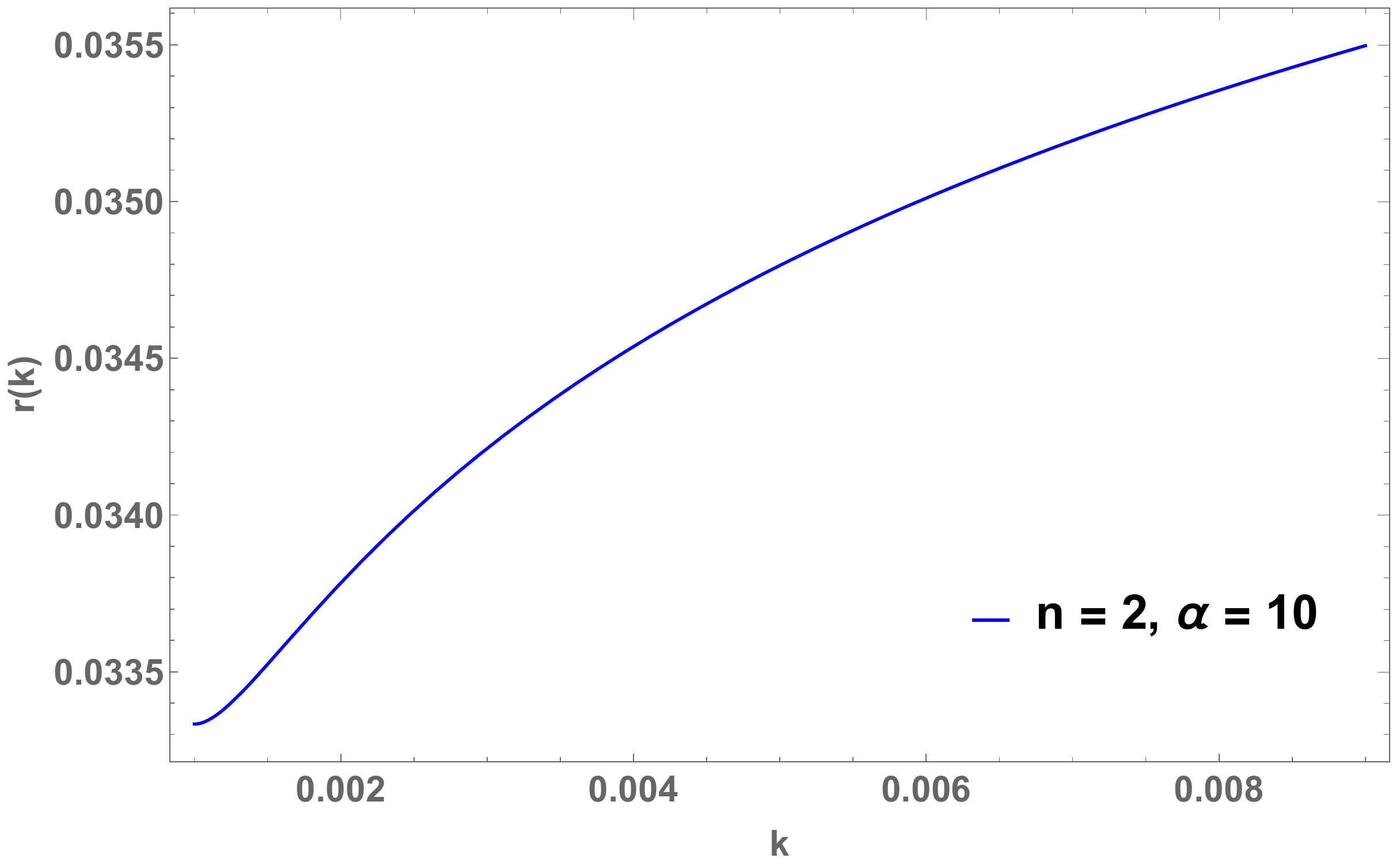}
   \subcaption{}
\end{subfigure}%

\begin{subfigure}{1\textwidth}
  \centering
   \includegraphics[width=80mm,height=80mm]{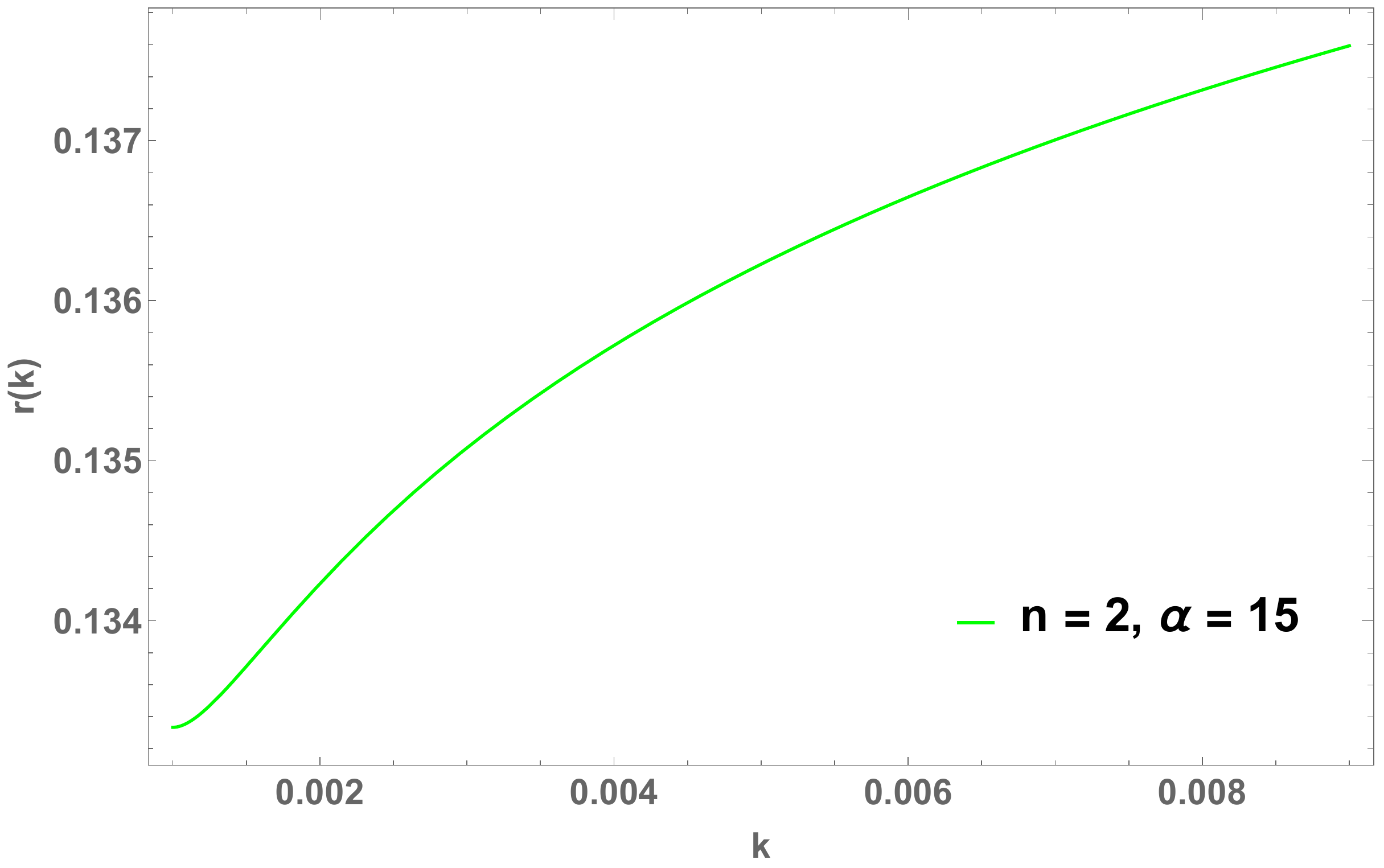}
   \subcaption{}
   \label{fig:ratio_T_2}
\end{subfigure}%
    \caption{Tensor to scalar ratio, $r(k)$, for the $\alpha$ attractor $T$-model potentials, for one value of $n$ and three values of $\alpha$. The values of $r(k)$ increase with the increase in $\alpha$. }
    \label{fig: tensor to scalar ratio_T}
\end{figure}

The variations of $r(k)$ with respect to $\alpha$ for the $E$ and the $T$-models, as shown in Figures \ref{fig: tensor to scalar ratio_E}, \ref{fig: tensor to scalar ratio_T}, are consistent with  the features of Figures \ref{fig: scalar power spectrum_E} - \ref{fig:fig_14}. 
\subsection{\texorpdfstring{$n_s$}{n} and \texorpdfstring{$r$}{r} graphs for \texorpdfstring{$\alpha=\frac{1}{3}, \frac{2}{3}, \frac{4}{3}, \frac{5}{3}, 2, \frac{7}{3}$}{a}, constrained by the  \texorpdfstring{$\mathcal{N}=1$}{m} and \texorpdfstring{$\mathcal{N}=8$}{n} supergravity theories}
As stated earlier in Sections \ref{sec:intro} and  \ref{sec:chaotic attractors}, the origin of the $\alpha$ attractor potentials relates to the hyperbolic geometry of Poincar\'{e} disk \cite{Ferrara:2016fwe,Carrasco:2015uma,Kallosh:2017ced}. Along these lines of maximal $\mathcal{N}=8$ supergarvity, M-theory and superstring theory  a discrete set values of $\alpha$  have been suggested:$1/3, 2/3, 1,4/3,$ $5/3, 2, 7/3$. Here, we calculate, within our framework, $n_{s}(k)$ and $r(k)$ with these values of $\alpha$ and verify that they conform to the Planck-2018 data within 68\% CL. 
\begin{figure}[H]
\begin{subfigure}{0.5\textwidth}
   \centering
    \includegraphics[width=75mm,height=65mm]{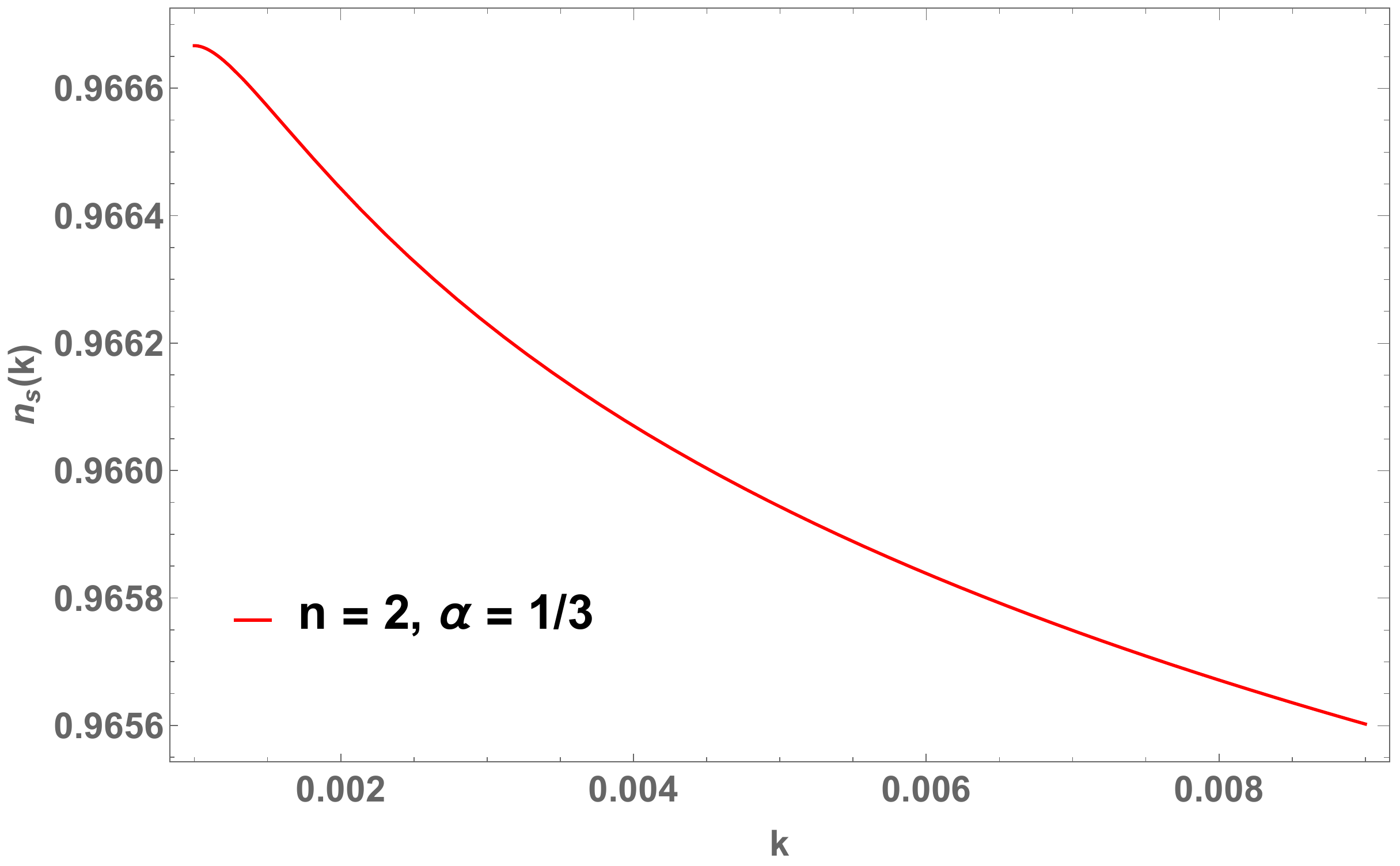} 
\end{subfigure}%
\begin{subfigure}{0.5\textwidth}
   \centering
    \includegraphics[width=75mm,height=65mm]{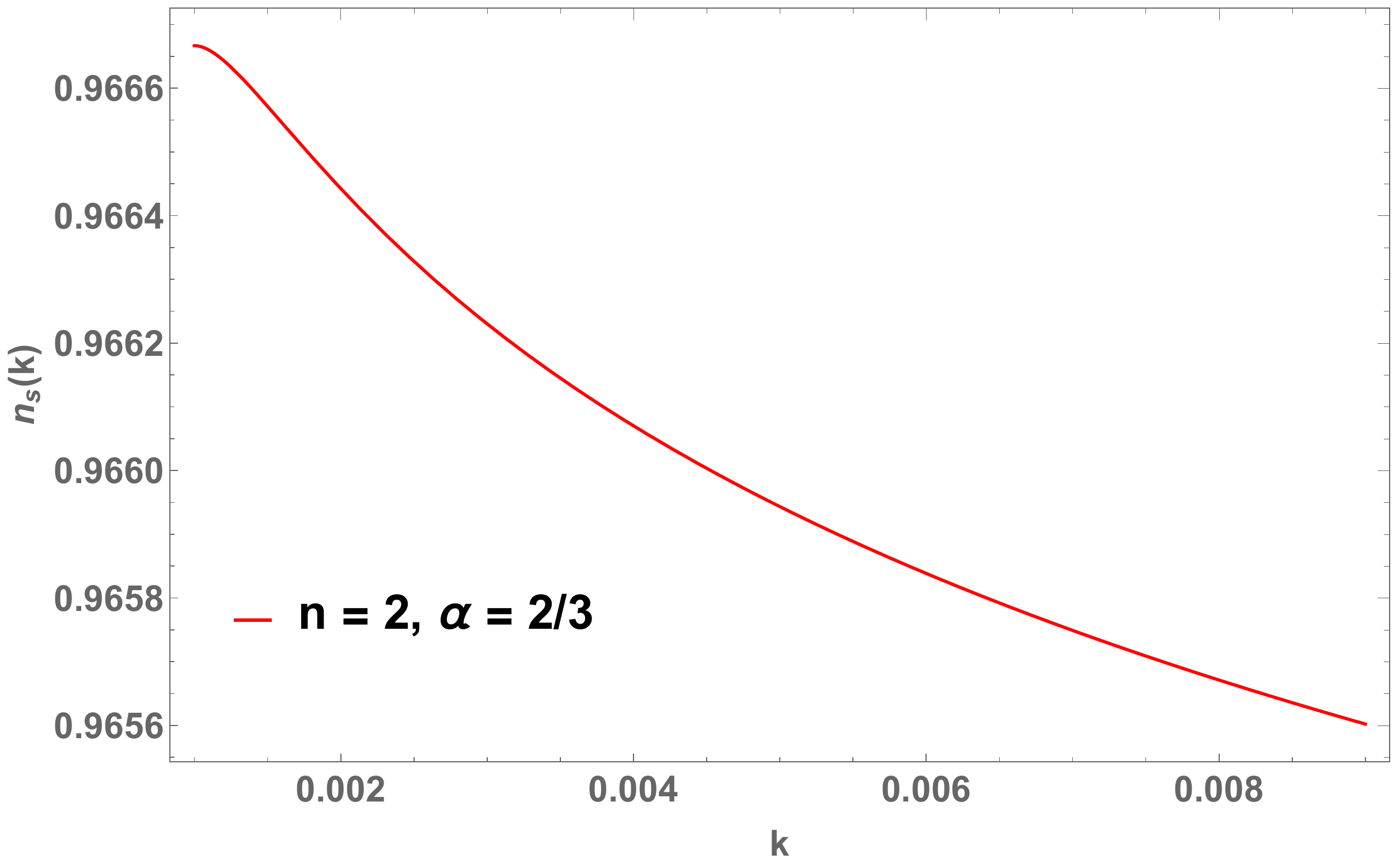} 
    \end{subfigure}%
    
\begin{subfigure}{0.5\textwidth}
   \centering
    \includegraphics[width=75mm,height=65mm]{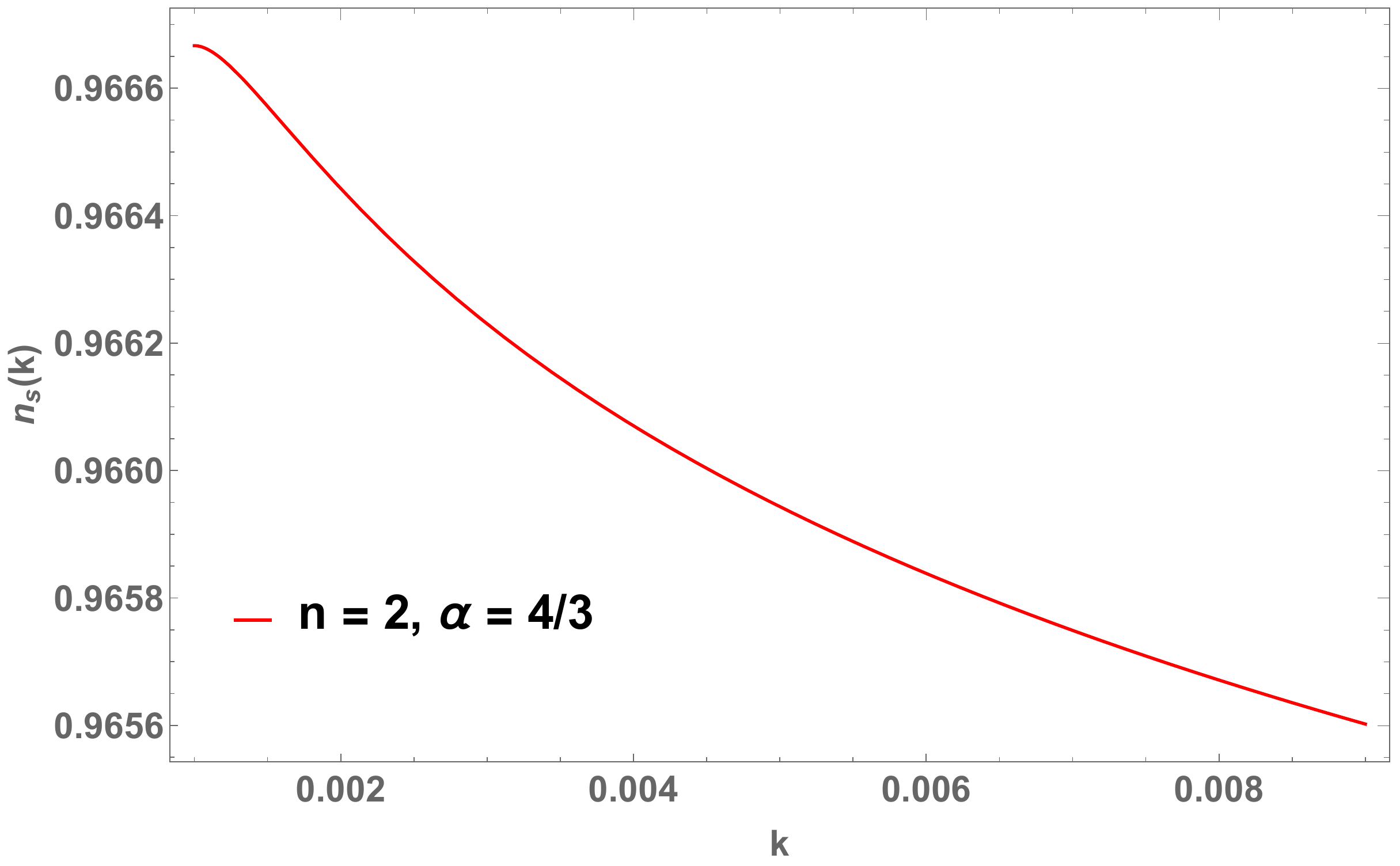} 
\end{subfigure}%
\begin{subfigure}{0.5\textwidth}
   \centering
    \includegraphics[width=75mm,height=65mm]{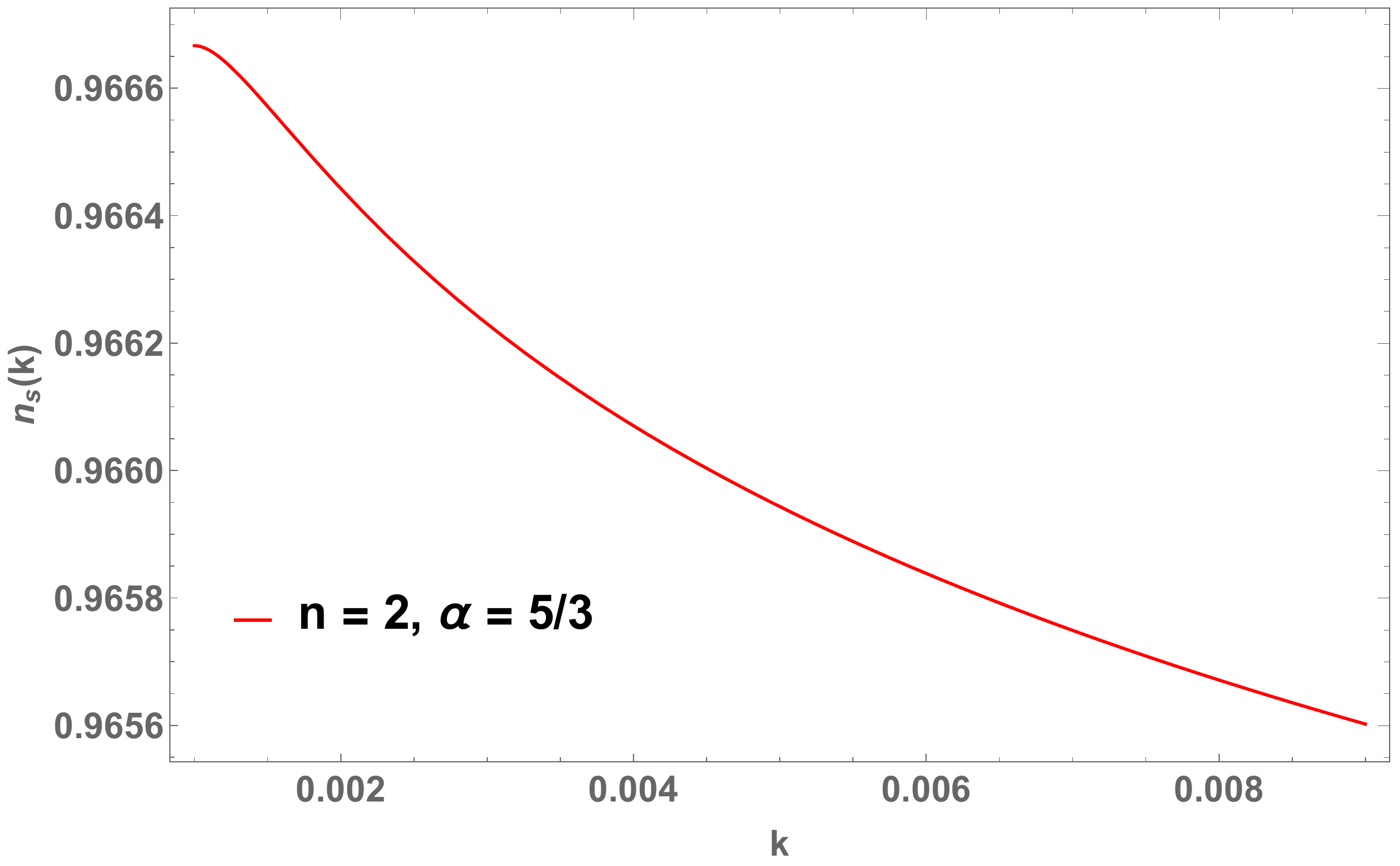}
\end{subfigure}%

\begin{subfigure}{0.5\textwidth}
   \centering
    \includegraphics[width=75mm,height=65mm]{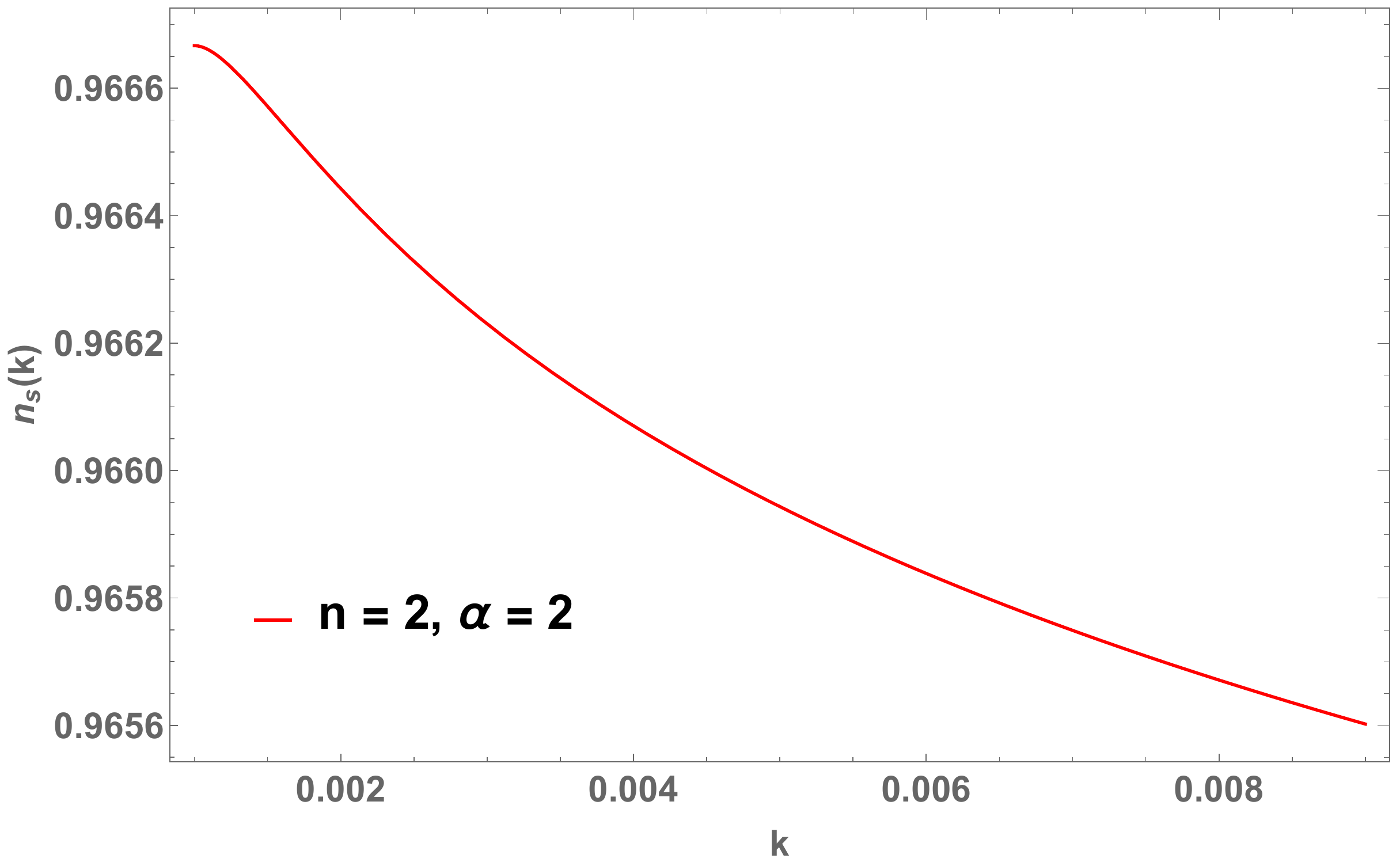}
    \end{subfigure}%
    \begin{subfigure}{0.5\textwidth}
   \centering
    \includegraphics[width=75mm,height=65mm]{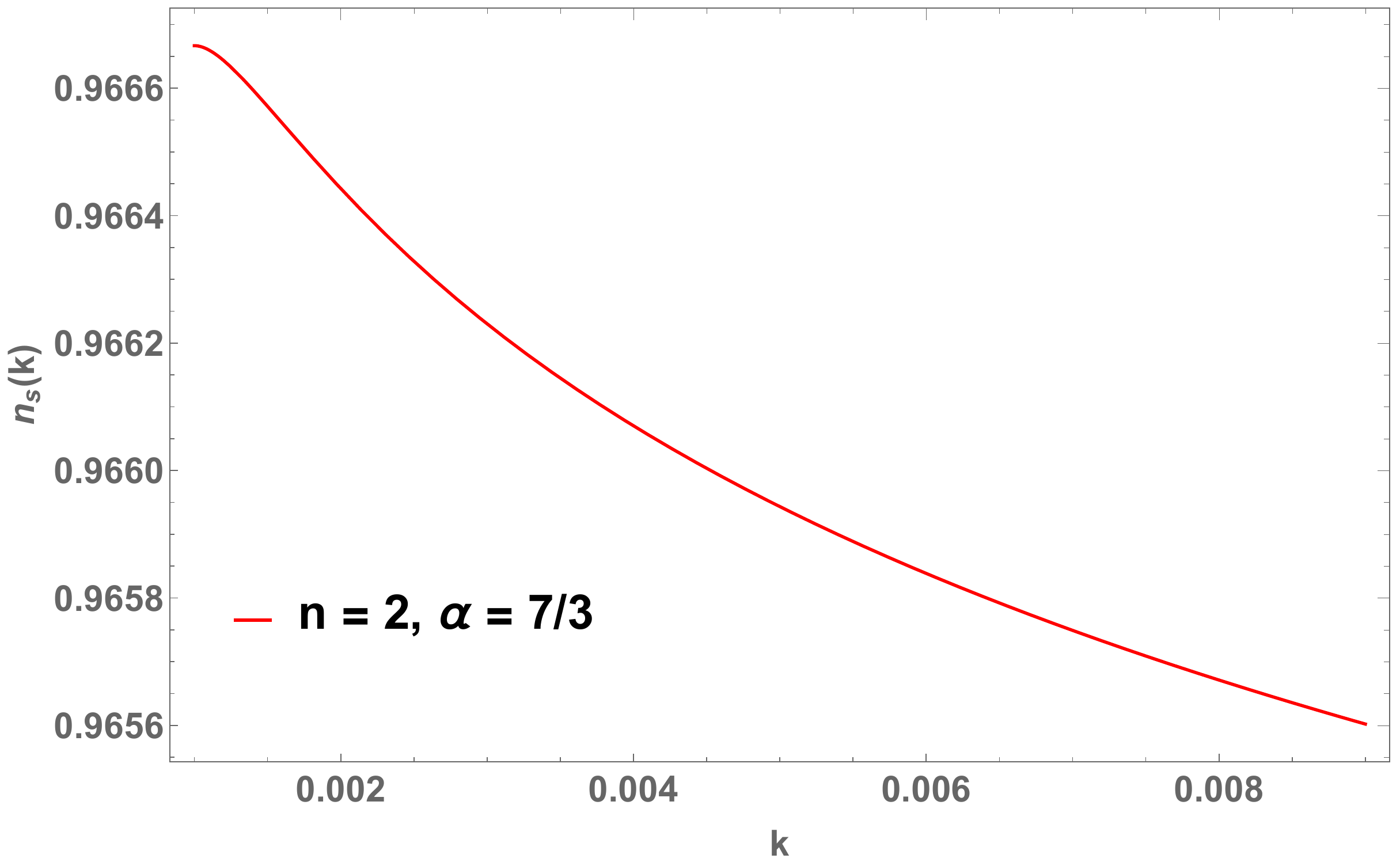} 
\end{subfigure}%
\caption{The $n_{s}(k)$ graphs in the $E$-model for the $\alpha$ values constrained by the $\mathcal{N}=1$ and the $\mathcal{N}=8$ supergravity.}
\end{figure}
\begin{figure}[H]
\begin{subfigure}{0.5\textwidth}
   \centering
    \includegraphics[width=75mm,height=65mm]{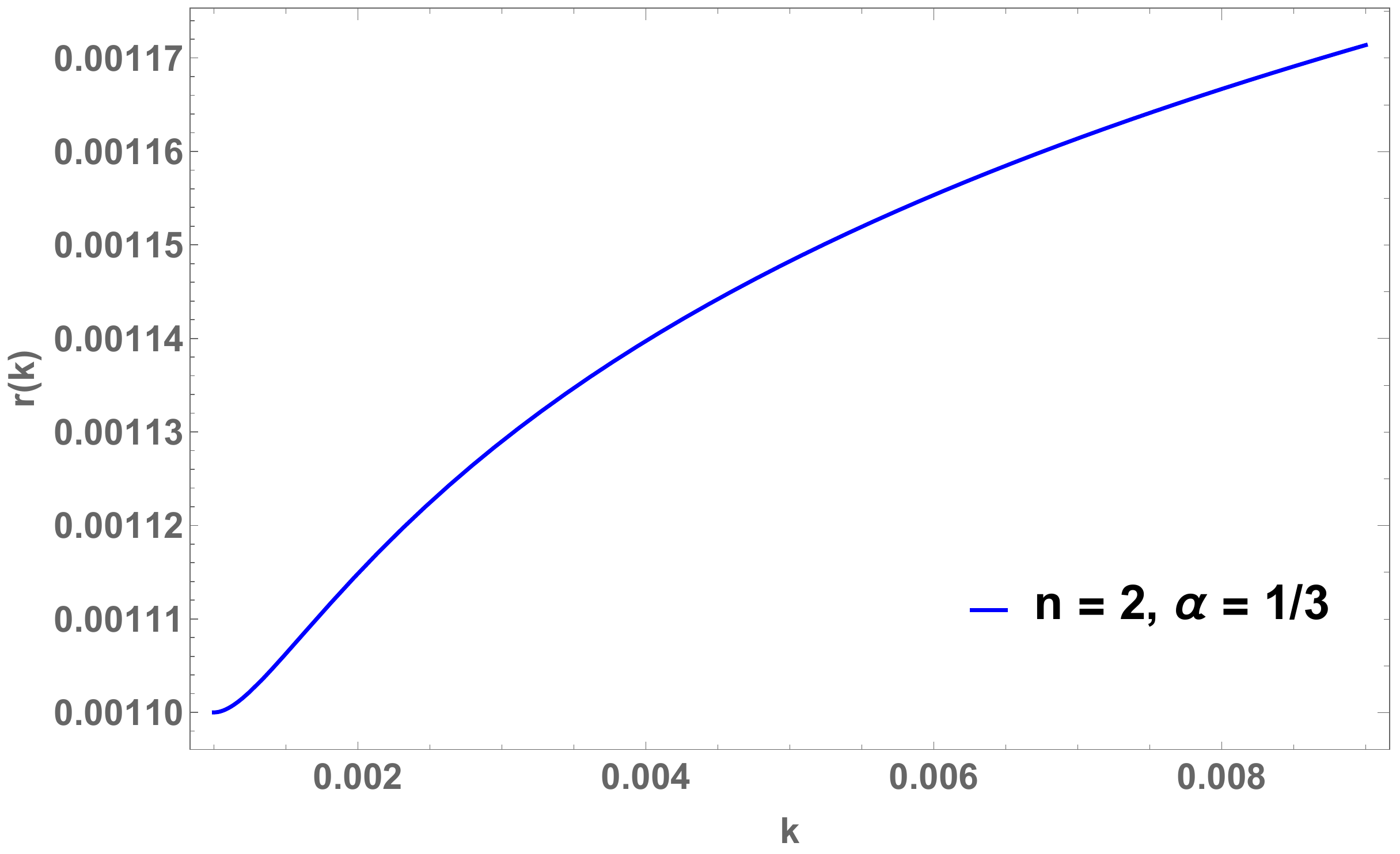} 
\end{subfigure}%
\begin{subfigure}{0.5\textwidth}
   \centering
    \includegraphics[width=75mm,height=65mm]{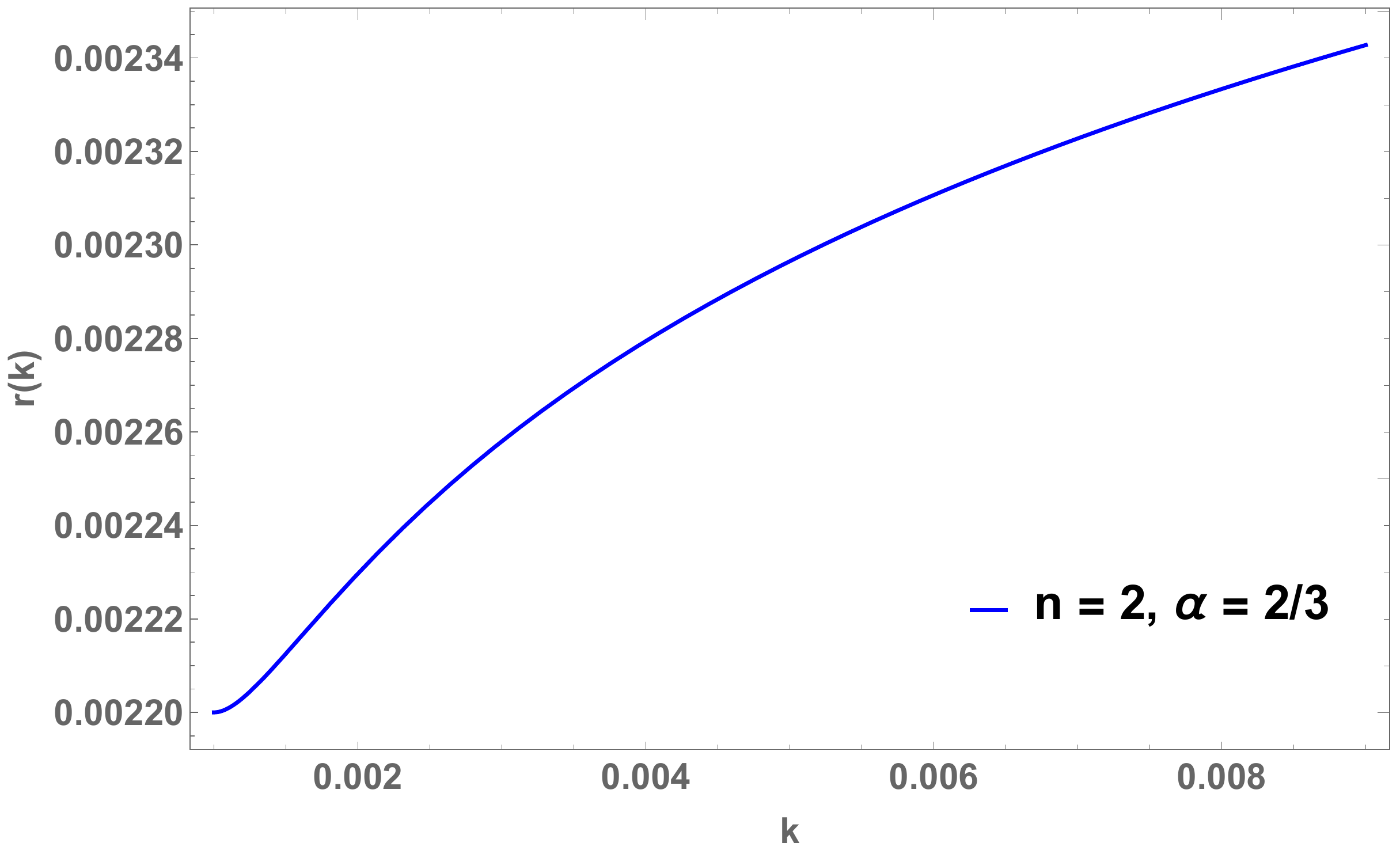} 
\end{subfigure}%

\begin{subfigure}{0.5\textwidth}
   \centering
    \includegraphics[width=75mm,height=65mm]{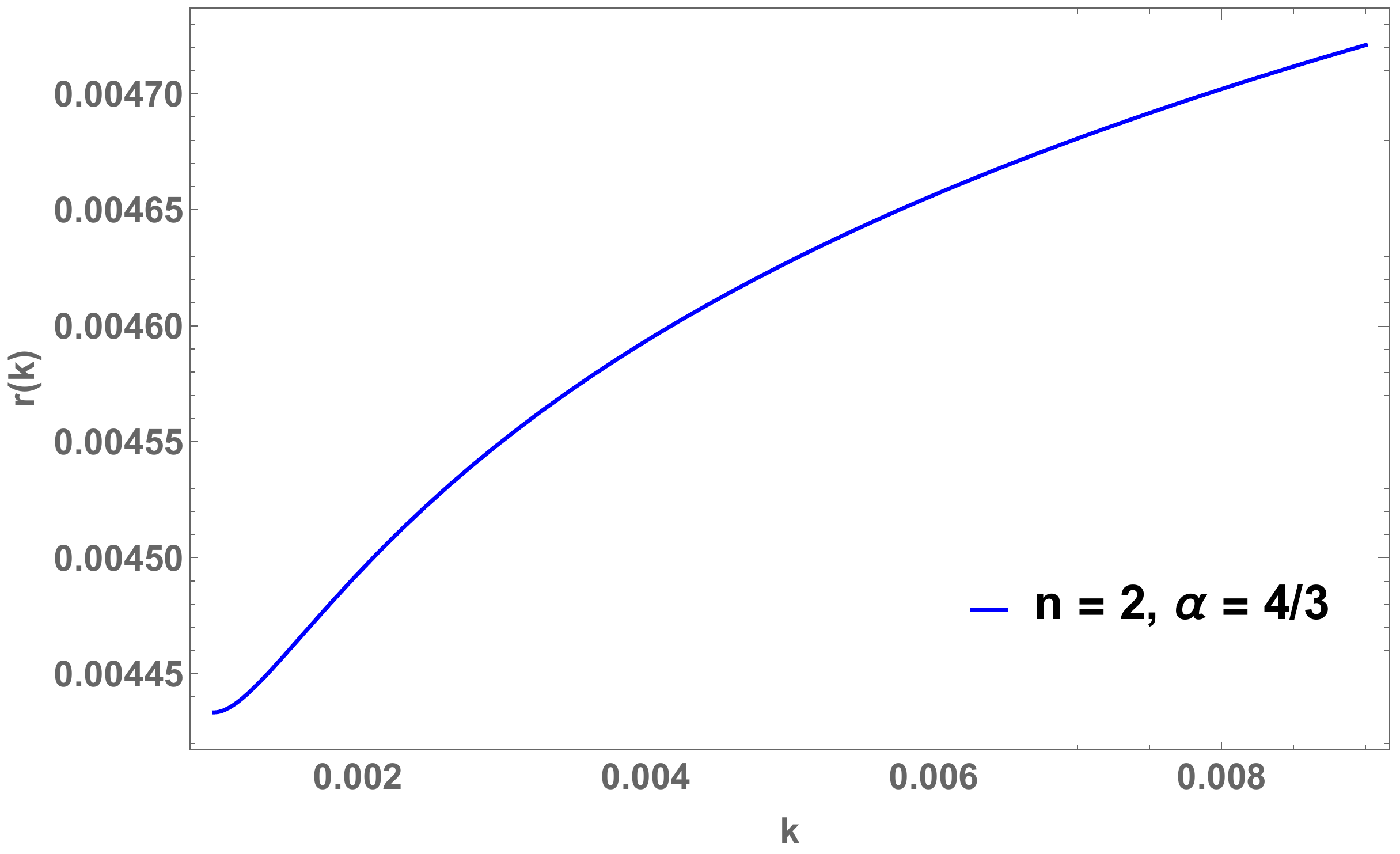} 
\end{subfigure}%
\begin{subfigure}{0.5\textwidth}
   \centering
    \includegraphics[width=75mm,height=65mm]{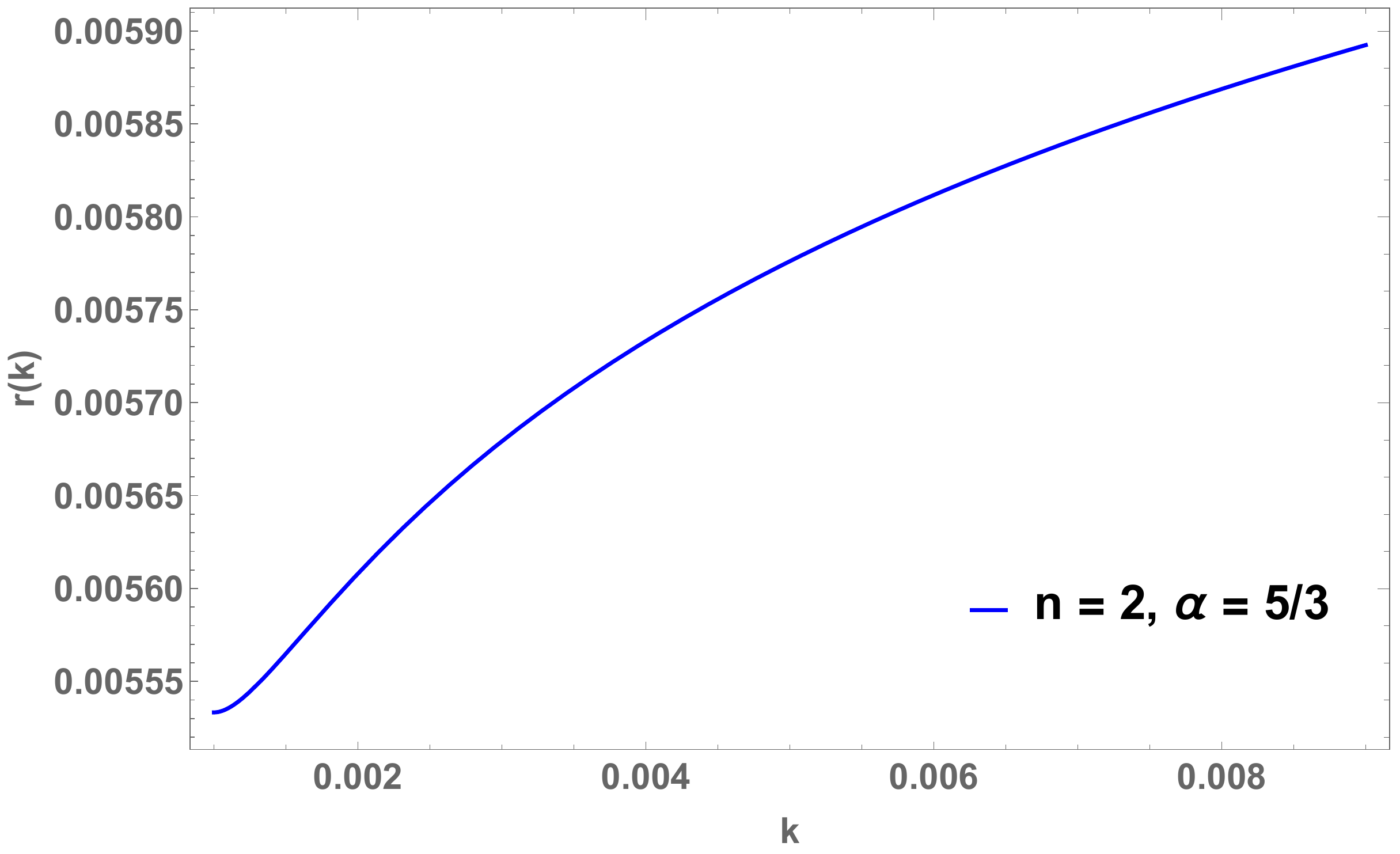}
\end{subfigure}%

\begin{subfigure}{0.5\textwidth}
   \centering
    \includegraphics[width=75mm,height=65mm]{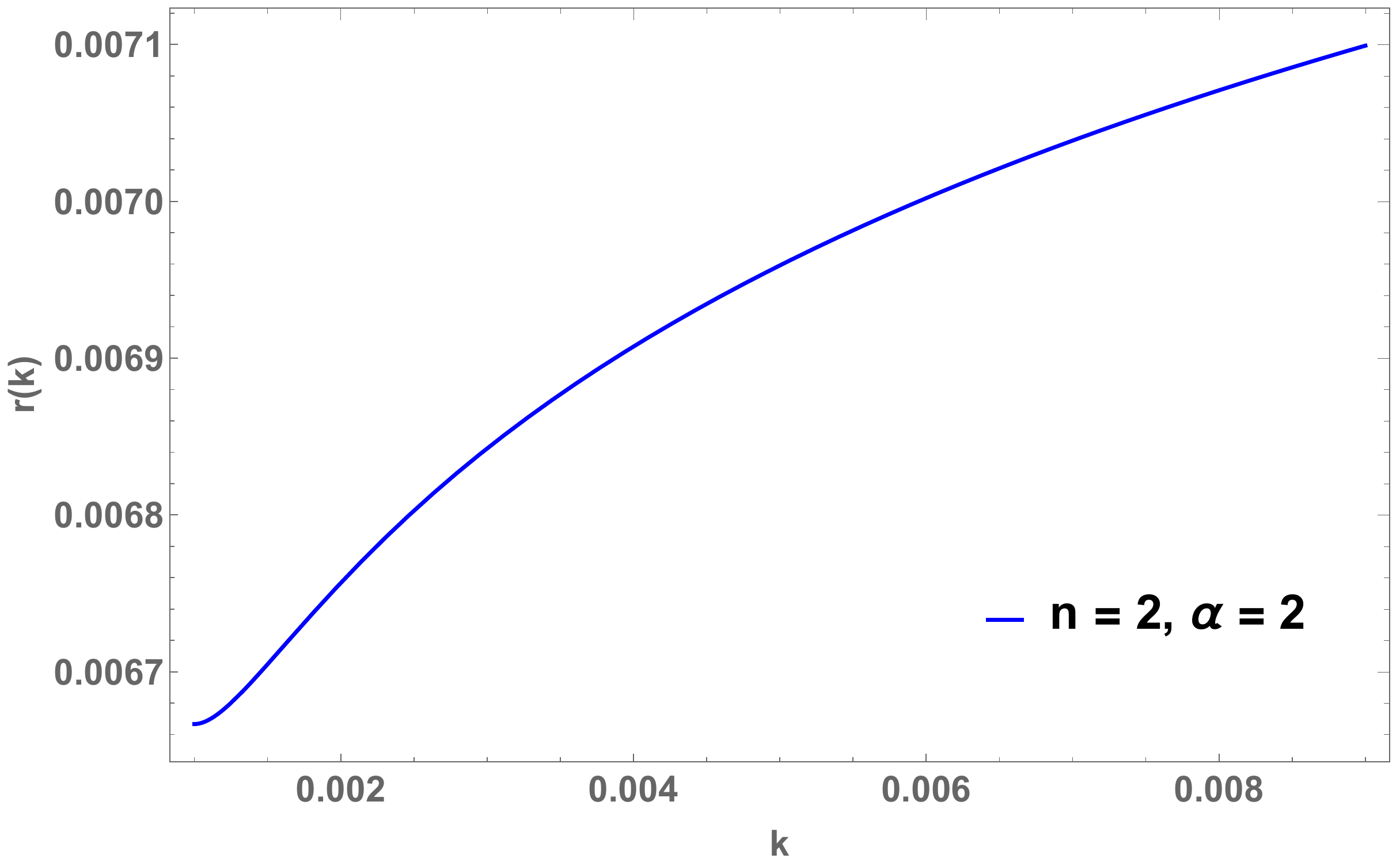}
    \end{subfigure}%
    \begin{subfigure}{0.5\textwidth}
   \centering
    \includegraphics[width=75mm,height=65mm]{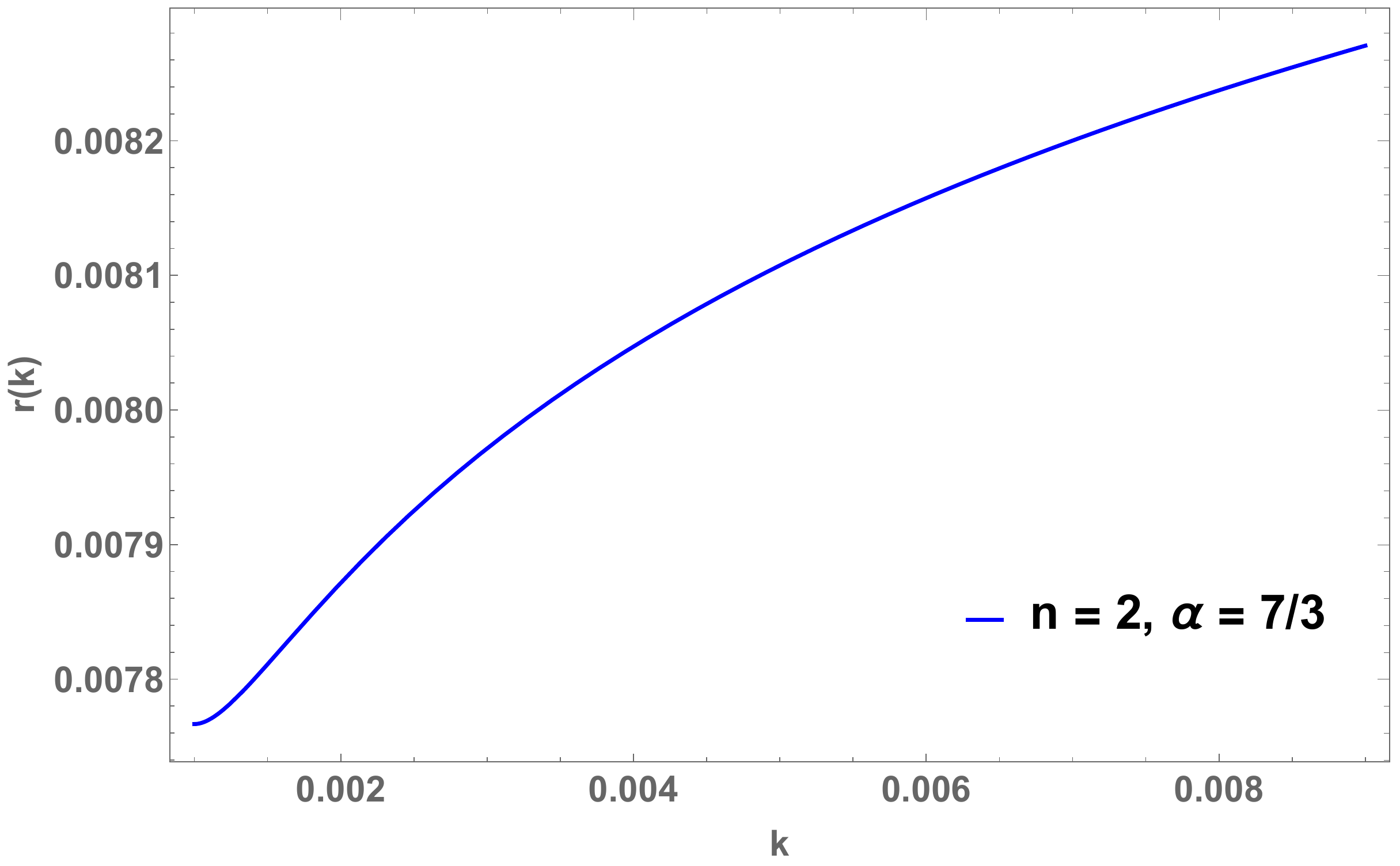} 
\end{subfigure}%
\caption{The $r(k)$ graphs in the $E$-model for the $\alpha$ values constrained by the $\mathcal{N}=1$ and the $\mathcal{N}=8$ supergravity.}
\end{figure}
\begin{figure}[H]
\begin{subfigure}{0.5\textwidth}
   \centering
    \includegraphics[width=75mm,height=65mm]{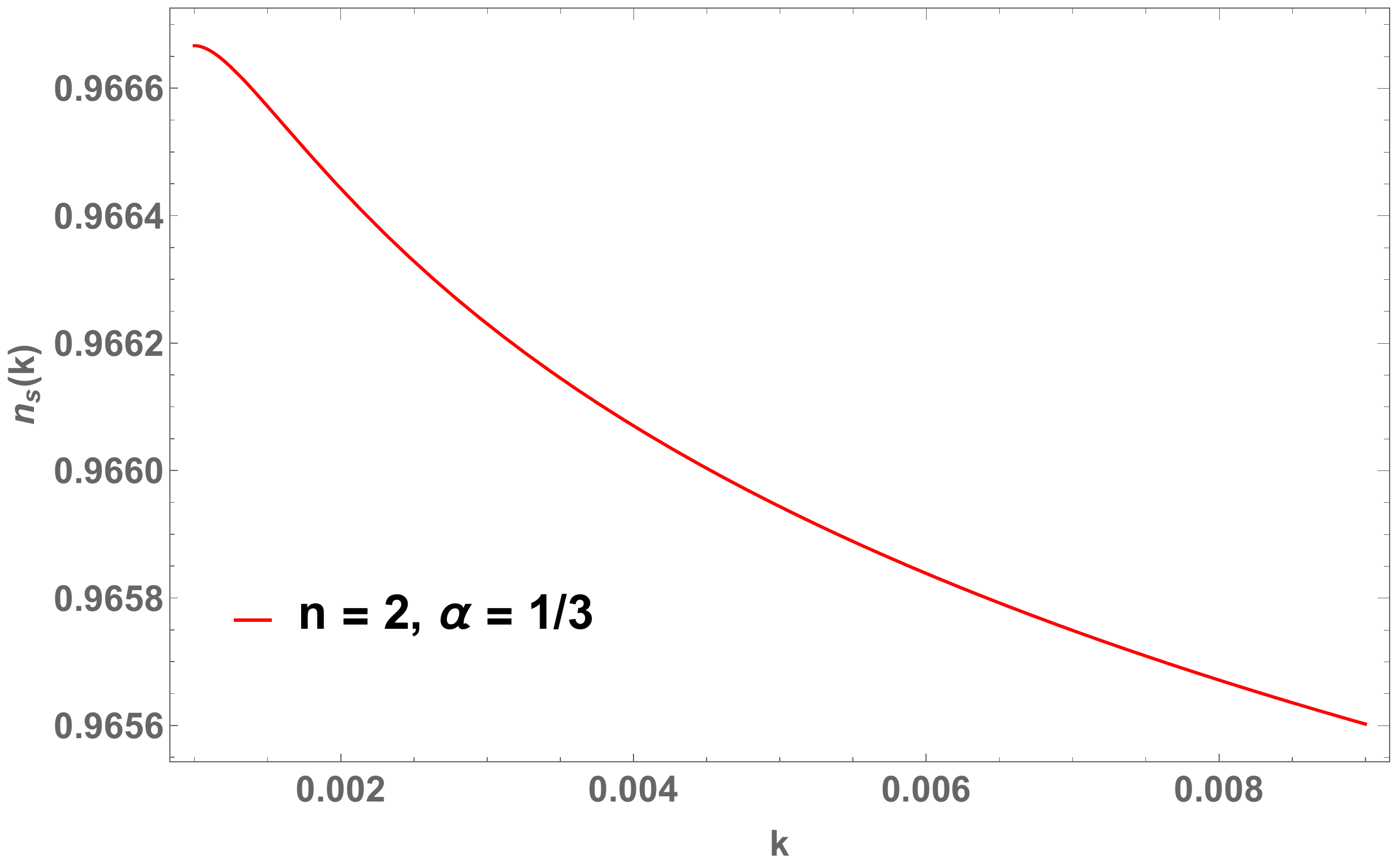} 
\end{subfigure}%
\begin{subfigure}{0.5\textwidth}
   \centering
    \includegraphics[width=75mm,height=65mm]{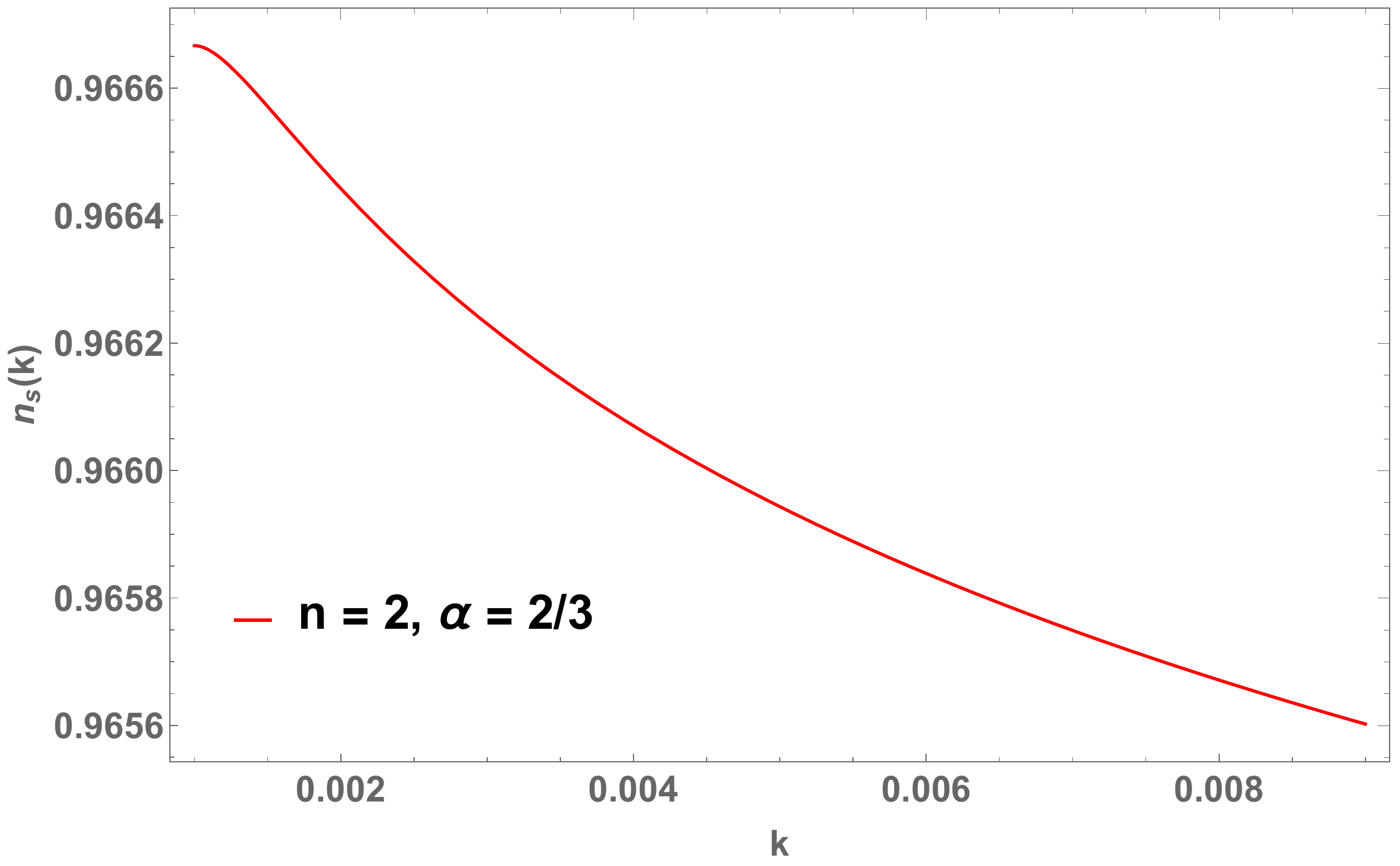} 
\end{subfigure}%

\begin{subfigure}{0.5\textwidth}
   \centering
    \includegraphics[width=75mm,height=65mm]{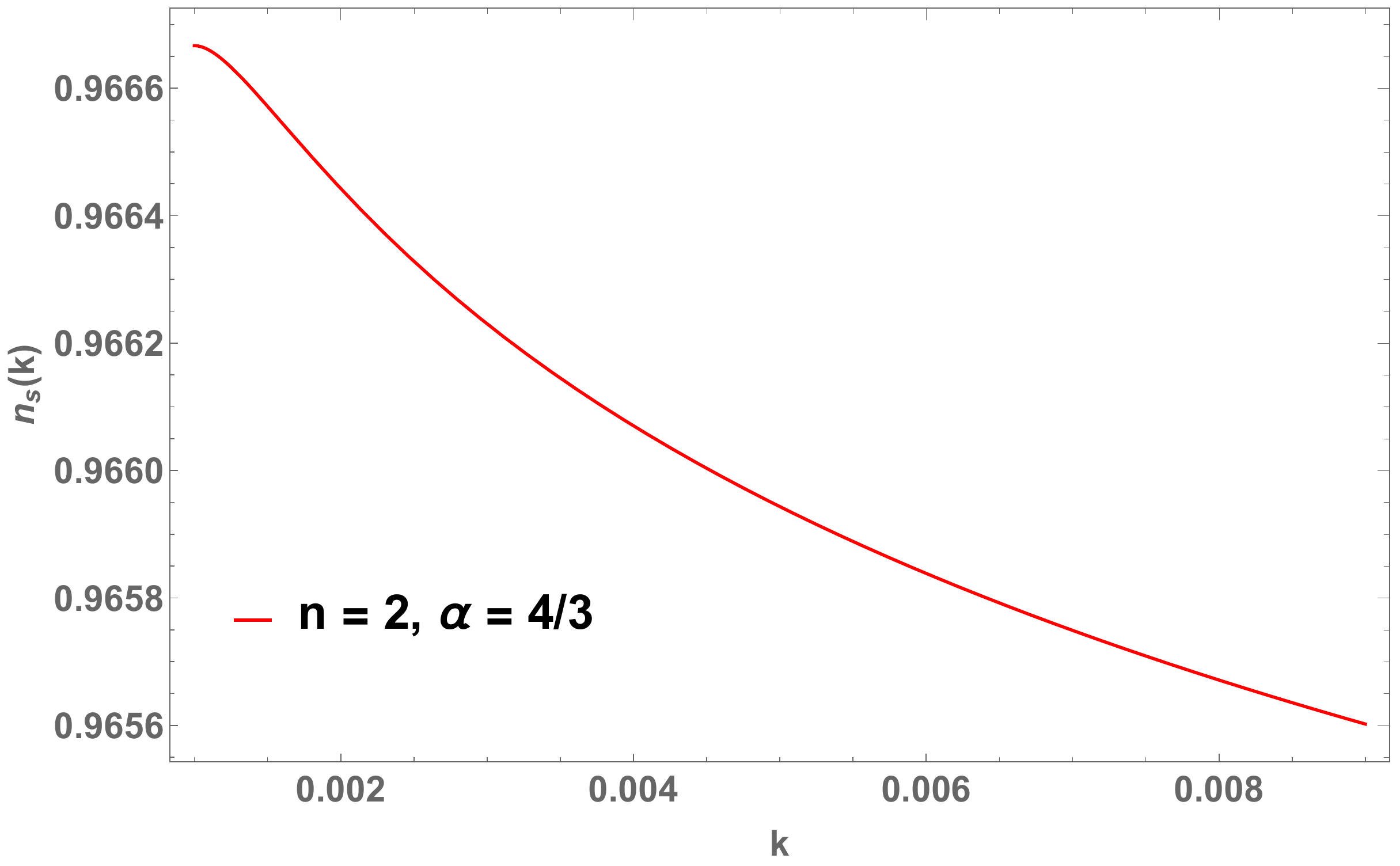} 
\end{subfigure}%
\begin{subfigure}{0.5\textwidth}
   \centering
    \includegraphics[width=75mm,height=65mm]{E_ED/E_Ed_ns_a_166.pdf}
\end{subfigure}%

\begin{subfigure}{0.5\textwidth}
   \centering
    \includegraphics[width=75mm,height=65mm]{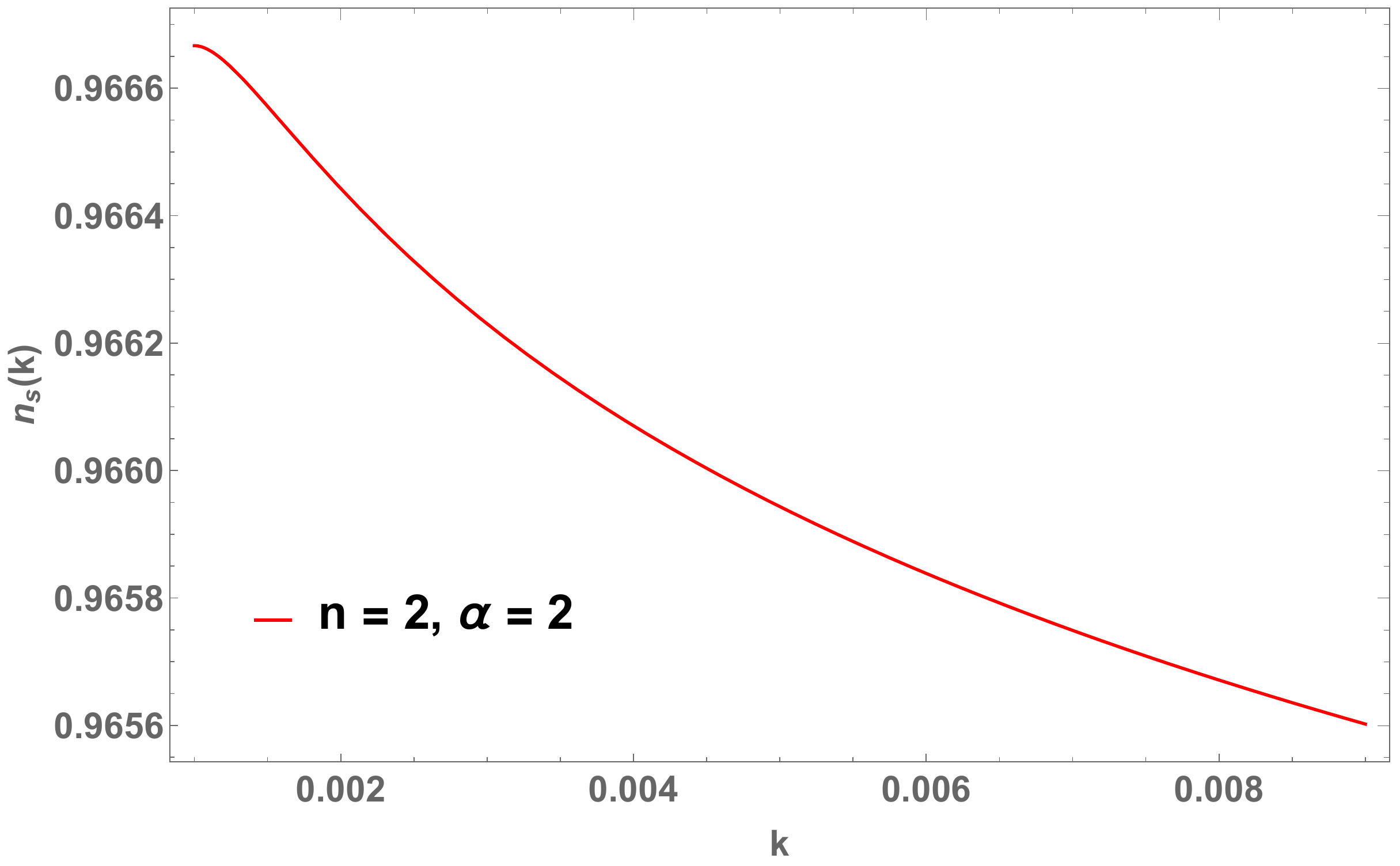}
    \end{subfigure}%
    \begin{subfigure}{0.5\textwidth}
   \centering
    \includegraphics[width=75mm,height=65mm]{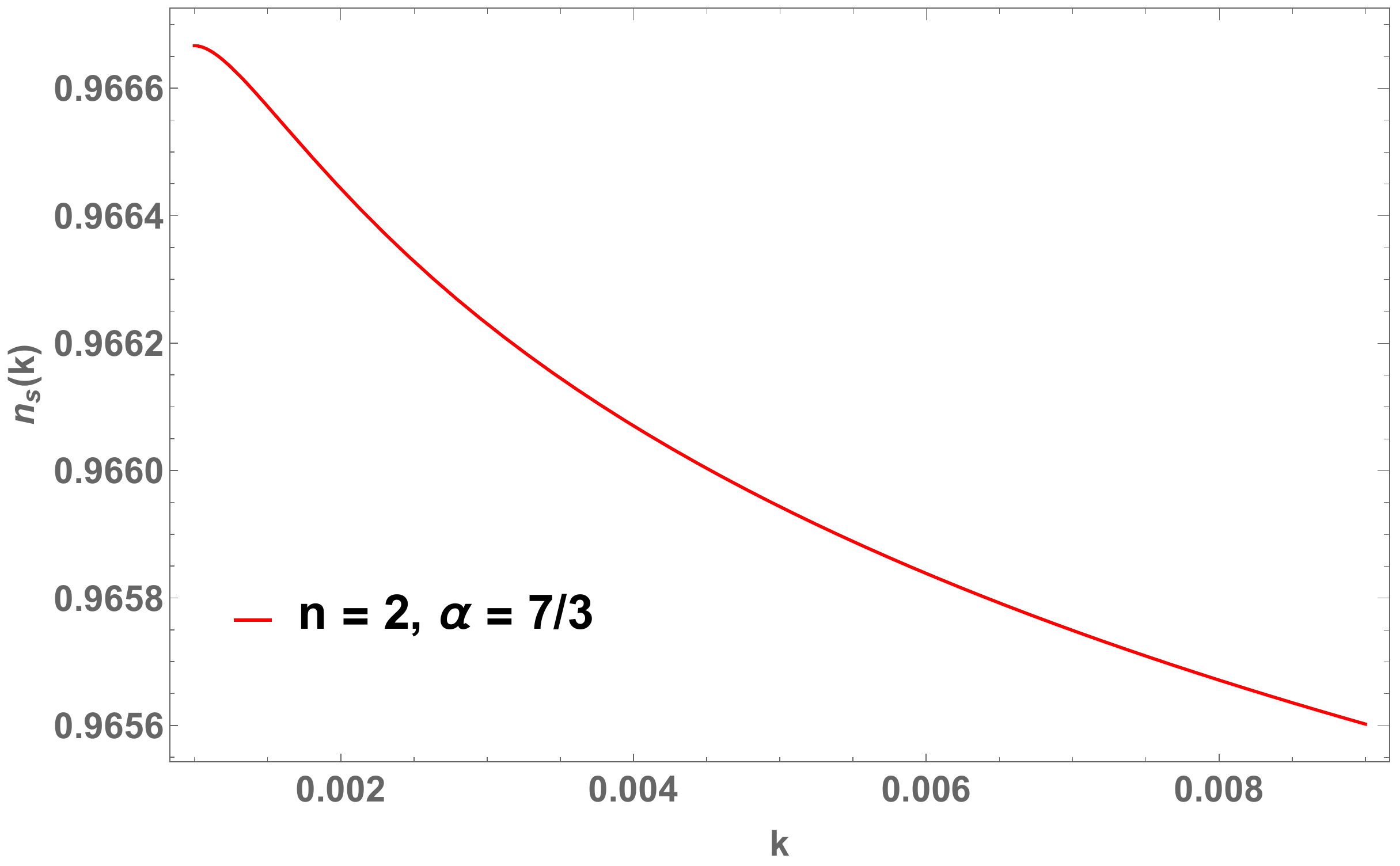} 
\end{subfigure}%
\caption{The $n_{s}(k)$ graphs in the $T$-model for the $\alpha$ values constrained by the $\mathcal{N}=1$ and the $\mathcal{N}=8$ supergravity.}
\end{figure}
\begin{figure}[H]
\begin{subfigure}{0.5\textwidth}
   \centering
    \includegraphics[width=75mm,height=65mm]{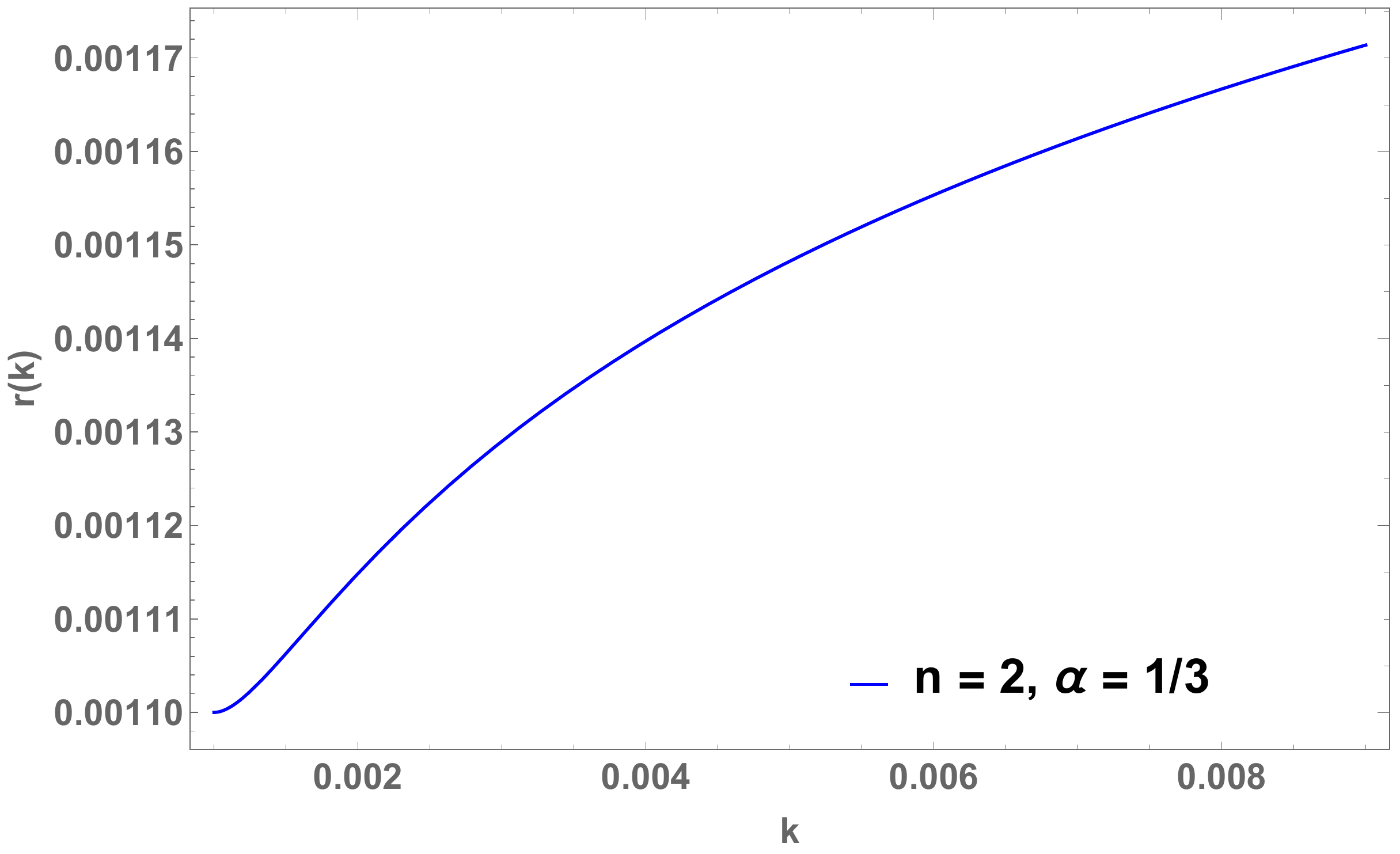} 
\end{subfigure}%
\begin{subfigure}{0.5\textwidth}
   \centering
    \includegraphics[width=75mm,height=65mm]{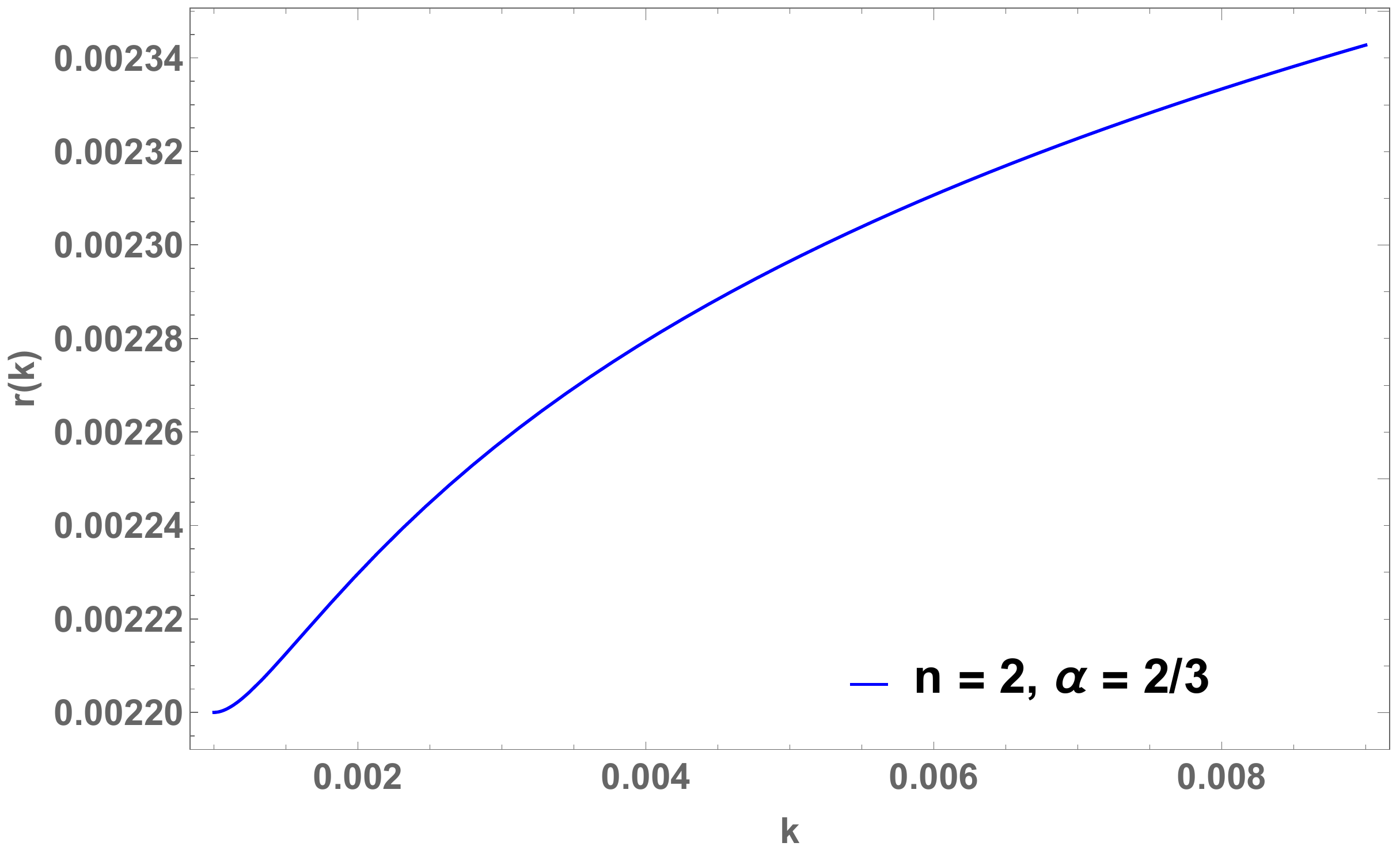} 
\end{subfigure}%

\begin{subfigure}{0.5\textwidth}
   \centering
    \includegraphics[width=75mm,height=65mm]{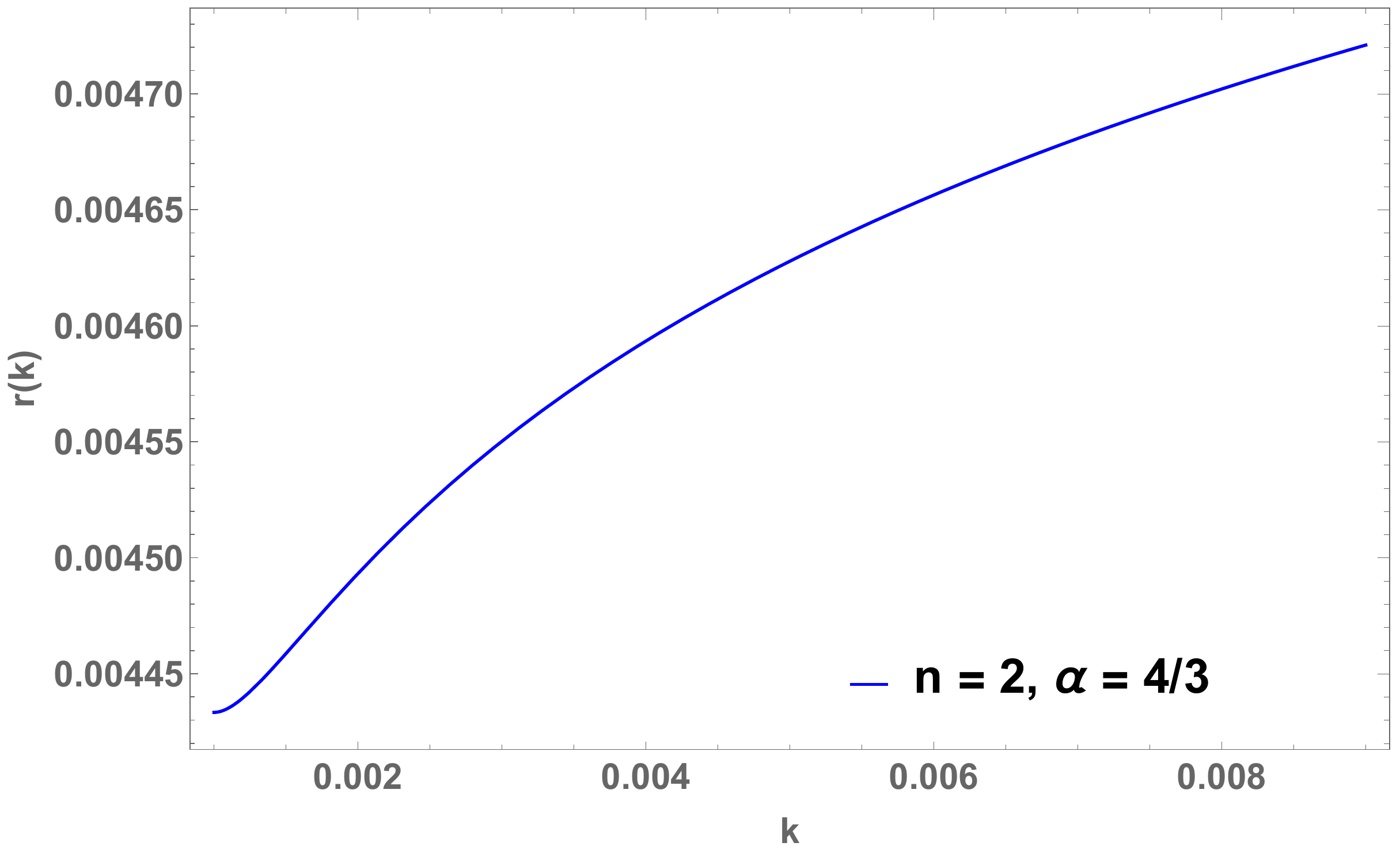} 
\end{subfigure}%
\begin{subfigure}{0.5\textwidth}
   \centering
    \includegraphics[width=75mm,height=65mm]{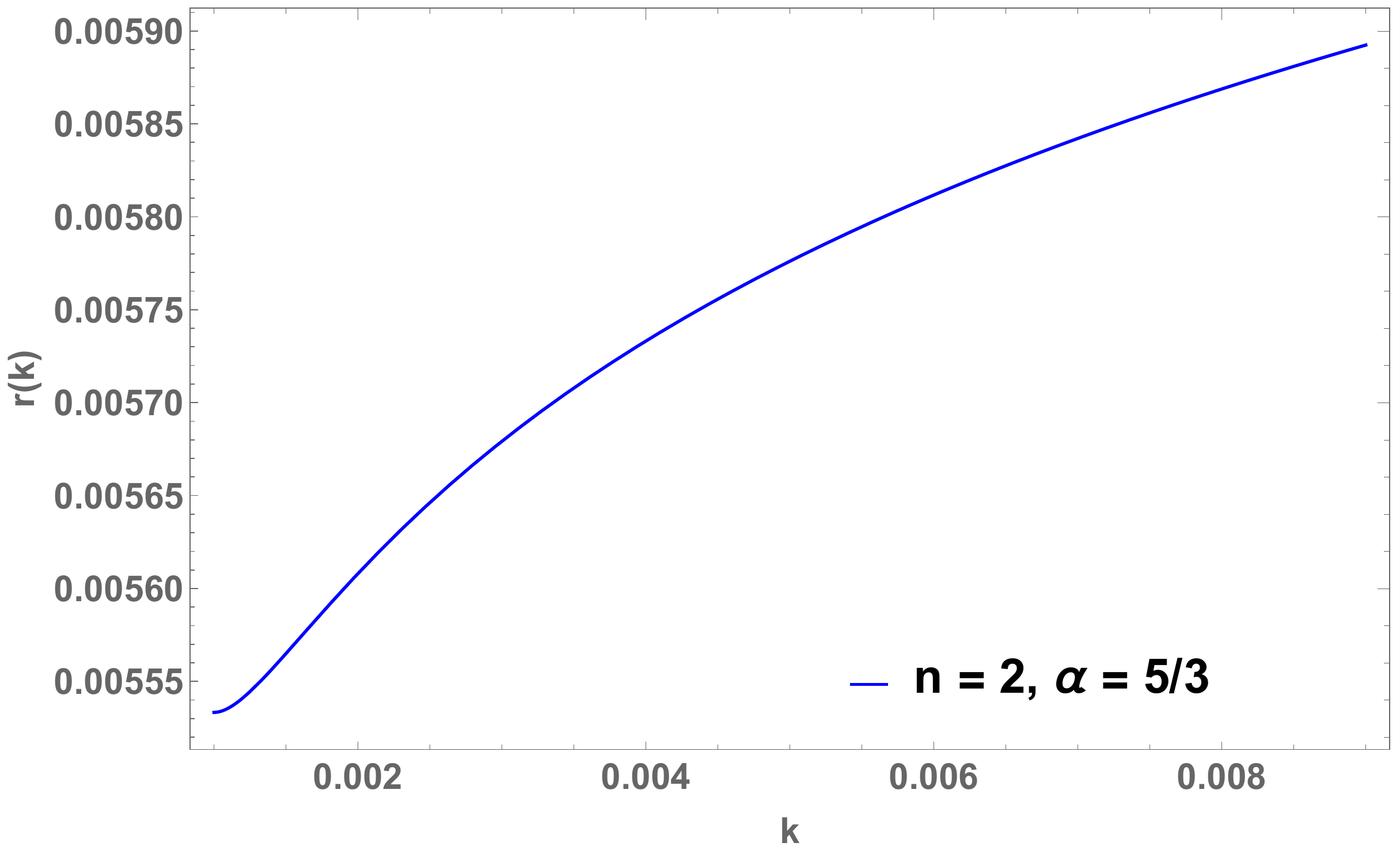}
\end{subfigure}%

\begin{subfigure}{0.5\textwidth}
   \centering
    \includegraphics[width=75mm,height=65mm]{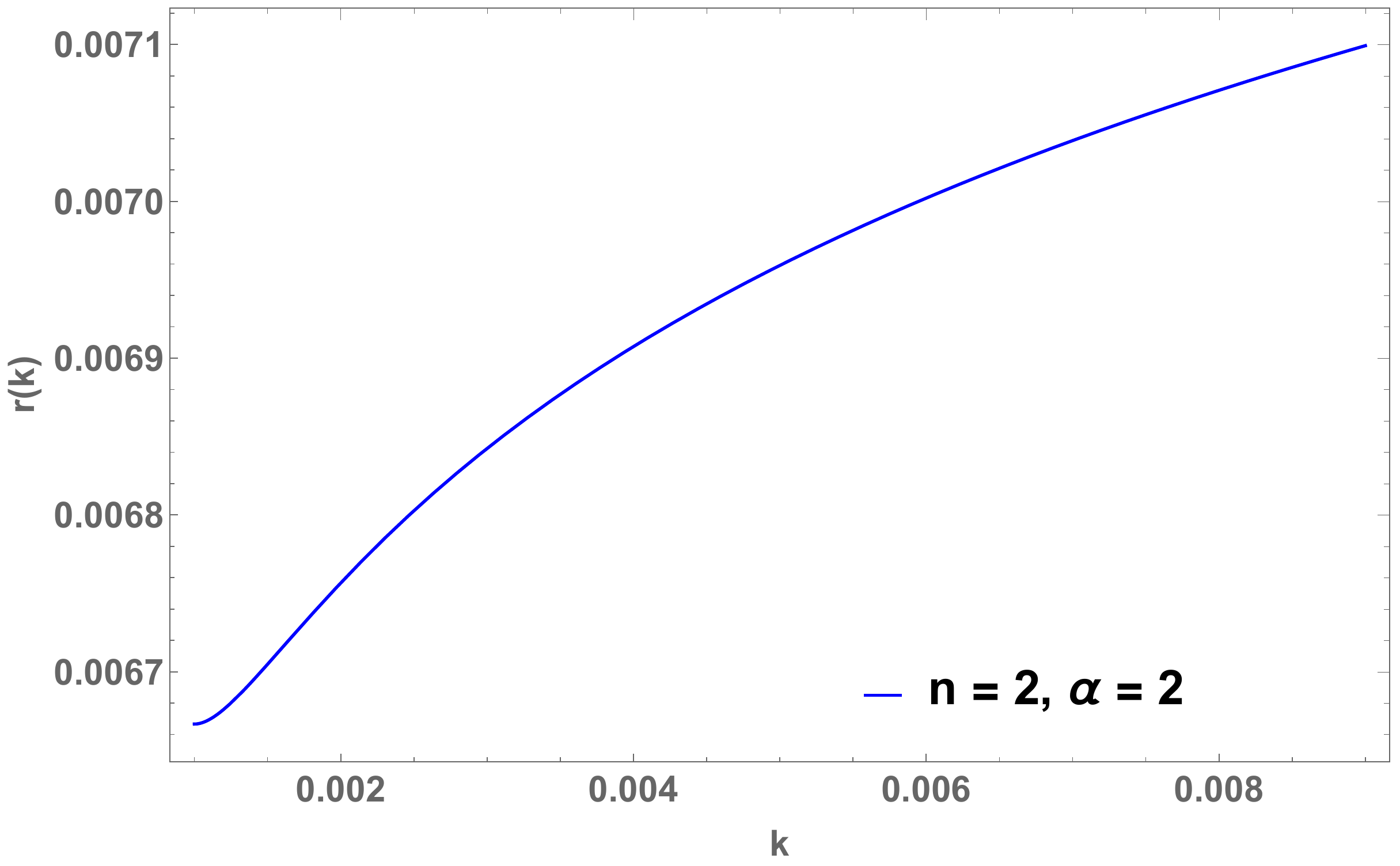}
    \end{subfigure}%
    \begin{subfigure}{0.5\textwidth}
   \centering
    \includegraphics[width=75mm,height=65mm]{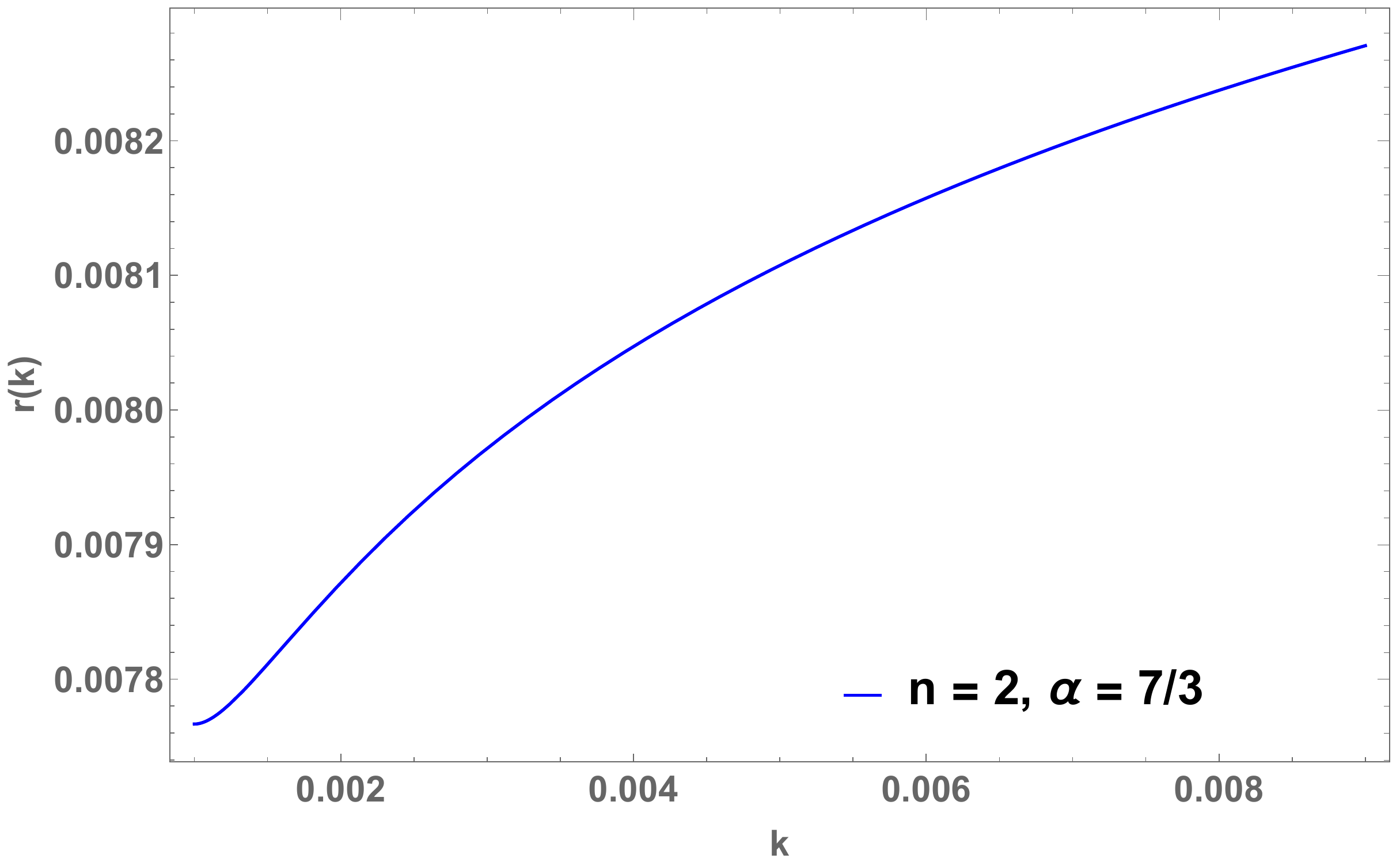} 
\end{subfigure}%
\caption{The $r(k)$ graphs in the $T$-model for the $\alpha$ values constrained by the $\mathcal{N}=1$ and the $\mathcal{N}=8$ supergravity models.}
\end{figure}
\subsection{Comparison with PLANCK-2018 data}
\begin{figure}[H]
\begin{subfigure}{0.37\textwidth}
  \centering
   \includegraphics[width=57mm,height=65mm]{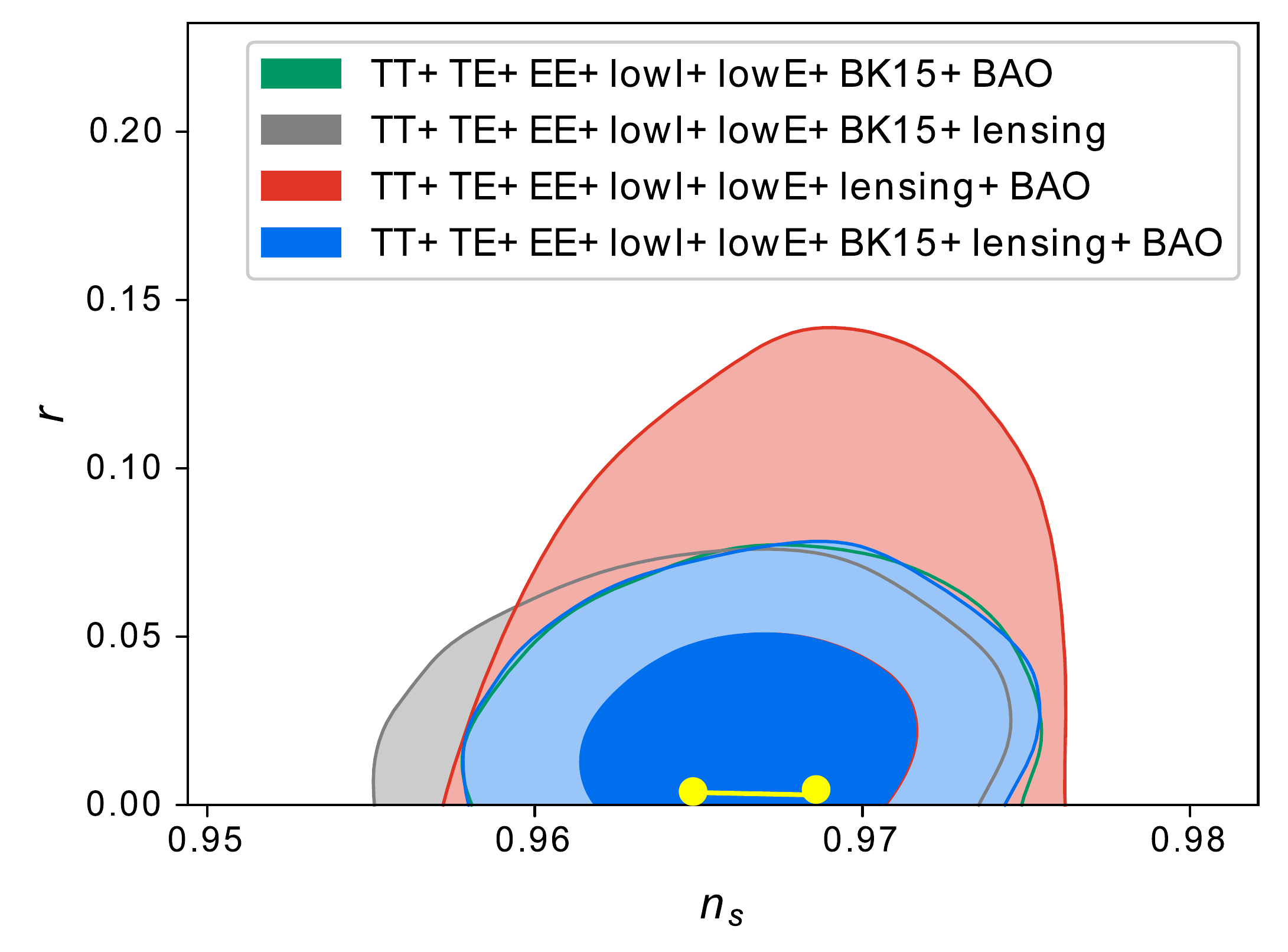}
    \caption{E model with $\alpha=1$}
    \label{fig:Planck_E_1}
\end{subfigure}%
\begin{subfigure}{0.37\textwidth}
  \centering
   \includegraphics[width=57mm,height=65mm]{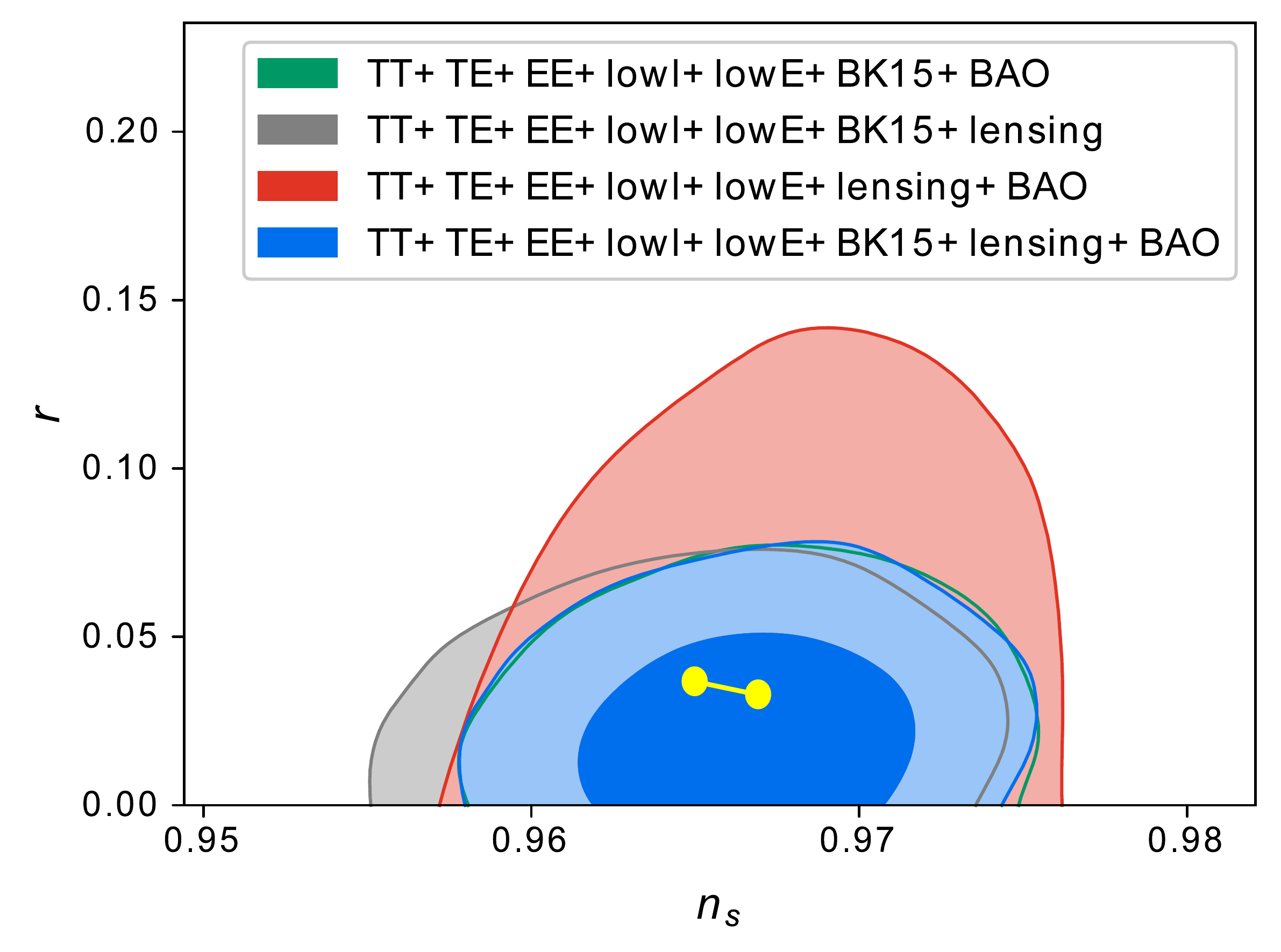}
    \caption{E model with $\alpha=10$}
     \label{fig:Planck_E_2}
\end{subfigure}%
\begin{subfigure}{0.37\textwidth}
  \centering
   \includegraphics[width=57mm,height=65mm]{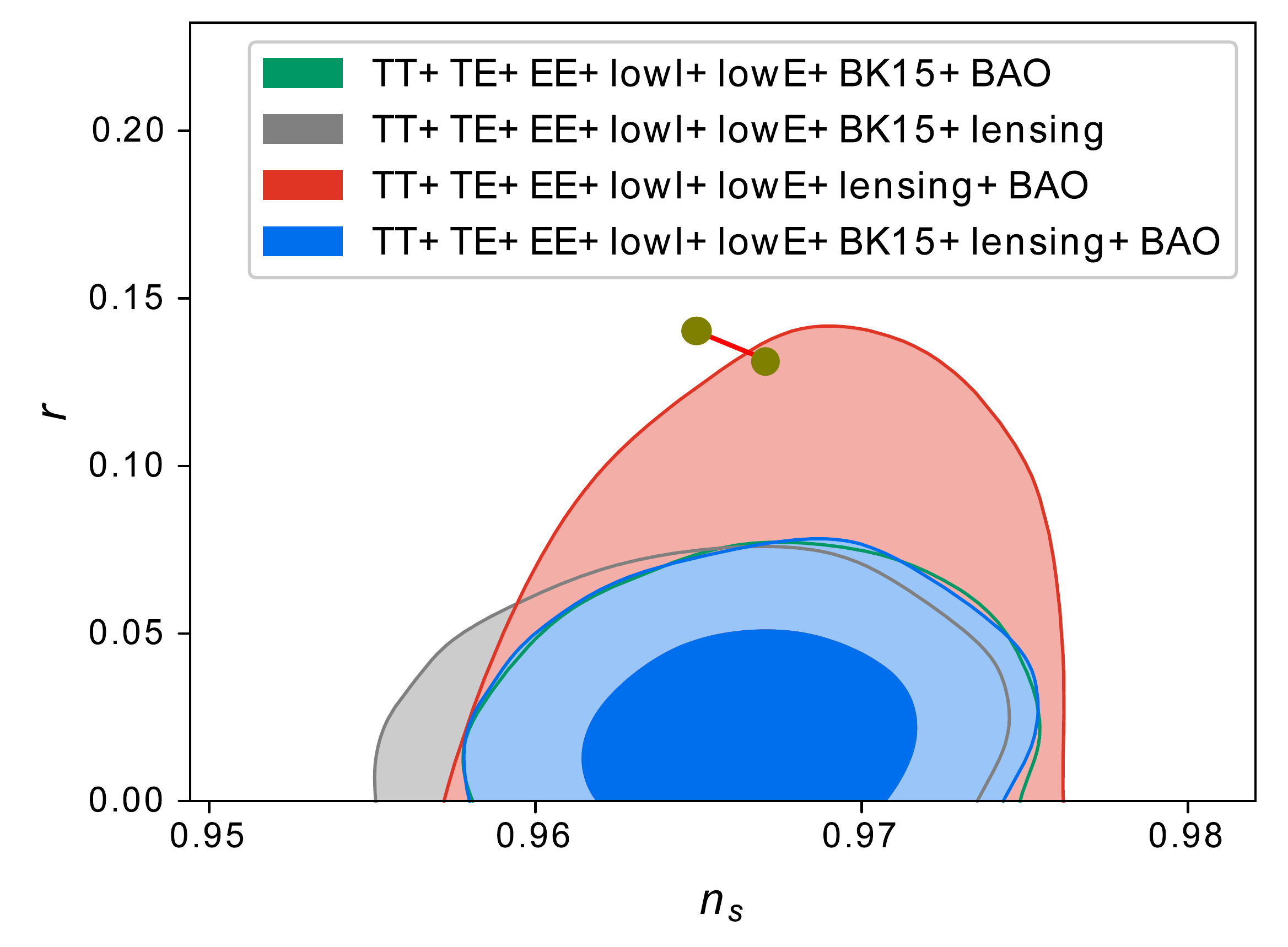}
    \caption{E model with $\alpha=15$}
     \label{fig:Planck_E_3}
\end{subfigure}

\begin{subfigure}{0.37\textwidth}
  \centering
   \includegraphics[width=57mm,height=65mm]{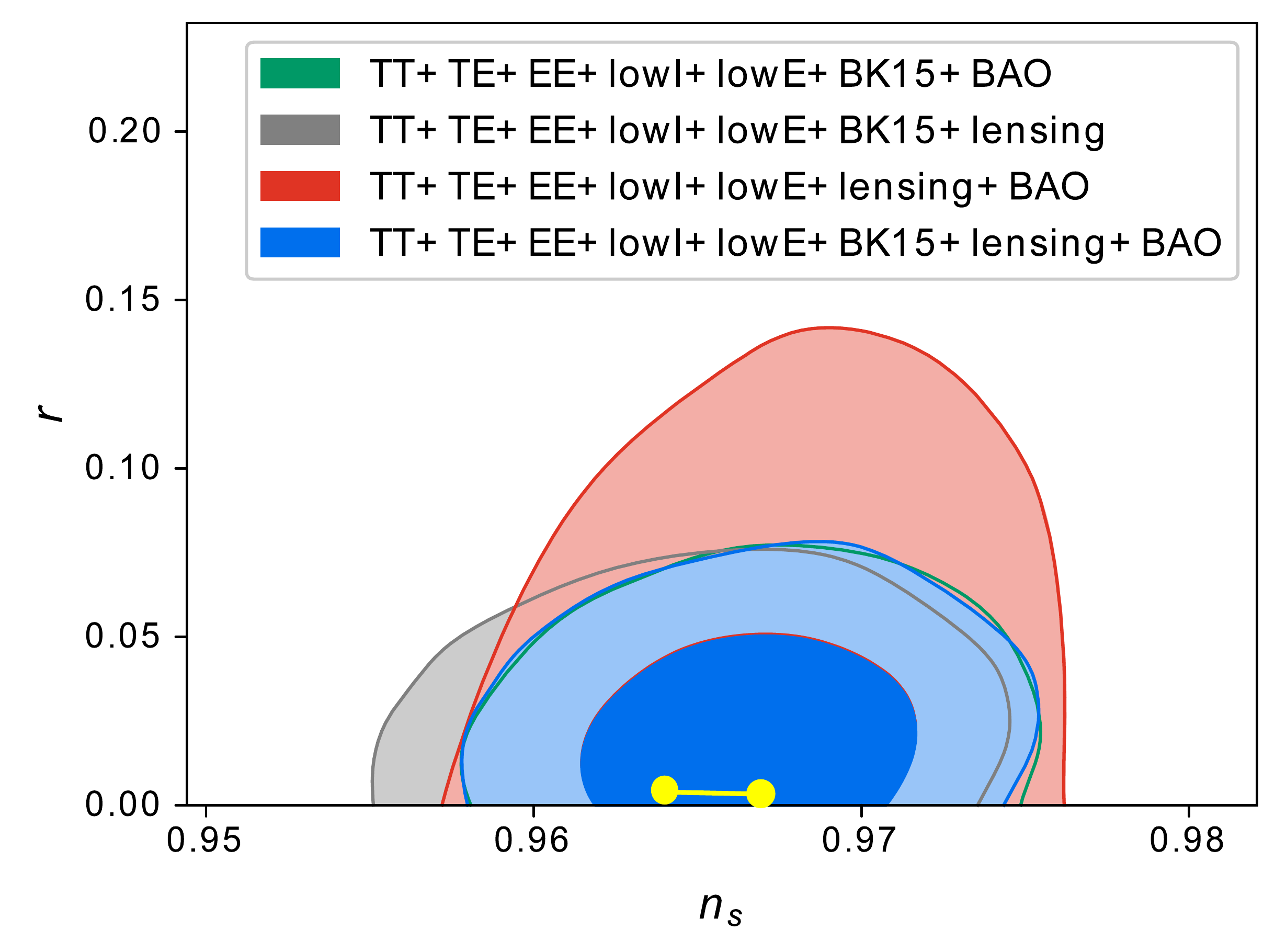}  
    \caption{T model with $\alpha=1$}
     \label{fig:Planck_T_1}
\end{subfigure}%
\begin{subfigure}{0.37\textwidth}
  \centering
   \includegraphics[width=57mm,height=65mm]{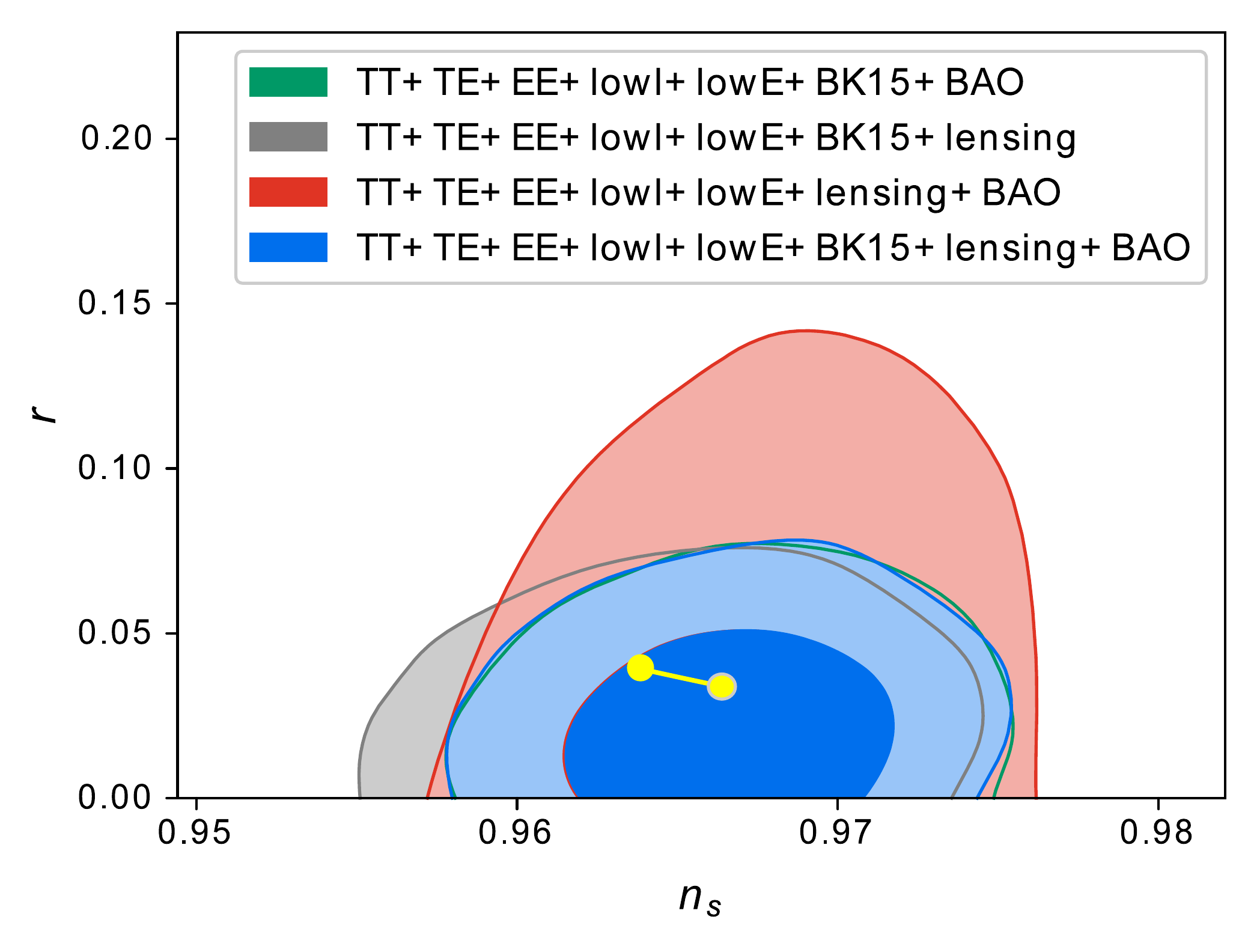} 
    \caption{T model with $\alpha=10$}
     \label{fig:Planck_T_2}
\end{subfigure}%
\begin{subfigure}{0.37\textwidth}
  \centering
   \includegraphics[width=57mm,height=65mm]{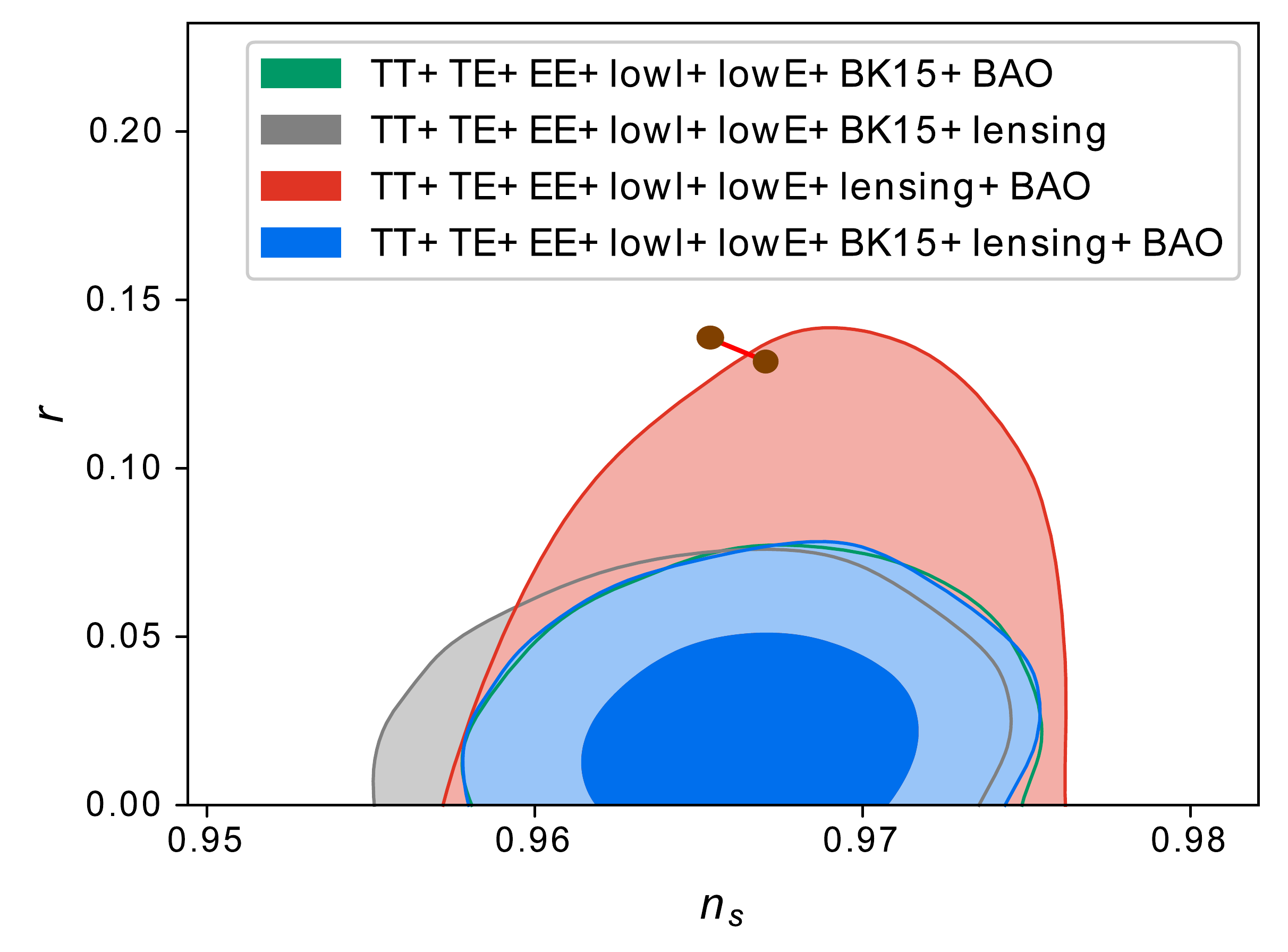} 
   \caption{T model with $\alpha=15$}
    \label{fig:Planck_T_3}
\end{subfigure}%
\caption{Comparison of  theoretical results with the Planck-2018 data on the $n_{s}$-$r$ plane for the $E$ and the  $T$ models ($n=2$). The theoretical graph is shown by yellow line for $\alpha=1,10$ and red line for $\alpha=15$ with two dots at the ends. For the $E$ model (Figures \ref{fig:Planck_E_1}, \ref{fig:Planck_E_2} and \ref{fig:Planck_E_3}): $\alpha = 1, 10,$ lie within $68\%$ CL and $\alpha=15$ lie near 95\% CL. For the $T$ model (Figures \ref{fig:Planck_T_1}, \ref{fig:Planck_T_2} and \ref{fig:Planck_T_3}): $\alpha = 1, 10$ lie within $68\%$ CL and $\alpha=15$ lie near $95$\% CL.}
    \label{fig:planck data}
\end{figure}
\begin{figure}[H]
\begin{subfigure}{0.55\textwidth}
  \centering
   \includegraphics[width=75mm,height=70mm]{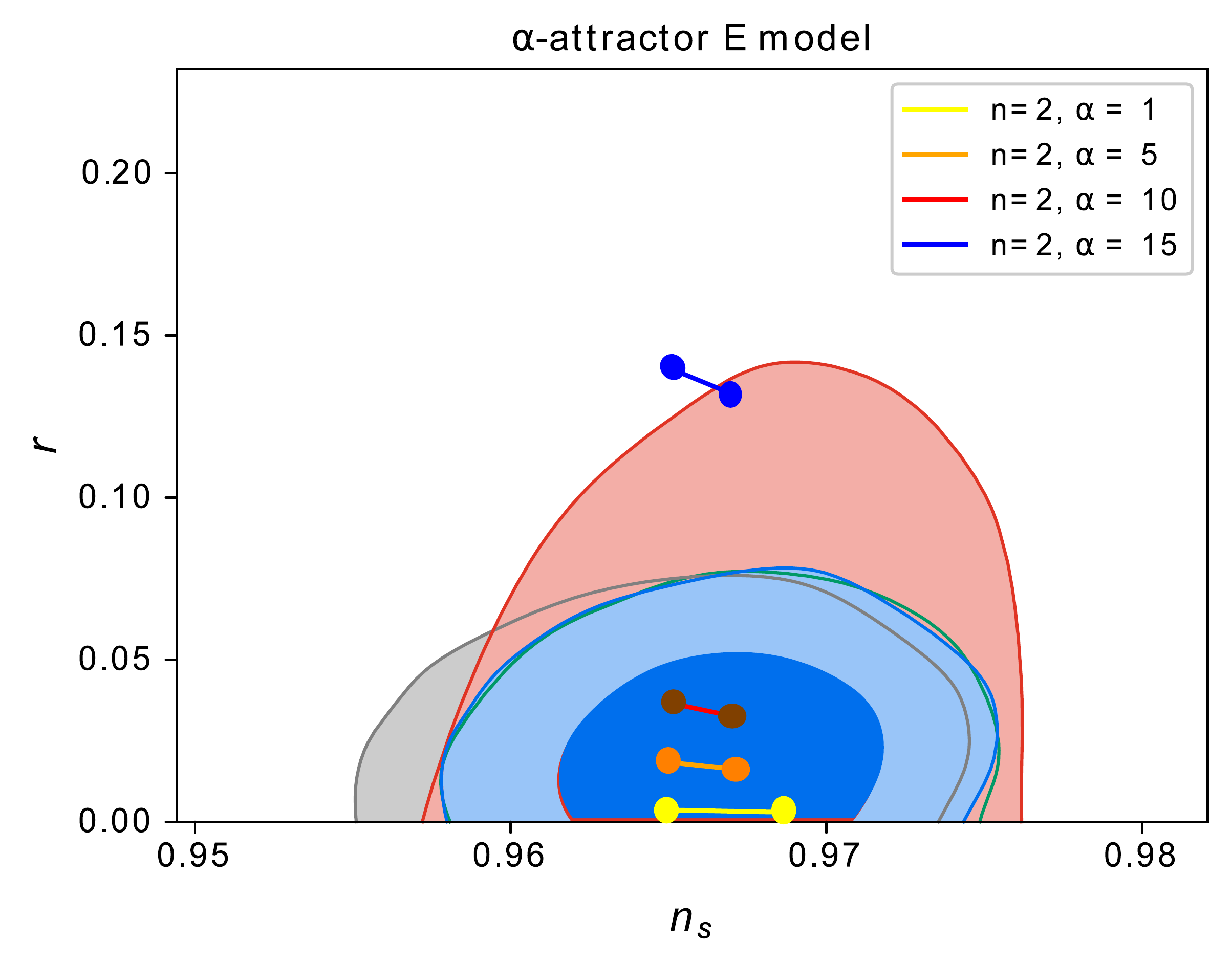}
   \subcaption{}
   \label{fig:combined_Planck_E}
\end{subfigure}
\begin{subfigure}{0.55\textwidth}
  \centering
   \includegraphics[width=75mm,height=70mm]{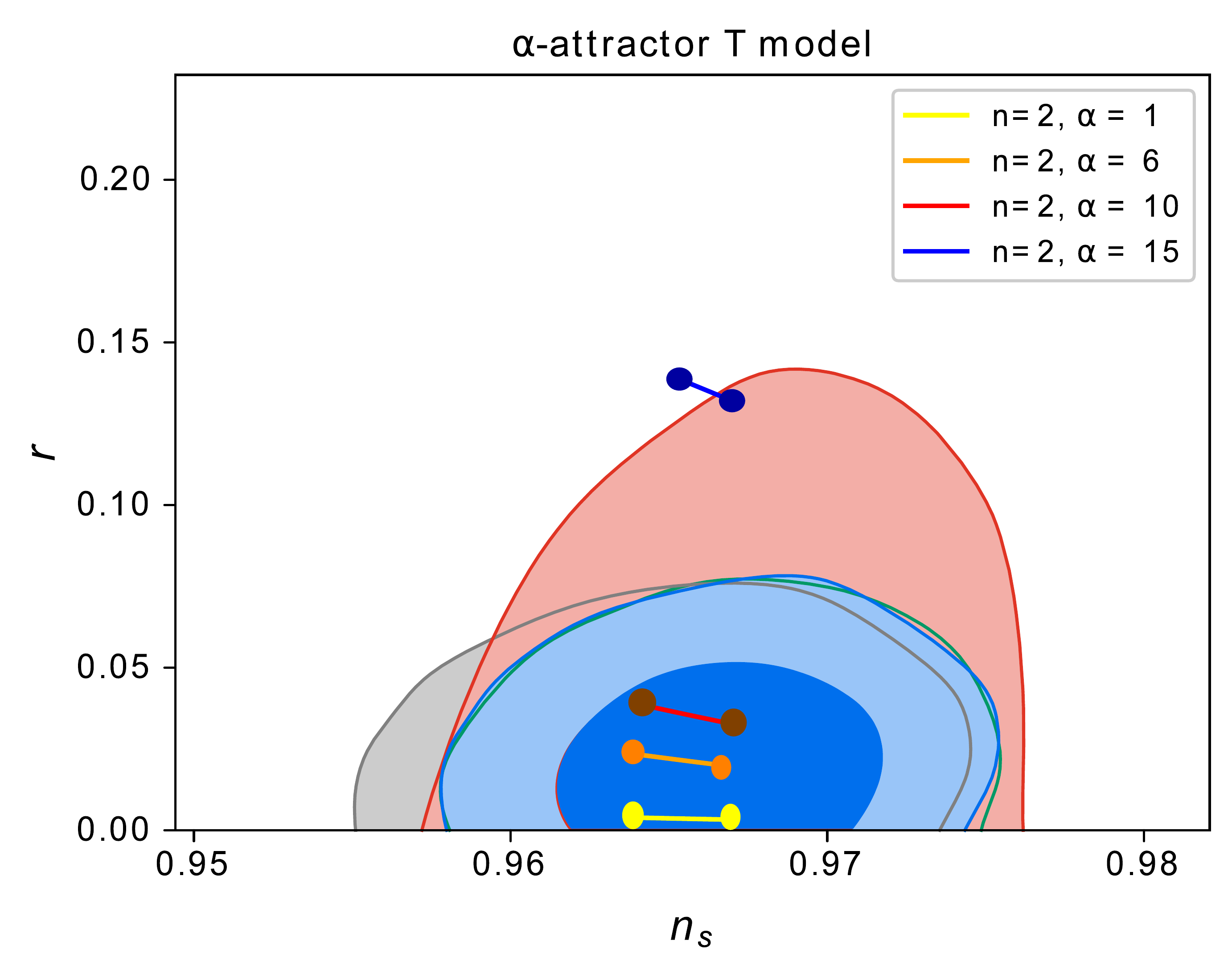}
   \subcaption{}
    \label{fig:combined_Planck_T}
\end{subfigure}
    \caption{Comparison of theoretical results with the Planck-2018 data for the $E$ and the $T$ models ($n=2$) on the $n_{s}$-$r$ plane. The theoretical graphs are shown by colored lines with two dots at the ends. (a) $E$ models: $\alpha=1$ (yellow), $5$ (orange) and $10$ (red) lie within $68\%$ CL and $\alpha=15$ (blue) lie within $95$\% CL. (b) $T$ models: $\alpha=1$ (yellow), $6$ (orange) and $10$ (red) lie within $68\%$ CL and $\alpha=15$ (blue) lie near $95$\% CL. }
    \label{fig:chaotic planck}
\end{figure}
\begin{figure}[H]
\begin{subfigure}{0.5\textwidth}
  \centering
   \includegraphics[width=75mm,height=70mm]{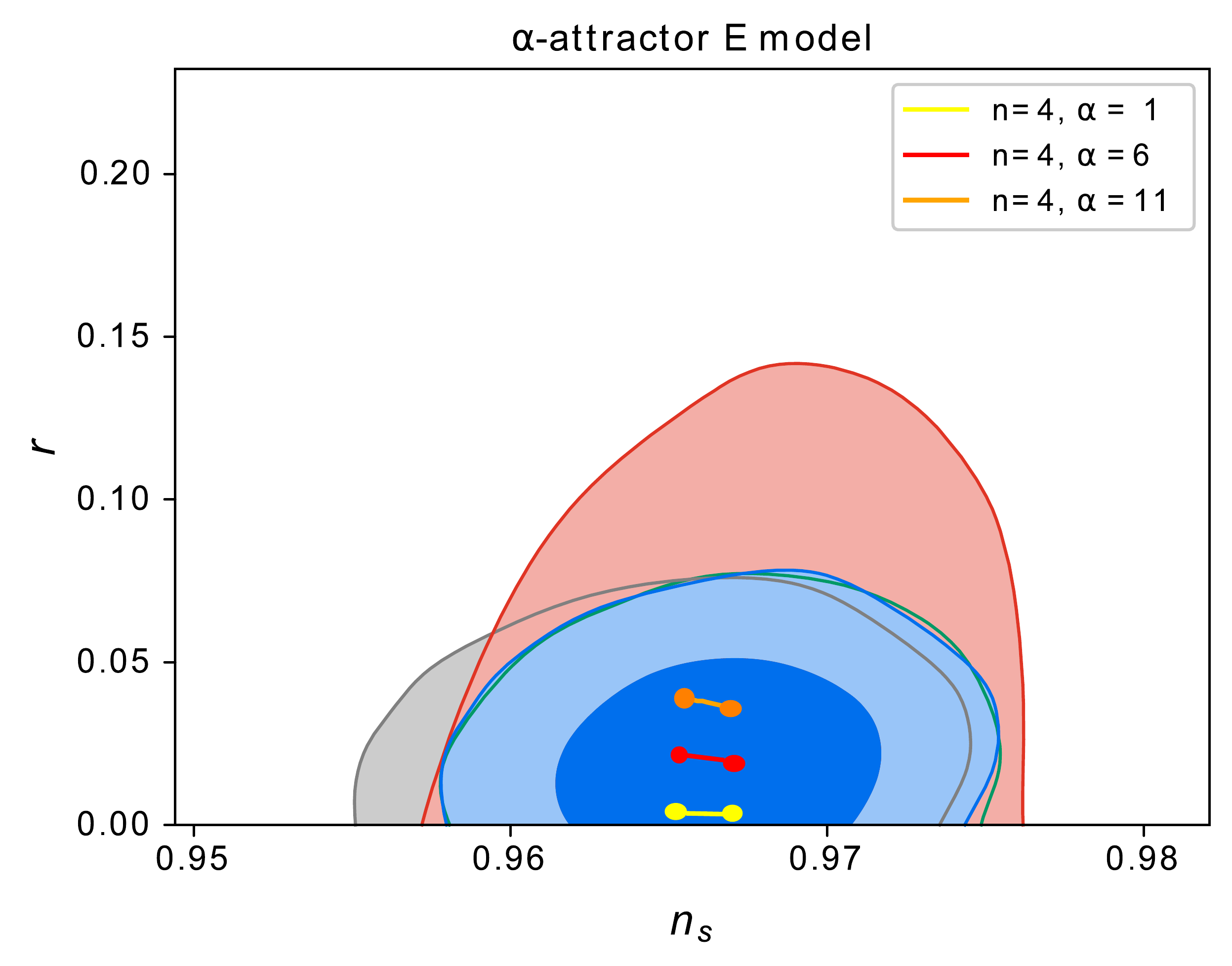}
   \subcaption{}
   \label{fig:E_n_4_a_1611}
\end{subfigure}%
\begin{subfigure}{0.5\textwidth}
  \centering
   \includegraphics[width=75mm,height=70mm]{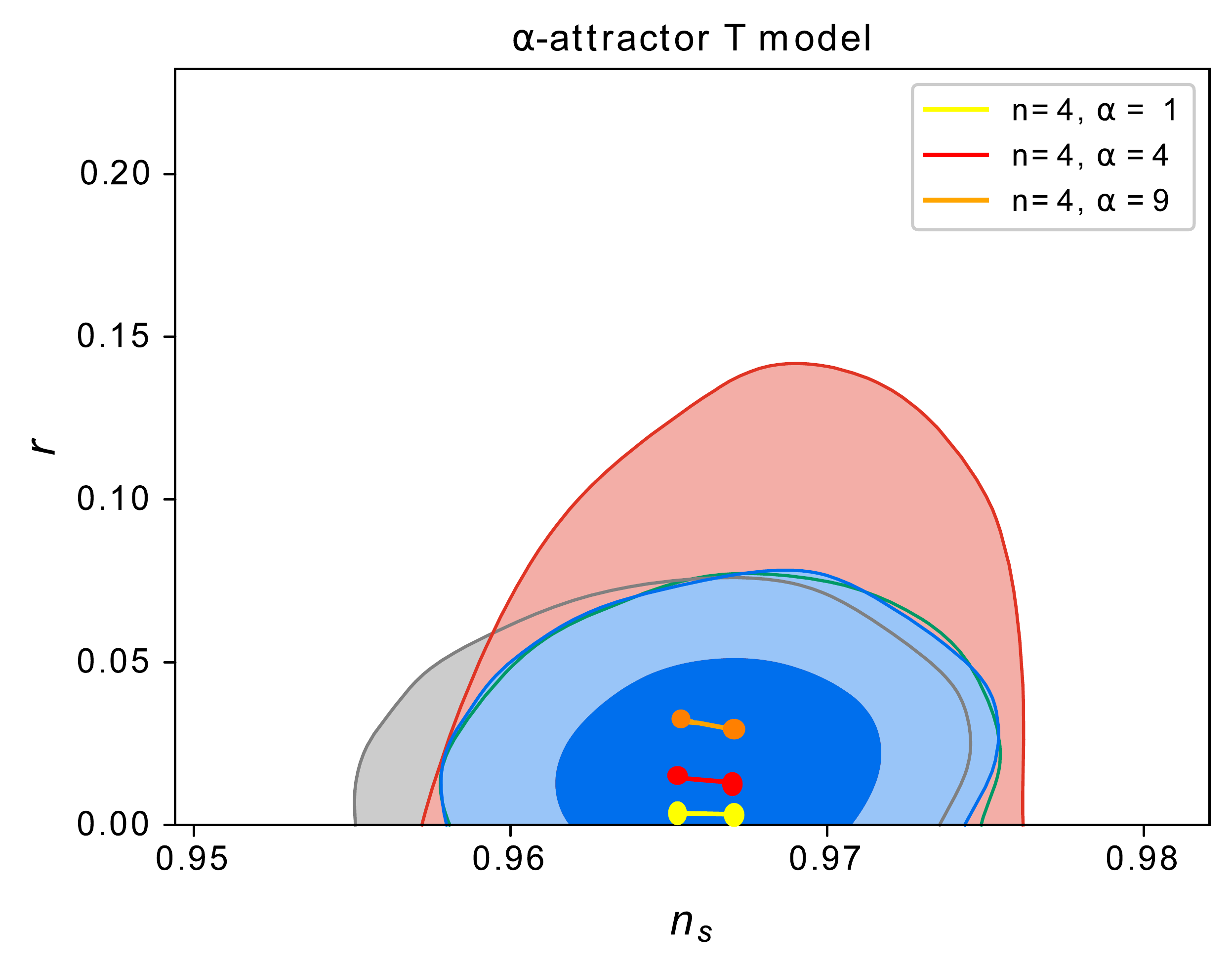}
   \subcaption{}
    \label{fig:T_n_4_a_149}
\end{subfigure}
    \caption{Comparison of  theoretical results with the Planck-2018 data on the $n_{s}$-$r$ plane for the $E$ and the  $T$ models ($n=4$). The theoretical graph is shown by colored lines with two dots at the ends. For the $E$ model (Figures \ref{fig:E_n_4_a_1611}): $\alpha = 1$ (yellow), $6$ (red) and $11$ (orange) lie within $68\%$ CL. For the $T$ model (Figures \ref{fig:T_n_4_a_149}): $\alpha = 1$ (yellow), $4$ (red) and $9$ (orange) lie within $68\%$ CL.}
    \label{fig:Constrain_n_E}
\end{figure}

In Figures \ref{fig:planck data}, \ref{fig:chaotic planck} and \ref{fig:Constrain_n_E}, we have shown how the results of our calculation fit the Planck-2018 data \cite{Akrami:2018odb, Aghanim:2018eyx} in the $n_{s}$-$r$ plane. The lengths of the lines, representing the calculated values, correspond to the range of $k$ values, $0.001$ Mpc$^{-1}$ - $0.009$ Mpc$^{-1}$, from right to left. The experimental data have been obtained as the public chains from the \href{https://pla.esac.esa.int/#cosmology}{Planck Legacy Archive} by running the \href{https://getdist.readthedocs.io/en/latest/}{GetDist}  plotting utility along with the \href{https://jupyter.org/}{Python Jupyter notebook} and \href{https://docs.anaconda.com/anaconda/user-guide/tasks/integration/spyder/}{Spyder integrated development environment}.  
The Planck-2013 \cite{Planck:2013jfk} and Planck-2018 \cite{Akrami:2018odb, Aghanim:2018eyx} data fit well with the idea of the single field inflationary models. The Planck data not only promotes the single field inflaton model but also select the class of potentials in this category, namely the concave potentials \cite{Chen:2017qeh}, with some confidence level. The analysis of these data incorporates effects like gravitational lensing, Baryon Acoustic Oscillations(BAO) and $E$-modes to construct an image of the present universe. We have shown  that our microscopic calculations during inflation in the $k\longrightarrow 0$ limit match with the Planck-2018 data in the $n_s$-$r$ plane within $68\%$  and near $95\%$ CL. This supports the idea that the seed of the present large scale structures of the universe was planted during inflation. In this work we take the $\alpha$-attractor potential and try to find the values of $\alpha$ which are capable to fit the Planck data. In Figure \ref{fig:planck data}, we observe that both in the $E$-model and the $T$-model, $\alpha=1$ and $\alpha=10$ ($n_{s}$-$r$) data lie  within $68\%$ CL. For these values of  $\alpha$  we have made use of (\ref{eq:final_tensor_scalar_ratio}), the low-$\alpha$ limit in the $k$-space. The $\alpha=15$ ($n_{s}-r$) data lie near the $95\%$ CL. For this case we have used (\ref{eq:high_alpha}), which works as the high-$\alpha$ formula in the $k$-space.\par The curvature of the potentials, we use here, are as follows: for the $E$-model, for $\alpha=1$(10),      $\frac{d^{2}V}{d\phi^{2}}=-10.38$ (-0.017) at $\phi=4$, and for the $T$-model,  for $\alpha=1$(10),      $\frac{d^{2}V}{d\phi^{2}}=-0.04$ $(-0.0002)$ at $\phi=5$, showing that, the potentials are concave at these values of $\alpha$. However, for the $E$-model ($T$-model), for $\alpha=15$, $\frac{d^{2}V}{d\phi^{2}}=+0.0116$ $(+0.0049)$ at $\phi=4 (5)$. The $\alpha=15$ case lies in the the convex region with $r=0.1340$ ( $0.1342$) for the $E$-model ($T$-model) at $k=0.002$. These values of $r$ lie above the upper limit of Planck \textit{viz}., $r_{0.002}<0.056$ \cite{Akrami:2018odb}. Thus our results are in conformity with Planck-2018 data which favour single field concave potentials. \par In Figure \ref{fig:Constrain_n_E}, we have repeated our calculations for $n=4$ for $\alpha=1, 6, 11$ ($E$-model) and $\alpha=1, 4, 9$ ($T$-model). The results lie within $68\%$ CL showing that the potentials are concave in this case also. 
Besides the above-quoted values of $\alpha$, we have also got solutions for $\alpha=3, 4$ for $n=2$  and $\alpha=  3, 4, 7$ for $n=4$ for the $E$-model and $\alpha=3, 4, 7$ for $n=2$ and $\alpha=4, 9$ for $n=4$,  for the $T$-model, using the same set of intial conditions as mentioned in Section \ref{sec:Validity}. It may be possible to obtain soultions for more combinations of $\alpha$ and $n$, using different sets of initial conditions, but we have not tried that. The fact that solutions of all $\alpha$'s are not obtained, for the some set of initial conditions, indicates a possible mismatch between the profiles of the inflaton potential perturbations and the metric perturbations, in the first-order perturbation theory.  
\section{Conclusions}
\label{sec:conclusions}
In conclusion, we have done detailed mode analysis of cosmological perturbations including quantum fluctuations during inflation in the sub-Planckian region, with the chaotic $\alpha$-attractor potentials. We have mainly focused on the long wavelength ($k\longrightarrow 0$) regions of the modes which drive  the late time evolution  of the universe \cite{Brandenberger:1990wu}. The role of the short wavelength modes, which are responsible for reheating, is not examined in this paper.\par 

In this paper, we project out the possible experimentally favoured values of the $\alpha$ parameter through chaotic inflationary mode analysis. In addition to selecting the participating modes during inflation, our calculation has another aspect of constraining $\alpha$, which may have quite intricate relationship (see (\ref{eq:final_tensor_scalar_ratio})) with high-scale microscopic scenario, as follows.\par Geometrically $\alpha$ is the inverse curvature of $SU(1,1)/U(1)$ K\"ahler manifold (see (\ref{eq:geometry})), as it also controls the tensor-to-scalar ratio $r$. If $\alpha\longrightarrow 0$, i.e. for an infinitely curved manifold, $r\longrightarrow 0$ and if $\alpha \longrightarrow \infty$, i.e. for a flat manifold $r\longrightarrow \infty$. Therefore the values of $\alpha$ have direct correspondence to those of $r$ and thereby to the geometry (shape and size) of the internal manifold. Now this interplay is complicated both from the theoretical and the experimental point of view. Finding out the exact volume of the internal manifold is an outstanding task for string theoretic models such as Large Volume Scenario (LVS) in the context of moduli stabilisation, since it is related to finding a particular Calabi-Yau manifold out of a large number of possibilities, which are themselves related by dualities such as mirror symmetry. And as stated in Section \ref{sec:intro}, finding $r$ is the aim of the ongoing and the future CMB experiments (Background Imaging of Cosmic Extragalactic Polarization(BICEP/Keck), Suborbital polarimeter for Inflation, Dust, and Epoch of Reionization(Spider), POLARization of the Background Radiation(PolarBear), Atacama Cosmology Telescope(ACT), South Pole Telescope(SPT), Simons Observatory(SO), Lite (Light) satellite for the studies of B-mode polarization and Inflation from cosmic background Radiation Detection(LiteBIRD), CMB-S4)   \cite{LiteBIRD:2020khw, remazeilles2016sensitivity, Fidler:2014oda, Williams:2020hqk}, through the possible  detection of  $B$-mode polarisation of gravitational waves with accuracy never achieved before. It is challenging enough because primordial gravitational waves are very weak and $B$-modes are also very feeble to detect. So we believe that the results of our calculation may throw some light on these issues in future. Interestingly, we have reproduced, in the present study, the values of $\alpha$ and corresponding values of $r$ as predicted by maximal supersymmtery models with $B$-mode target \cite{Kallosh:2017ced}.\par In contrast to our microscopic approach, the statistical model calculations yield a continuous range of values of $\alpha$. For example, $\alpha<19.95$ ($E$-Model) and $\alpha< 10$ ($T$-Model)  obtained by Bayesian analysis           \cite{Akrami:2018odb} and $\alpha=7.56\pm 5.15$ by MCMC analysis \cite{Rodrigues:2021olg}. However, our findings of discrete values of $\alpha$ conform to the predicted values from the supergravity \cite{Kallosh:2017ced,Ferrara:2016fwe} and the type IIB string theory \cite{Kallosh:2021vcf} models, at least partially.


\acknowledgments
The authors acknowledge the grants provided by the University Grants Commission, Government of India, through the CAS-II program and by the Department of Science and Technology, Government of India, through the FIST-II program, for carrying out this research. A.S. and C.S. acknowledge the Government of West Bengal for granting them the Swami Vivekananda fellowship. The authors thank Dr. Amitabha Choudhuri for useful discussions. We sincerely acknowledge a number of illuminating correspondences from Dr. Yashar Akrami.


\newpage
\bibliographystyle{utcaps}
\bibliography{biblio}

\end{document}